\documentclass[prb,superscriptaddress,twocolumn,eqsecnum]{revtex4-1}

\usepackage{hyperref}
\usepackage{amsthm}
\usepackage{graphicx}
\usepackage{amsfonts}
\usepackage[figuresright]{rotating}  
\usepackage{amssymb}
\usepackage{amsmath}
\usepackage{psfrag}
\usepackage{subfigure}
\usepackage{multirow}
\usepackage{tabularx}
\usepackage{bm}

\def\beq{\begin{eqnarray}}
\def\eeq{\end{eqnarray}}

\newtheorem*{criterion}{Criterion}
\newtheorem*{definition}{Definition}

\begin{document}

\title{A discretized abelian Chern-Simons gauge theory on arbitrary graphs}
\author{Kai Sun}
\affiliation{Department of Physics, University of Michigan, 450 Church Street, Ann Arbor, MI 48109, USA}

\author{Krishna Kumar}
\affiliation{Department of Physics and Institute for Condensed Matter Theory, University of Illinois at Urbana-Champaign, 1110 West Green Street, Urbana, IL 61801, USA}

\author{Eduardo Fradkin}
\affiliation{Department of Physics and Institute for Condensed Matter Theory, University of Illinois at Urbana-Champaign, 1110 West Green Street, Urbana, IL 61801, USA}

\begin{abstract}
In this paper, we show how to discretize the abelian Chern-Simons gauge theory on generic planar lattices/graphs (with or without translational
symmetries) embedded in arbitrary 2D closed orientable manifolds. 
We find that, as long as a one-to-one correspondence between vertices and faces can be defined on the graph such that each face is paired up with 
a neighboring vertex (and vice versa), a discretized abelian Chern-Simons theory can be constructed consistently.
We further verify that all the essential properties of the Chern-Simons gauge theory are preserved in the discretized setup.
In addition, we find that the existence of such a one-to-one correspondence is not only a sufficient condition for discretizing a Chern-Simons
gauge theory but, for the discretized theory to be nonsingular and to preserve some key properties of the
topological field theory, this correspondence is 
 also a necessary one. A specific example will then be provided, in which we discretize the abelian Chern-Simons gauge theory on a tetrahedron. 
\end{abstract}
\date{\today}
\maketitle 

\section{Introduction}
\label{sec:introduction}

As the prototypical topological field theory, the Chern-Simons gauge theory,  has had a deep and broad impact on 
a wide range of physics research, ranging from knot theory\cite{Witten1989} and parity anomalies in quantum field theory\cite{Deser-1982} to the theory 
of the integer and fractional quantum Hall effects\cite{Zhang-1989,Lopez-1991,Wen-1995} and  the effective field theory description of chiral spin liquids\cite{Chen-1989,Wen-1989} in condensed matter physics (for a review see, {\it e.g.} Ref. [\onlinecite{Fradkin-2013}]). 
Although well understood as a continuum field theory, there is still  limited understanding on how to discretize this topological 
field theory on 2D lattices or graphs.\cite{Fradkin1989} This task turns out to be highly nontrivial. In particular, the topological  and gauge-theoretic nature of the Chern-Simons gauge theory
enforces strong constraints on the dynamics of the gauge fields. These constrains, if not treated carefully, can  result in inconsistencies
in the discretized theory, making the theory ill-defined.\cite{Eliezer1992a,Eliezer1992b} 
Until recently, the discretization has only been done  only for a very special case, i.e., on a square lattice (with only nearest bonds) 
embedded in a torus.\cite{Eliezer1992a,Eliezer1992b} 
It remains highly unclear whether similar construction can be extended for other lattices, or for lattices embedded 
on other 2D manifolds aside from the torus, or in any discretized systems without translation symmetries (e.g. a graph). In a recent publication,\cite{Kumar-2014}
we presented a consistent construction of the Chern-Simons gauge theory on one of the simplest non-bipartite lattices in two-dimensions, the Kagome lattice,
and used it to study the magnetizations plateaus of the spin-1/2 frustrated quantum Heisenberg antiferromagnet on this  lattice. 

The main purpose of the present paper is to develop a consistent discretization of the (abelian) Chern-Simons abelian U(1) gauge theory on general planar lattices and graphs. There are several motivations to search for a discretized Chern-Simons gauge theory on generic lattices/graphs. 
For example, it has been known that the Chern-Simons gauge theory plays a crucial role in the study of chiral spin 
liquid. 
Such an exotic state of matter can only be stabilized in the presence of strong geometric frustration. Much of the work in frustrated antiferromagnets uses the fact that these systems are equivalent to a system of (generally interacting) hard-core bosons on the same lattice. The hard-core bosons are then mapped into a system of fermions coupled to  a discretized Chern-Simons gauge field.\cite{Fradkin1989} 
Except for some very special exactly solvable models, in the study of such frustrated systems the dynamics and quantum fluctuations of the effective gauge fields are typically ignored, and  frustrated quantum antiferromagnets are frequently described only at the level of
 the average field approximation.\cite{Yang1993,Misguich2001}
  However, such a approximation is unreliable, and has a strong and obvious bias towards time-reveal breaking ground states.
As shown in Ref.[\onlinecite{Lopez1994}], to correctly address the competition
between different quantum ground states,  it is necessary to go beyond the average field approximation
by carefully introducing the correct quantum dynamics.

It was recently realized that, in addition to the well known case of two-dimensional electron gases  (in the continuum) in strong magnetic fields, 
the fractional quantum Hall effect can also been stabilized in lattices even with zero net magnetic
field.\cite{Tang2011,Sun2011,Neupert2011,Sheng2011,Regnault2011,Parameswaran2013}
Fractional quantum Hall states have also been explored on lattice systems earlier on\cite{Kol1993,Moller2009}  and, more recently, also in optical lattice systems.\cite{Sorensen2005,Palmer2006,Hafezi2007,Palmer2008}
This type of (discrete) fractional topological states is now often referred to as the fractional Chern insulators or 
the fractional quantum anomalous Hall state.
In particular, it has been shown that these (discrete) fractional Chern insulators are adiabatically connected to 
the corresponding fractional quantum Hall states in the continuum.\cite{Wu2012,Scaffidi2012,Wu2015}

The Chern-Simons gauge theory is known to be the low-energy, hydrodynamic, theory of topological phases such as the 
fractional quantum Hall fluids,\cite{Wen-1995} they are also expected to describe the low-energy and long-distance limit of  topological Chern insulators, fractional or not.\cite{Qi2008,Cho2011,Chan2013,Hansson2015,You2013} 
Although Chern-Simons theories yield a natural description of the hydrodynamic behavior of topological phases, apriori they are not absolutely necessary in the microscopic construction of a theory of such states.\cite{Wen-1995} However, all known fractionalized phases are in inherently strongly coupled systems and, notably in the case of the fractional quantum Hall fluids, the use of Chern-Simons gauge theory in the microscopic derivations has been invaluable.\cite{Zhang-1989,Lopez-1991}
Aside from some recent and promising work,\cite{Murthy2011,Murthy2012} it is not yet clear  what role does Chern-Simons gauge theory  play in the theory of fractional topological Chern insulators. Although adiabatic continuity strongly implies that the theory of factional Chern insulators should be smoothly related to the theory of the fractional Hall effect on lattice systems,\cite{Kol1993,Moller2009} where discrete Chern-Simons gauge theory is expected to be generally applicable. 
The general answer to these questions has remained  problematic in view of  the fact that the Chern-Simons gauge theory has not been discretized on most of the lattices
on which lattice fractional Chern insulators are known to occur (e.g. the checkerboard lattice, the Kagome lattice and the multiple-orbital square lattice). Although we will not give an answer for systems on lattices as general as it is needed, we will give an explicit construction for a large class of lattices, which include some of clear physical interest.

In this paper, we propose and study a discretized Chern-Simons gauge theory on generic planar graphs embedded in arbitrary 2D closed and orientable 
manifolds.\cite{note_orientable}
Same as in a lattice gauge theory, we will define the space components of the  gauge field to live on the nearest-neighbor bonds of the lattice and the time components on the sites (vertices) of the lattice. We will consider only planar lattices (and hence with only non-crossing bonds).
This is a lattice gauge theory\cite{Wilson-1974,Kogut-1975,kogut-1979,Creutz-1983} but one with a broken time reversal and parity invariance. Here we will be interested in a version of Chern-Simons theory on a a class of spatial lattices with continuous time. Earlier work focused on the square lattice,\cite{Fradkin1989,Kantor1991,Eliezer1992a,Eliezer1992b} and recently we discussed the case of the Kagome lattice.\cite{Kumar-2014} Discretized versions of Chern-Simons gauge theory have been discussed both in Euclidean space-time lattices,\cite{Frohlich1989} which suffer from the species doubling problem analogous to those of lattice fermions.
By enlarging the scope of investigation from periodic lattices to graphs (with or without
translational symmetries), our conclusions are generally applicable for a wide range of systems.

We will require the discretized theory to retain the central features a topological field theory. Chern-Simons gauge theory on a continuous space-time manifold has several key features.\cite{Witten1989} It is gauge theory which means that it has a local symmetry under local (in space-time) gauge transformations. At the quantum level this requires that the quantum states of the physical Hilbert space be gauge invariant,\cite{Dirac-1966} and hence that the generators of local time-independent gauge transformations must generate superselection sectors, i.e. the Gauss law is satisfied as constraint on the physical space of states. For this requirement to be consistently implemented, the generators of local gauge transformations must commute with each other on different spatial locations. In the case of the Chern-Simons theory this implies that the local magnetic flux must commute with each other (since they are the generator of time-independent gauge transformations). This condition imposes stringent constraints on the possible form of the discretized theories,\cite{Eliezer1992a,Eliezer1992b} and it is the main focus of this work.

On the other hand, at the classical level, the Chern-Simons theory is topological in the sense that the action is invariant under general coordinate transformations and hence it is independent of the metric. A consequence of this feature is that the energy-momentum tensor  classically vanishes and, consequently, the Hamiltonian is also zero. Clearly, any lattice discretization implies a choice of coordinates, i.e. a fixed spatial metric. Furthermore, a change of the lattice stricture leads to a change in the form of the metric. Therefore, a lattice version of Chern-Simons theory cannot be explicitly independent of the metric and, in this sense, it cannot be formally topological. However, we will show below, that one can construct a Chern-Simons gauge theory for a large class of lattices a U(1) lattice gauge theory with continuous time (i.e. in ``Hamiltonian'' form\cite{Kogut-1975}) which is gauge-invariant. We will see that the resulting discretized theory nevertheless has a vanishing Hamiltonian since the content of the action reduces to a set of (reasonably local) equal-time commutation relations and a set of local and commuting constraints. This theory is topological in the sense that it does not have local excitations, and that only the global degrees of freedom (non-trivial Wilson loops) matter. In the long-wavelength limit the discretized theory becomes (formally) the continuum Chern-Simons gauge theory.

\begin{figure}[t]
\includegraphics[width=.64\linewidth]{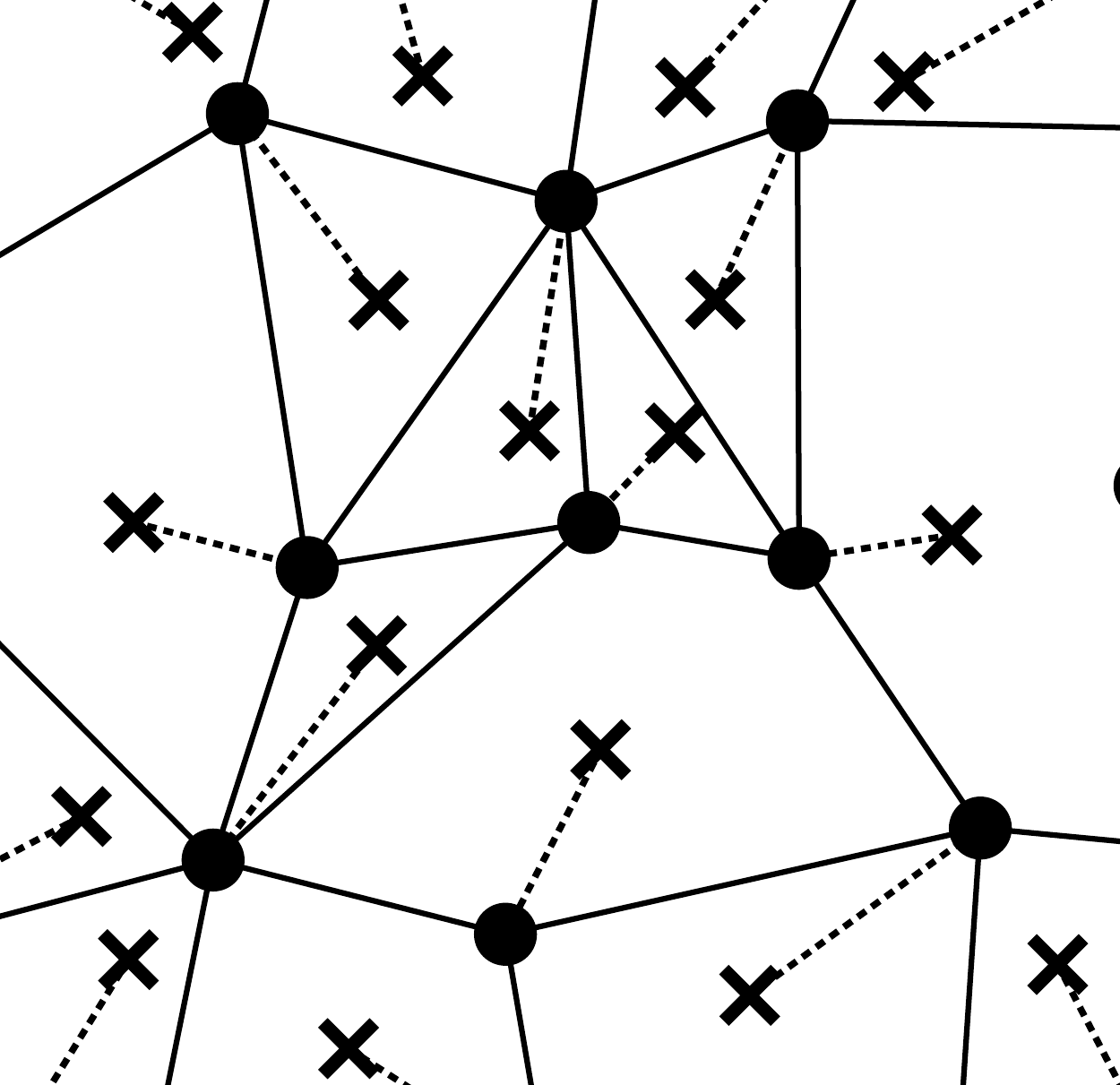}
\caption{Part of a planar graph, on which a local vertex-face correspondence is defined. 
The disks and solid lines represent vertices and edges of the graph respectively. 
Each face is marked by a cross. The local vertex-face correspondence is indicated by dotted lines, 
that pair up each face with one (and only one)  adjacent vertex.}
\label{fig:local_correspondence}
\end{figure}

We find that such a Chern-Simons gauge theory can be constructed for 
arbitrary 2D planar  graphs (lattices) as long as a {\it local vertex-face correspondence} can be defined on the graph/lattice. We adopt the following definition:
\begin{definition}
A local vertex-face correspondence is a one-to-one correspondence between faces and vertices defined on a graph 
such that every vertex is adjacent to its corresponding face (and vice versa).
\label{def:lvfc}
\nonumber
\end{definition}
\noindent
An example of such a correspondence is shown in Fig.~\ref{fig:local_correspondence}.

The relevance of this correspondence to Chern-Simons theory lies in the nature of this gauge theory. In the continuum the (abelian) Chern-Simons Lagrangian of a gauge field $\mathcal{A}_\mu$ in 2+1 dimensions is (coupled to a matter current $J_\mu$)
\begin{equation}
\mathcal{L}_{CS}[\mathcal{A}]= \frac{k}{4\pi} \epsilon^{\mu \nu \lambda} \mathcal{A}_\mu \partial_\nu \mathcal{A}_\lambda- J_\mu \mathcal{A}^\mu
\label{eq:SCS}
\end{equation}
The Chern-Simons (CS) gauge theory is a topological field theory.\cite{Witten1989} At the classical level the CS action is independent on the metric of the manifold on which it is defined. The content of this Lagrangian, Eq\eqref{eq:SCS}, is seen in  Cartesian components
\begin{align}
\mathcal{L}_{CS}[\mathcal{A}]=&  \frac{k}{2\pi} \mathcal{A}_0 \mathcal{B} - J_0 \mathcal{A}_0
\nonumber\\
&- \frac{k}{4\pi} \epsilon_{ij} \mathcal{A}_i \partial_t \mathcal{A}_j - {\bm J} \cdot {\bm {\mathcal{A}}}
\label{eq:SCS-components}
\end{align}
At the quantum level, the first term of the r.h.s. becomes the requirement that the states in the physical Hilbert space, $\{ \vert \textrm{Phys}\rangle \}$,  obey the ``Gauss law'' as a local constraint. Thus, the physical states are gauge-invariant and are annihilated  by the generator of local gauge transformations,
\begin{equation}
\Big[\frac{k}{2\pi} \mathcal{B}(\bm x)-J_0(\bm x)\Big] \vert \textrm{Phys} \rangle=0
\label{eq:GaussCS}
\end{equation}
Hence, the physical states are those in which the local charge density $J_0$ to the local magnetic flux  $\mathcal{B}=\epsilon_{ij} \partial_i \mathcal{A}_j$ are precisely related, i.e. flux attachment. The second term of the r.h.s. of Eq.\eqref{eq:SCS-components} implies that the components of the gauge field obey the equal-time commutation relations,
\begin{equation}
\left[ \mathcal{A}_i(\bm x), \mathcal{A}_j(\bm y)\right]=i \frac{2\pi}{k} \epsilon_{ij}\delta({\bm x}-{\bm y})
\end{equation}
Further, the Hamiltonian of this system is zero unless sources are present, i.e. 
\begin{equation}
\mathcal{H}={\bm J} \cdot {\bm {\mathcal{A}}}
\end{equation}
which is a consequence that a topological field theory does not have any  excited states with finite energy.

As with any lattice gauge theory, in a  discretized Chern-Simons gauge theory the gauge fields (which are connections and hence are 1-forms) are naturally defined on the links of the lattice while the matter fields are defined on the sites of the lattice.\cite{Wilson-1974,Kogut-1975} The field strength is a 2-form and it is defined on the elementary plaquettes of the lattice. While in a conventional  lattice gauge theory the lattice is generally hypercubic (i.e. square in 2D), here we will consider more general (and translationally invariant) planar lattices. For instance, in Ref.[\onlinecite{Fradkin1989}] (and in Refs. [\onlinecite{Eliezer1992a,Eliezer1992b}]) the Chern-Simons theory was defined on a square lattice and in Ref.[\onlinecite{Kumar-2014}] the theory was defined on a Kagome lattice. In both cases the Gauss law of Eq.\eqref{eq:GaussCS} is naturally implemented as a constraint that relates the occupation number of a site (or vertex) to the gauge flux through a (uniquely defined) adjoint plaquette (or face). While in the case of the square lattice all plaquettes are identical (squares), in the case of the Kagome lattice has three inequivalent sites in its unit cell and, correspondingly, three faces (two triangles and a hexagon) in its unit cell. Nevertheless, the correspondence of vertices to faces is one-to-one in both lattices. 

We will see here that this correspondence  is a key feature which will allow us to impose the constraint (and hence gauge invariance) in a unique way which, in addition, does not break the point group (or space group) symmetries of the lattice. 
Below we will find a construction of the Chern-Simons gauge  theory on lattices for which for a charge located at a vertex, 
the magnetic field attached to it by the Chern-Simons gauge theory is located at the face that is naturally paired up with this vertex. 

Whether or not a local vertex-face correspondence can be defined for a graph is fully determined by the connectivity of the graph.
In Sec.~\ref{sec:lvfc}, we will provide an sufficient and necessary condition, which can be used to decide whether
such a correspondence exists or not for an arbitrary graph.  
In Fig.~\ref{fig:lattices_CS}, we show some examples of lattices that support such a correspondence 
(i.e. a discretized Chern-Simons gauge theory can be constructed on these lattices). 
These examples include some of the lattices used in the study of chiral spin liquids and the lattice fractional quantum Hall effect 
(e.g. the Kagome lattice).

It is also worthwhile to emphasized that in the continuum, the Chern-Simons gauge theory
can be defined on arbitrary 2D closed and orientable manifolds. This  plays a critical role in the phenomenon 
of topological degeneracy in fractional quantum Hall systems~\cite{Wen1990}. In addition, it is also known that all the essential physics 
of the Chern-Simons gauge theory (in the continuum) is stable against the explicit breaking of the translational symmetry, 
which is the underlying reason for the stability of the quantum Hall states against weak disorders. 
On the discretized side, however, it is still  unclear whether the Chern-Simons gauge theory can be defined on any 
manifold aside from a torus and/or on a discrete graph without translational symmetries. Our study will provide an answer to these questions.

In addition to those geometric considerations, a key consistency requirement of the gauge theory is that the lattice version of the local constraints of Eq.\eqref{eq:GaussCS} must commute with each other and hence act as superselection rules on the Hilbert space~\cite{Eliezer1992a,Eliezer1992b} 
(otherwise, these constraints cannot be simultaneously satisfied). 
This consistency condition places restrictions on the commutation relations satisfied by the gauge fields defined on the links. For the square lattice this problem was solved by Eliezer and Semenoff,\cite{Eliezer1992a,Eliezer1992b} and was more recently generalized by us to the case of the Kagome lattice.\cite{Kumar-2014} In this paper we will show that the commutation relations can be defined consistently on any lattice (and graph) which obeys the one-to-one correspondence between vertices and faces. We will show that this restriction is implemented in terms of a suitably defined non-singular (and hence invertible) matrix. Therefore, the lattice Chern-Simons theory can be defined as a consistent gauge theory at the quantum level on these planar lattices and graphs.


This paper is organized in the following way. In Sec.~\ref{sec:lvfc}, we present a necessary and sufficient criterion 
for determining whether 
a local vertex-face correspondence can be defined for an arbitrary graph/lattice. In Sec.~\ref{sec:CS-action}, we write down 
the action of the discretized Chern-Simons gauge theory for generic graphs with a local vertex-face correspondence. 
In Secs.~\ref{sec:gauge_invariance}--\ref{sec:locality}, 
we prove that our discretized gauge theory preserves all key features of the Chern-Simons gauge theory, 
including the gauge invariance, flux attachment, commutation relations, 
duality transformation and the locality condition. In Sec.~\ref{sec:Nv_Nf}, we show that the existence of a local
vertex-face correspondence is the necessary condition for discretizing the Chern-Simon theory, if we want the theory to be nonsingular and to preserve the  key properties of the Chern-Simons gauge theory.
In Sec.~\ref{sec:tetra} we present a simple example by discretizing the Chern-Simons gauge theory on a tetrahedron, 
which is a 2D planar graph on a sphere. 
 In Sec.~\ref{sec:discussion} we conclude our paper by discussing open problems and applications of this theory to a number of systems of interest.  Details of the calculations are presented in several appendices.

\section{the local vertex-face correspondence}
\label{sec:lvfc}

\begin{figure}[t]
        \subfigure[]{\includegraphics[width=.48\linewidth]{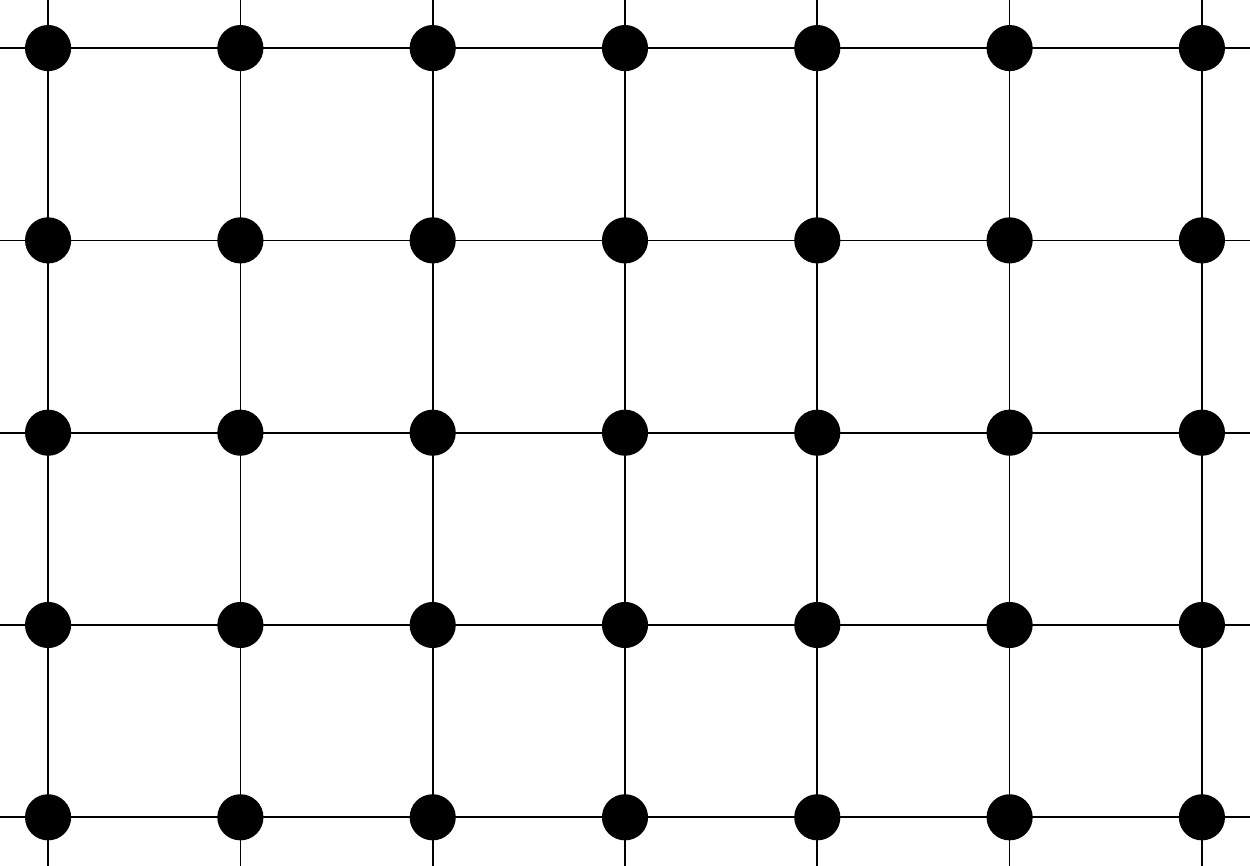}
                \label{fig:square}}
        \subfigure[]{\includegraphics[width=.48\linewidth]{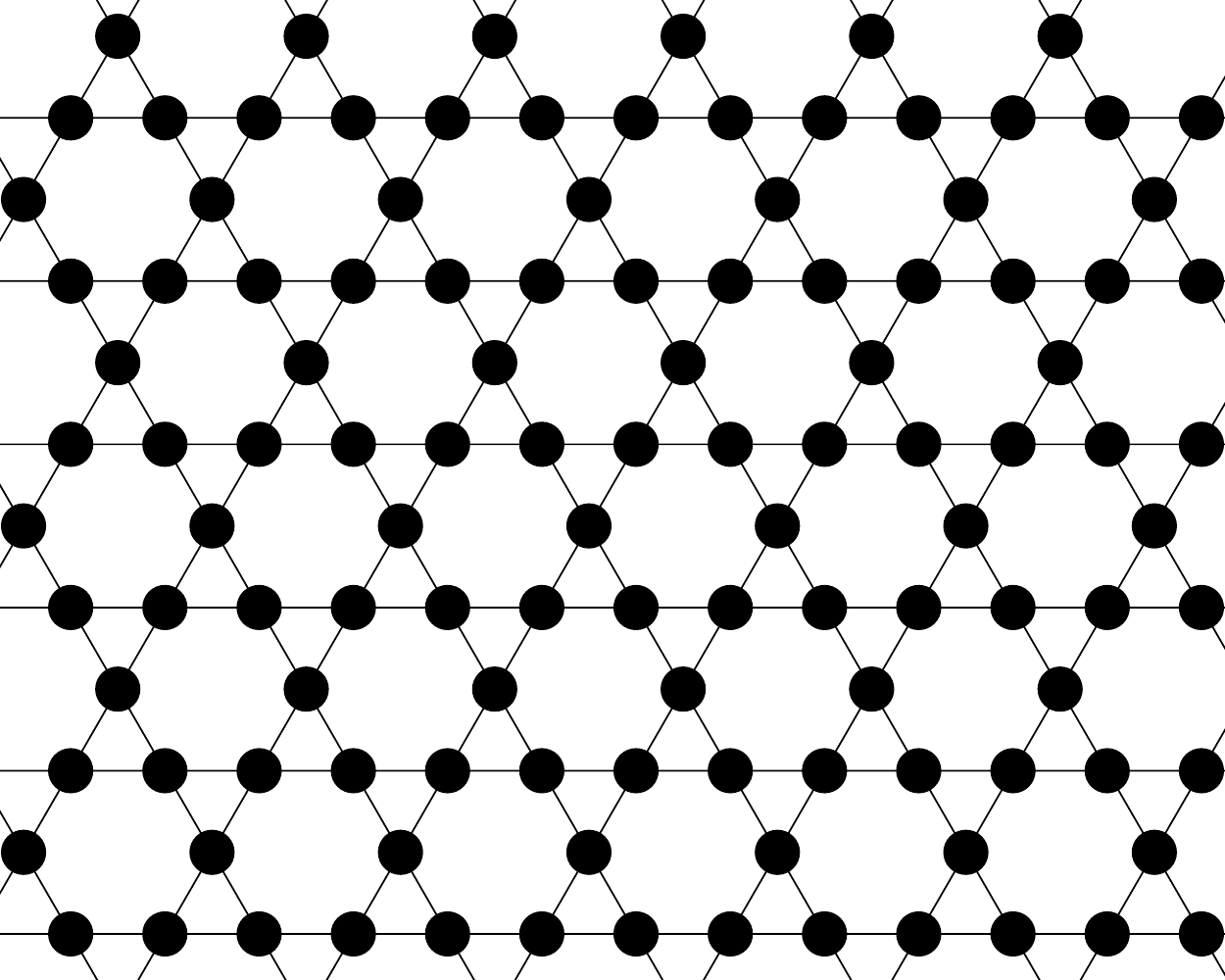}
                \label{fig:kagome}}
        \subfigure[]{\includegraphics[width=.48\linewidth]{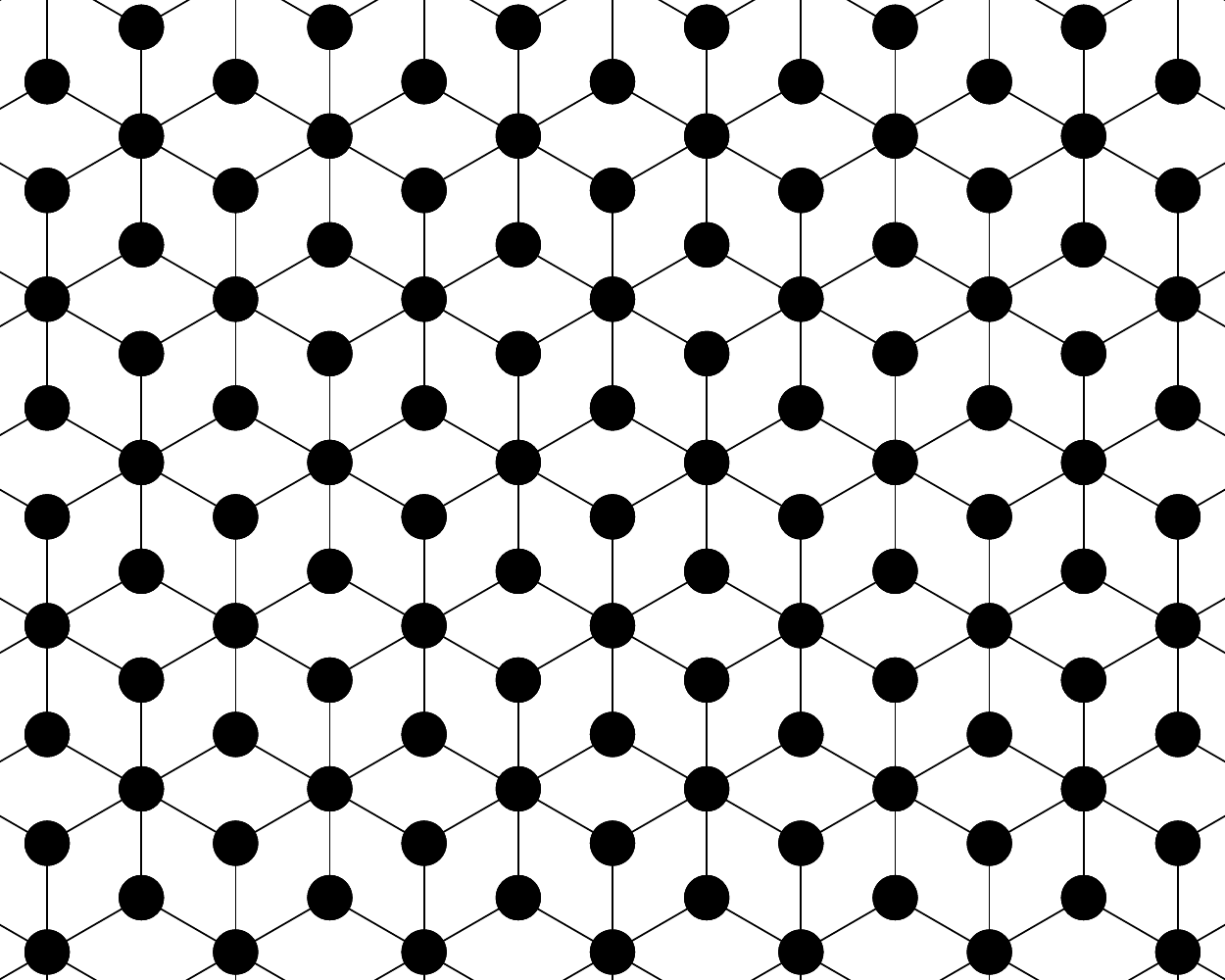}
                \label{fig:dice}}
        \subfigure[]{\includegraphics[width=.48\linewidth]{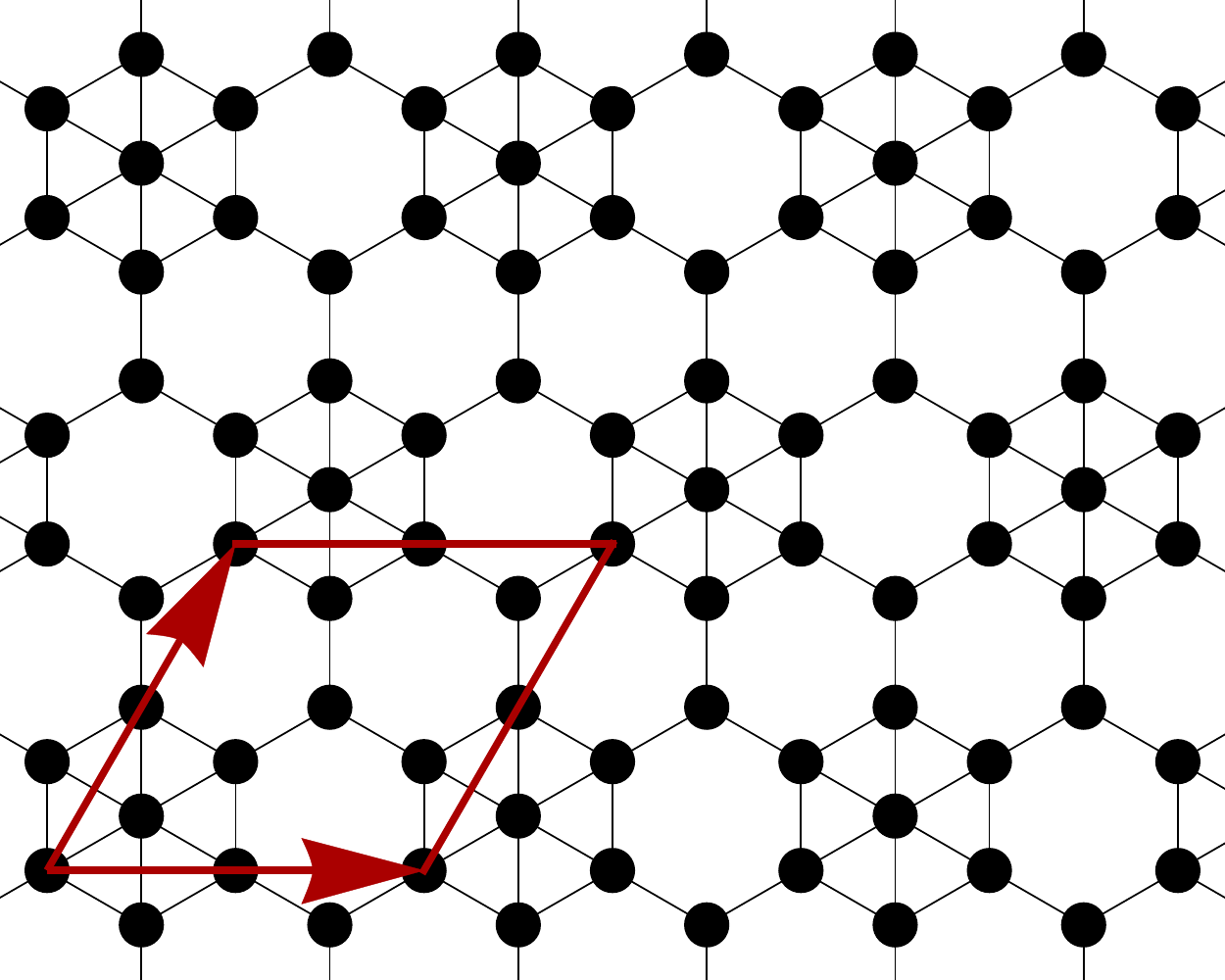}
                \label{fig:honeycomb1}}
\caption{(Color online) Examples of lattices and graphs that support local vertex-face correspondences. 
(a) a square lattice with $1$ vertex and $1$ face per unit cell, (b) a Kagome lattice with $3$ vertices and $3$ faces per unit cell, (c) a dice lattice with 
$3$ vertices and $3$ faces per unit cell and (d) a lattice that contains $9$ vertices $18$ edges and $9$ faces per unit cell. 
The (red) parallelogram marks a unit cell with lattice vectors indicated by the two (red) arrows. 
It is easy to verify that for all these lattices $N_v=N_f$ and for any subgraphs the number of faces never exceeds the number of vertices,
which is a sufficient and necessary condition for the existence of (at least) one local vertex-face correspondence.}
\label{fig:lattices_CS}
\end{figure}

\begin{figure}[hbt]
        \subfigure[$N_v<N_f$]{\includegraphics[width=.48\linewidth]{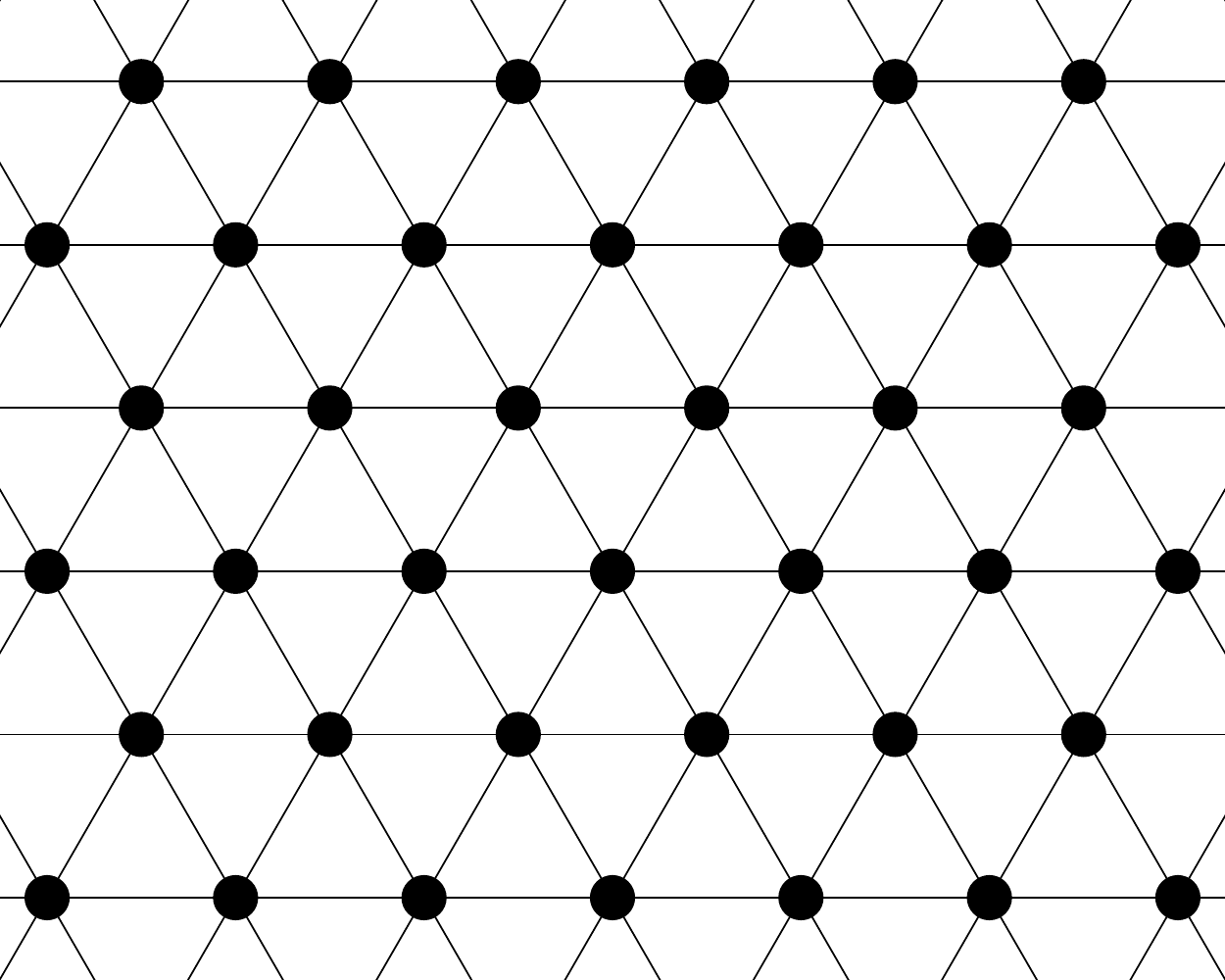}
                \label{fig:triangular}}
        \subfigure[$N_v>N_f$]{\includegraphics[width=.48\linewidth]{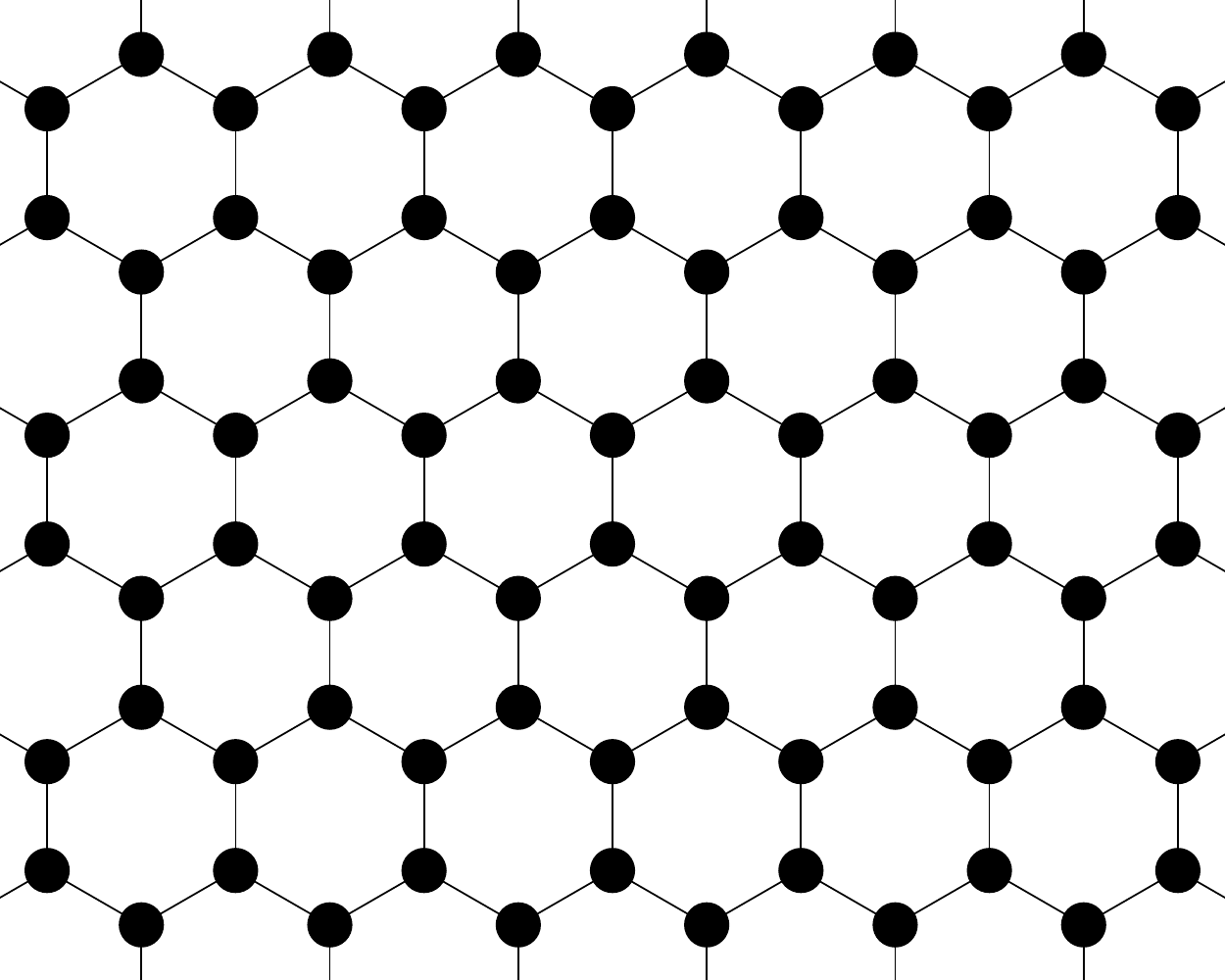}
                \label{fig:honeycomb}}
	\subfigure[$N_f=N_v$]{\includegraphics[width=.48\linewidth]{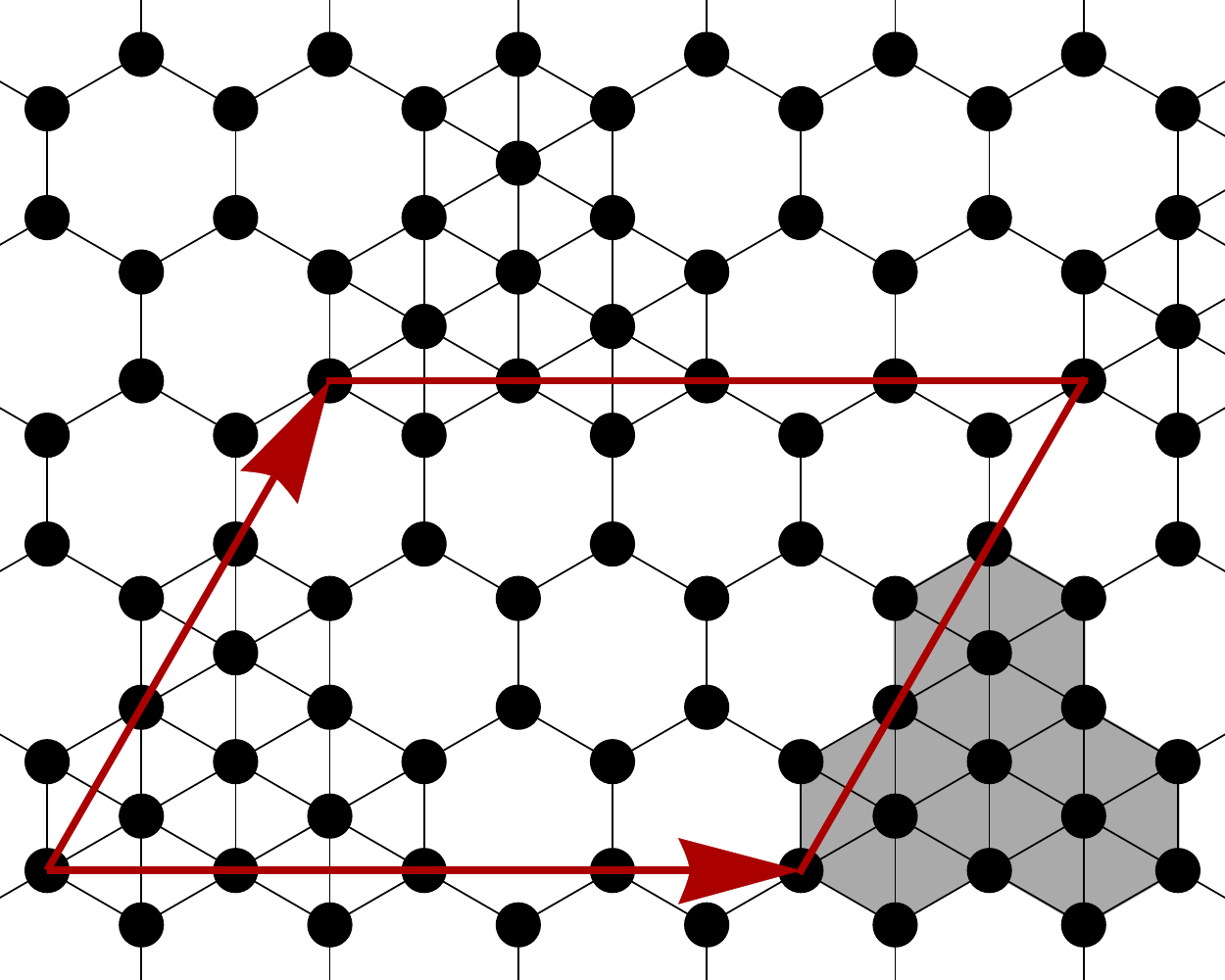}              
                \label{fig:honeycomb2}}
\caption{(Color online) Examples of lattices/graphs that do not support a local vertex-face correspondence. (a) a triangular lattice, which has $1$ 
vertex and $2$ faces per unit cell ($N_v<N_f$), 
(b) a honeycomb lattice, which has $2$ vertices and $1$ face per unit cell ($N_v>N_f$) 
and 
(c) a lattice with $N_v=N_f$ but some of the subgraph has more faces than vertices, e.g. 
 the dark area, which has $18$ faces and $16$ vertices. 
The (red) parallelogram marks a unit cell with lattice vectors indicated by the two (red) arrows. 
Each unit cell of this lattice contains $27$ vertices, $54$ edges and $27$ faces ($N_v=N_f$).}
\label{fig:lattices_no_CS}
\end{figure}

We start our discussion by presenting all constrains and assumptions that will be enforced on the graphs (and lattices) that we will consider.
In this paper, we study generic {\it planar simple} graphs embedded on arbitrary closed and orientable 2D manifolds. 
Here, ``planar'' indicates that the graph can be drawn on a 2D manifold without any crossing bonds, 
while ``simple'' means no multiple bonds connecting the same pair of sites and 
no bond connecting a site to itself (See Fig.~\ref{app:fig:not_simple_graphs} in Appendix \ref{app:simple} for an explicit  example).
The ``simple'' condition is automatically implied for most (if not all) lattices studied in physics, 
while the ``planar'' condition holds for many (but  not all) of them.

For a planar graph $G$, we can construct the dual graph $G^*$ by mapping vertices to faces, and vice versa. 
Because,  as will be discussed below, the dual graph will be needed for the dual gauge theory, we will also require 
the dual graph $G^*$ to be {\it simple}. For the original graph $G$, this condition implies that 
$G$ cannot contain any dangling bonds, and that two faces in $G$ can share at most one common edge.

From now on, we will focus our study on graphs, on which a local vertex-face correspondence can be defined.
Below, in Sec.~\ref{sec:Nv_Nf}, we will prove that this constrain is necessary
in order to preserve certain key defining properties of the Chern-Simons gauge theory.

To determine whether a graph can support such a local vertex-face correspondence, we will use the 
following criterion: 
\begin{criterion}
A local vertex-face correspondence can be defined on a 2D planar graph $G$, if and only if the graph 
has the same number of vertices and faces (i.e. $N_v=N_f$), and that for any subgraph of $G$ the number of faces 
never exceeds the number of vertices (i.e. $N_v' \ge N_f'$).
\label{criterion:lvfc}
\end{criterion}
That this criterion is a sufficient and necessary condition is  proved in Appendix \ref{app:criterion:lvfc}.
The proof  utilizes {\it Hall's marriage theorem} by mapping the local vertex-face correspondence to 
{\it  Hall's marriage problem}.~\cite{Wilson1998}
The marriage theorem is named after the British mathematician,
Philip Hall, who should not be confused with the physicist Edwin Hall, after whom the Hall effect is named.

Using this criterion it is straightforward to determine whether or not a graph or lattice can support a local vertex-face
correspondence. In Fig.~\ref{fig:lattices_CS} (Fig.~\ref{fig:lattices_no_CS}), 
we provide examples of lattices/graphs, on which such a local correspondence exists (does not  exist).
In Fig.~\ref{fig:lattices_no_CS}, the first two lattices do not support any one-to-one correspondence between vertices and faces, 
because the number of faces does not  match the number of vertices. 
The third example, shown in Fig.~\ref{fig:honeycomb2}, has the same number of faces and vertices
and thus, in principle, a one-to-one correspondence between vertices and faces could be defined. However, in this case
such a correspondence cannot be local, as proven in Appendix \ref{app:criterion:lvfc}, because this lattice contains some subgraph, whose number of faces 
exceeds the number of vertices. For example, the dark area in Fig.~\ref{fig:honeycomb2} shows a subgraph with $18$ faces 
and $16$ vertices.

In Figs.~\ref{fig:kagome} and~\ref{fig:dice}, the two lattices are dual to each other. 
Generically, if a graph $G$ has a local vertex-face correspondence, so does its dual graph $G^*$. This is because one can
construct such a correspondence for $G^*$ by simply swapping the vertices and faces in the original 
vertex-face correspondence defined on $G$. As a result, our discretized Chern-Simons gauge theory 
always arises in pairs (one on the graph  $G$ and the other on the dual graph $G^*$). 
In Sec.~\ref{sec:dual}, we will prove that these 
two gauge theories are dual to each other. 
This duality relation is different from the continuum, in which the Chern-Simons theory is self-dual.
A discretized Chern-Simons gauge theory is in general not self-dual, 
unless the underlying graph is self-dual. One example of a self dual graph is shown in 
Fig.~\ref{fig:square}, i.e., a square lattice. Another self-dual graph will be presented 
in Sec.~\ref{sec:tetra}, i.e., a tetrahedron.

We conclude this section by highlighting some conventions adopted in this paper. For a graph $G$, we label
the numbers of vertices, edges, and faces as $N_v$, $N_e$ and $N_f$ respectively,
and we use the subindices $v$, $e$ and, $f$ to label each vertex, edge and face, respectively, where $v$, $e$ and $f$ take integer values 
($1 \le v \le N_v$, $1\le e \le N_e$ and $1\le f \le N_f$). For the dual graph $G^*$, we will use 
the ``$*$'' symbol  to label every object. For example, vertices, edges, and faces of the dual graph are 
labeled as $v^*$, $e^*$ and $f^*$, respectively. In addition, for convenience, if a vertex $v$ in graph $G$ is mapped 
to the face $f^*$ in the dual graph, we will use the same integer to label them, i.e., $v=f^*$. 
Same is true for corresponding $e$ and $e^*$ ($f$ and $v^*$).
Throughout the paper, repeated indices will be summed over unless specified otherwise. 
For the gauge field, the time-component lives on vertices and thus will be labeled as $A_v$. 
The spatial components (i.e. the vector potential) are defined on edges, and thus will be shown as $A_e$. 
Because the vector potential is a vector, we must choose a positive direction for each edge 
(from one of its end to the other). 
The vector potential $A_e$ on an edge $e$ is positive (negative), if it is along (against) the direction of the edge $e$.
In graph theory, after a direction is assigned to each edge, the graph is called a directed graph (or a digraph).\cite{Wilson1998}

\section{The discretized Chern-Simons action}
\label{sec:CS-action}

In this section, we construct the action of the discretized Chern-Simons gauge theory.
We should emphasize that as long as the conditions discussed in the previous section are satisfied, 
our construction is applicable for arbitrary graphs.
In addition, as will be shown below, the discretized gauge theory obtained here is a topological field theory, 
whose action only relies on the connectivity of the graph without any free parameter, except for a quantized
topological index~$k$.
 
\subsection{The $M$ matrix and the $K$ matrix}
\label{sec:M-K}

In this section, we define two matrices for arbitrary graphs with a local vertex-face correspondence.
For a graph satisfying the criterion given in the previous section typically there is more than one way to 
define the local vertex-face correspondence, and different choices here will in general result in different $M$
and $K$ matrices and thus lead to slightly different actions. Here we choose an specific (albeit arbitrary) one, 
consistently throughout the lattice.

The vertex-face correspondence defines a matrix $M_{v,f}$ with dimensions $N_v\times N_f$.
The first index of this matrix runs over all vertices, while the second one indicates faces in the graph. 
If vertex $v$ and face $f$ are paired-up according to the vertex-face correspondence, then $M_{v,f}=1$. Otherwise, the matrix element is zero. Hence,
\begin{align}
M_{v,f}=\left\{
\begin{array}{cl}
	1 & \textrm{if $v$ is paired with $f$} \\
	0 &  \textrm{otherwise}
\end{array}\right.
\end{align}
Because the vertex-face correspondence requires $N_v=N_f$, the matrix $M$ is a square matrix. 
In addition, it is easy to realize that, by definition, $M$ is an invertible and orthogonal matrix, i.e.
 the inverse matrix $M^{-1}$ is the transpose matrix,  $M^T=M^{-1}$.

\begin{figure}[hbt]
        \subfigure[$\eta_1=+1$ and $\eta_2=+1$]{\includegraphics[width=.45\linewidth]{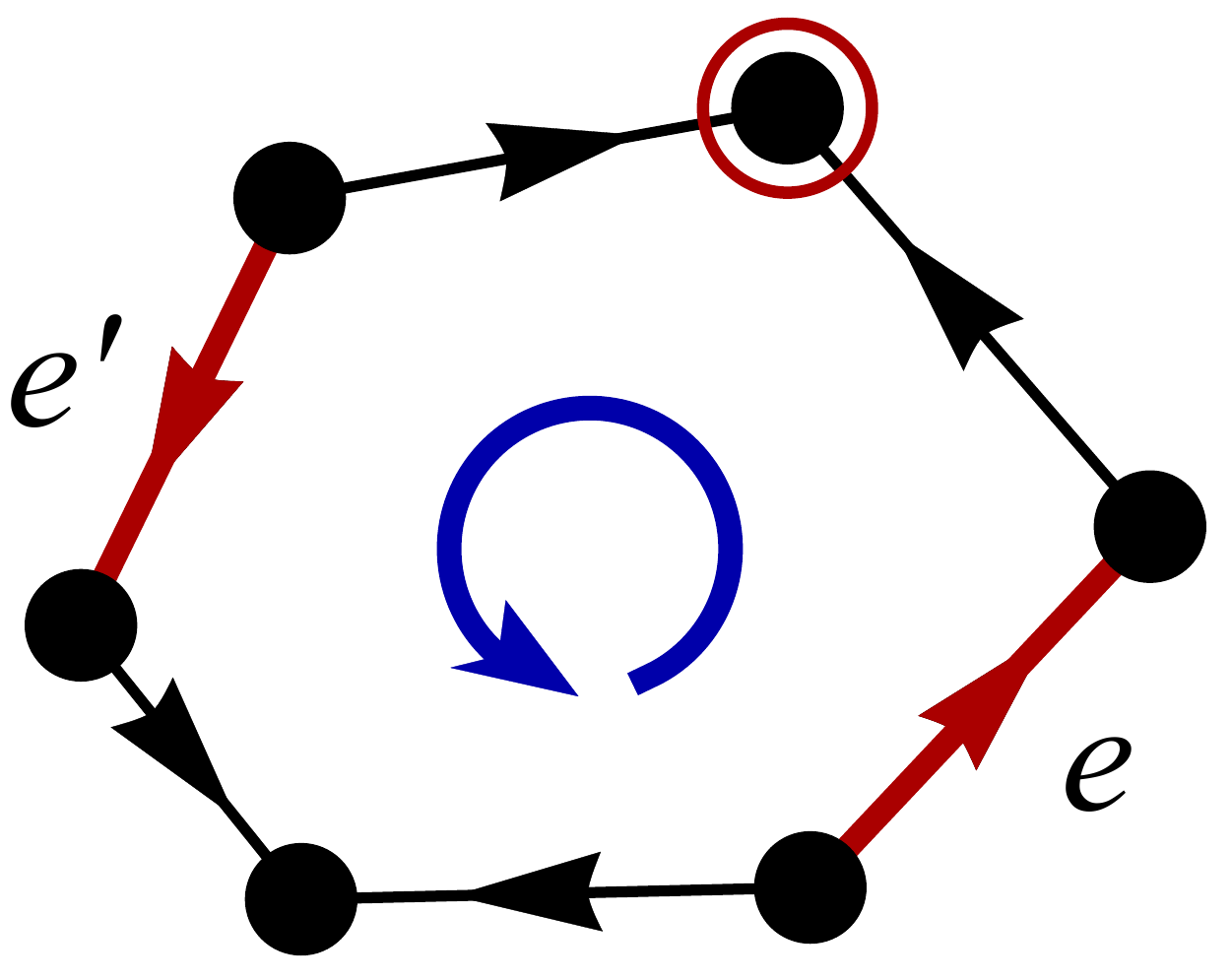}
                \label{fig:K_matrix_1}}
        \subfigure[$\eta_1=+1$ and $\eta_2=-1$]{\includegraphics[width=.45\linewidth]{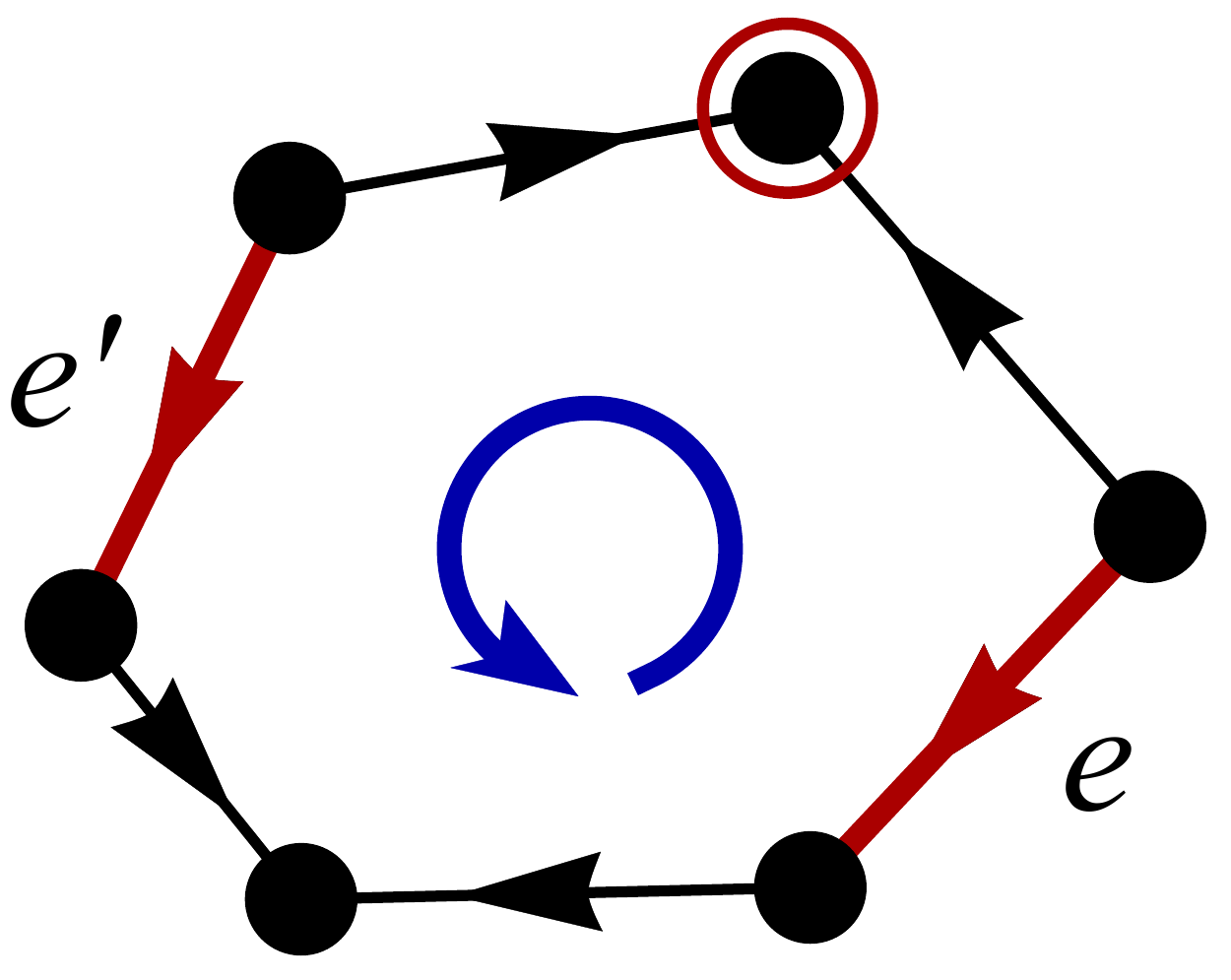}
                \label{fig:K_matrix_2}}
        \subfigure[$\eta_1=-1$ and $\eta_2=+1$]{\includegraphics[width=.45\linewidth]{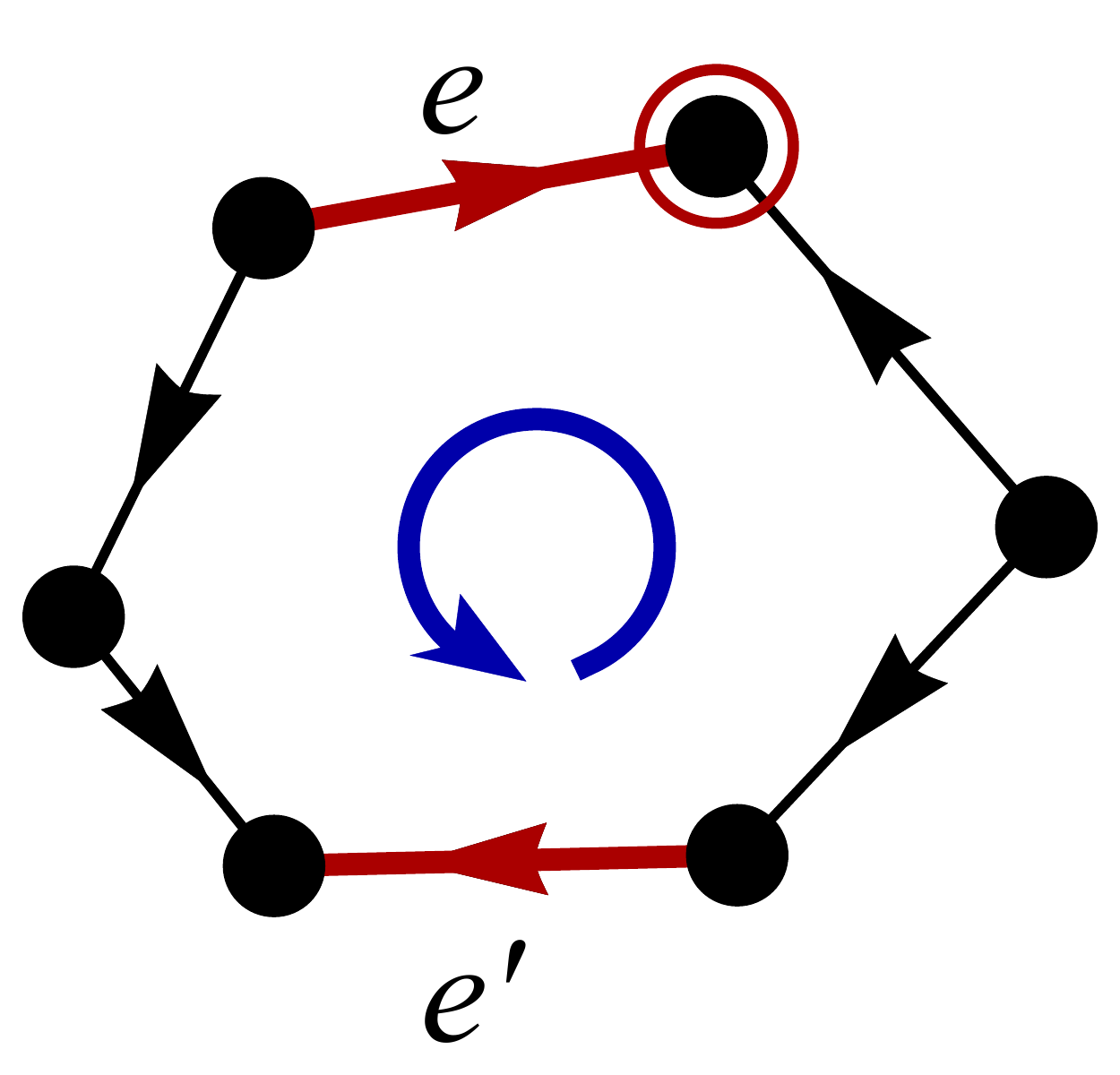}
                \label{fig:K_matrix_3}}
        \subfigure[$\eta_1=-1$ and $\eta_2=-1$]{\includegraphics[width=.45\linewidth]{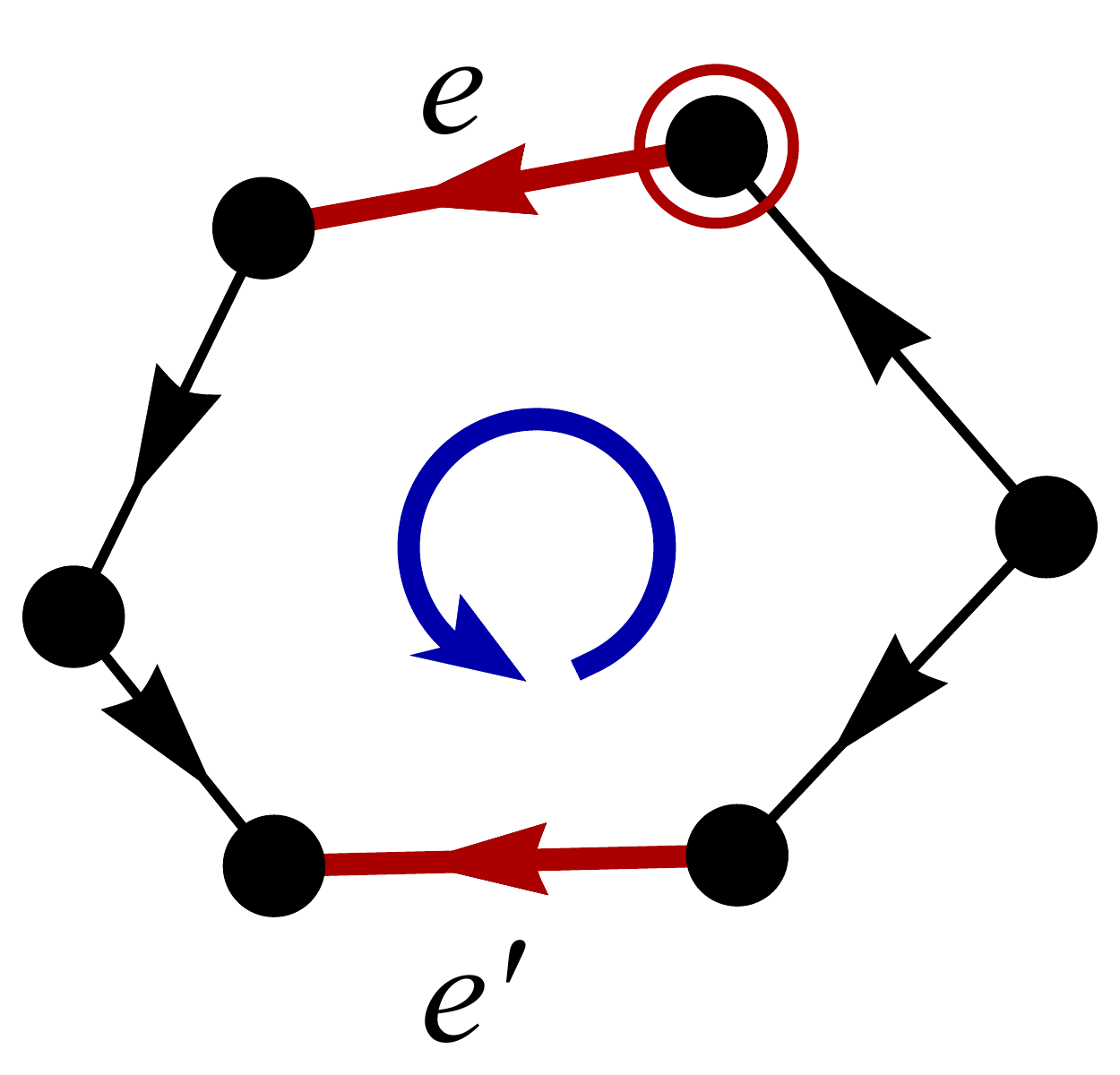}
                \label{fig:K_matrix_4}}
\caption{(Color online) Nonzero components of the $K$ matrix.
Here, we consider two edges $e$ and $e'$, which belongs to the same face $f$ (otherwise $K_{e,e'}=0$).
Based on the local vertex-face correspondence, the face $f$ is paired up with one of its vertices, 
which is marked by the (red) circle. We go around the face $f$ from $e$ to $e'$ by following the direction of 
the positive orientation marked by the (blue) circle at the center of the face. In Fig.~(a) and~(b), the path 
from $e$ to $e'$ goes through the special site (marked by the red circle), and thus $\eta_1=+1$. 
For Figs.~(c) and~(d), the special site is not on our path, and thus $\eta_1=-1$. 
The sign of $\eta_2$ is determined by the orientation of $e$ and $e'$. If their directions
are both along (or opposite) to the direction of the positive orientation [Figs~(a) and~(c)], $\eta_2=+1$.
Otherwise [Figs.~(b) and~(d)],  $\eta_2=-1$. Once $\eta_1$ and $\eta_2$ are determined, the value of $K_{e,e'}$
can be obtained as $K_{e,e'}=-\eta_1\times\eta_2/2=\pm1/2$.}
\label{fig:K_matrix}
\end{figure}

In addition to $M$, the local vertex-face correspondence can be used to define another $N_e\times N_e$ square matrix, which we will denote by $K$, 
whose two indices run over all edges of the graph (with $N_e$ being the number of edges),
\begin{align}
K_{e,e'}=\left\{
\begin{array}{cl}
	\pm\frac{1}{2} & \textrm{if $e$ and $e'$ belongs to the same face} \\
	0 &  \textrm{otherwise}
\end{array}\right.
\label{eq:K_matrix}
\end{align}
If there exists a face $f$ such that $e$ and $e'$ are both edges of this face,  
the component of the matrix $K_{e,e'}$ is $\pm1/2$. Otherwise the matrix element vanishes.
For nonzero $K_{e,e'}$, the $\pm$ sign is determined by the following formula,
\begin{align}
K_{e,e'}=-\frac{\eta_1\times\eta_2}{2}=\pm \frac{1}{2},
\end{align}
where $\eta_1=\pm 1$ and $\eta_2=\pm 1$ are two $\mathbb{Z}_2$ integers. 

The sign of $\eta_1$ is determined using the following rule.
As shown in Fig.~\ref{fig:K_matrix}, we first mark the vertex that is paired up with $f$ in the local vertex-face correspondence using a (red) circle. 
After that, we go from the edge $e$ to the edge $e'$ by moving counter-clockwise around the 
face $f$. If the path goes though the specially marked vertex (the red circle in  Fig.~\ref{fig:K_matrix}), 
$\eta_1=+1$, and otherwise $\eta_1=-1$

The sign of $\eta_2$ is determined by the directions of the two edges $e$ and $e'$. 
As discussed above, to define the vector potential, we must specify the direction for each edge.
When we goes around the face $f$ in the counter-clockwise direction, if both $e$ and $e'$ are pointing 
along (or opposite to) the direction of the path, $\eta_2=+1$. If one of them points along the path while the other is opposite, 
$\eta_2=-1$. 

With $\eta_1$ and $\eta_2$, their product (multiplied by -1), $-\eta_1 \times \eta_2=\pm 1$, determines the sign of $K_{e,e'}$ in Eq.~\eqref{eq:K_matrix}.
Some examples can be found in Fig.~\ref{fig:K_matrix}. 

\subsection{The Action}
\label{sec:action}

With the two matrices defined above, we can now write down the action of our discretized Chern-Simons gauge theory:
\begin{align}
S=\frac{k}{2\pi} \int dt \left[A_{v} M_{v,f} \Phi_{f}-\frac{1}{2} A_{e} K_{e,e'} \dot{A}_{e'}\right].
\label{eq:action}
\end{align}
Here, we sum over all repeated indices. The index $v$, $f$ and $e$ run over all vertices, faces and edges respectively.
$A_v$ is the time-component of the gauge field, which lives on vertices and $A_e$ represents the spatial components, which are
defined on edges. Here, $\dot{A}$ represents the time derivative, $K$ and $M$ are the two matrices defined in the previous subsection, and
$\Phi_{f}$ is the magnetic flux on the face $f$,  which equals to the loop integral of $A_e$ around $f$,
\begin{align}
\Phi_f= \xi_{f,e} A_e.
\label{eq:flux}
\end{align}
Here we sum over all edges and
\begin{align}
\xi_{f,e}=\left\{
\begin{array}{cl}
	+1 & \textrm{$e$ is an edge of $f$ with positive orientation} \\
	-1 & \textrm{$e$ is an edge of $f$ with negative orientation} \\
	0 &  \textrm{$e$ is not an edge of $f$}
\end{array}\right.
\label{eq:face_edges}
\end{align}
The sign of $\xi_{f,e}$ is determined by going around the face $f$ along the counter-clockwise direction. 
If the direction of the edge $e$ is along this path, $\xi_{f,e}=+1$. Otherwise, $\xi_{f,e}=-1$.
As can be seen from Eq.~\eqref{eq:flux}, the matrix $\xi_{f,e}$ is a discretized curl operator ($\nabla\times$) for planar graphs.

On a square lattice, the action that we constructed here reduces to the action found in Refs.~\onlinecite{Eliezer1992a} 
and~\onlinecite{Eliezer1992b},
which can be considered as a special situation of our generic construction. Similarly, for the Kagome lattice this general construction reduces to
the construction that we presented in Ref. [\onlinecite{Kumar-2014}].

We conclude this section by comparing our discretized theory with the Chern-Simons gauge theory in the continuum.
For comparison, we choose to write down the action in the continuum in a special form
\begin{align}
S=\frac{k}{2\pi}\int dt d{\bm x} \left( A_0 B -\frac{1}{2} A_i \epsilon_{i,j} \dot{A}_j \right).
\label{eq:action_continuous}
\end{align}
Here $A_0$ is the time component of the gauge field. $A_i$ and $A_j$ are the spatial component with $i$ and $j$
being $x$ or $y$. $\epsilon_{i,j}$ is the Levi-Civita symbol and $B$ is the magnetic field perpendicular to the 2D plane. 
The first term here enforces the flux attachment and
the second term dictates the dynamics of the vector potential $A_x$ and $A_y$.

By comparing Eq.~\eqref{eq:action} with Eq.~\eqref{eq:action_continuous}, we find that our discretized theory is 
in close analogy to the continuum case. Here, the $M$ matrix dictates the flux attachment (i.e. Gauss' law) and the $K$-matrix 
plays the role of the Levi-Civita symbol. It is worthwhile to highlight that,  just as the Levi-Civita symbol,
the $K$ matrix is  antisymmetric 
\begin{align}
K_{e,e'}=-K_{e',e},
\end{align}
This can be verified easily by noticing that $\eta_1 \to -\eta_1$ and $\eta_2 \to \eta_2$, if we swap $e$ and $e'$. 
This antisymmetry property is in fact expected. If we look at the second term in our action, 
Eq~\eqref{eq:action}, because $\int dt A_e \dot{A}_{e'}=-\int dt  \dot{A}_{e} A_{e'}$ (integration by part), 
only the antisymmetric part of $K$ contributes to the action.

In the next six sections, we will demonstrate that our action indeed offers a discretized Chern-Simons gauge theory on generic
graphs by showing that all the key properties of the Chern-Simons gauge theory are preserved by our action.

\section{Gauge invariance}
\label{sec:gauge_invariance}

For a gauge theory, the action must be gauge invariant. In this section, we will verify that our action [Eq.~\eqref{eq:action}]
preserves the gauge symmetry. In the case of Chern-Simons, this is also true provided the manifold has no boundaries. Furthermore, invariance under large gauge transformations (which wind around non-contractible loops of the systems) holds only if the index $k$ is an integer.\cite{Witten1989} These conditions are satisfied by our discretized Chern-Simons theory.

\subsection{Gauge transformation on a graph}
\label{sec:graph}

 For a graph/lattice, a gauge transformation takes the following form
\begin{align}
A_{v} \to &  A_{v} - \partial_t \phi_{v} 
\label{eq:gauge_transformation_A_v}\\
A_{e} \to & A_{e} - D_{v,e} \phi_{v}
\label{eq:gauge_transformation_A_e}
\end{align}
where $\phi_v$ is an arbitrary scalar function defined on vertices. 
The first formula [Eq.~\eqref{eq:gauge_transformation_A_v}] is the gauge transformation 
for the time component of the gauge field, while the second one [Eq.~\eqref{eq:gauge_transformation_A_e}]
is for the spatial components.
The matrix $D_{v,e}$ in Eq.~\eqref{eq:gauge_transformation_A_e} is the {\it incident matrix} of the graph~\cite{Biggs1974}
\begin{align}
D_{v,e}=\left\{
\begin{array}{cl}
	+1 & \textrm{if $v$ is the positive end of $e$} \\
	-1 & \textrm{if $v$ is the negative end of $e$}\\
	0 &  \textrm{otherwise}
\end{array}\right.
\label{eq:D_matrix}
\end{align}
Here, we called the vertex $v$ a positive (negative) end of the edge $e$, if $v$ is one of the two ends of $e$ and the direction of the edge $e$
is pointing towards (away from) $v$. The incident matrix contains all the information about the connectivity of the graph, as well
as the direction assigned to each edge.\cite{Biggs1974} The incident matrix plays the role of a (discretized) gradient operator, ${\bm \nabla}$,
which can be seeing easily by noticing that
\begin{align}
D_{v,e}\phi_{v}=\phi_{v_1}-\phi_{v_2}
\end{align}
where $\phi_v$ is an arbitrary scalar function and the edge $e$ points from $v_2$ to $v_1$. 
As a result, Eq.~\eqref{eq:gauge_transformation_A_e} can be considered as a discretized version of ${\bm A}\to {\bm A}-{\bm \nabla}\phi$.
Later, we will show that the incident matrix also serves as a discretized divergence, ${\bm \nabla} \cdot$.

\subsection{Gauge symmetry}
\label{sec:gauge-symmetry}

As proven in Appendix \ref{app:gauge_symmetry}, the sufficient and necessary condition for
the action of Eq.\eqref{eq:action} to be gauge-invariant is that the following identity is satisfied
\begin{align}
	M_{v,f}  \xi_{f,e}=K_{e,e'} D_{v,e'}
	\label{eq:gauge_invariance}
\end{align}
where $\xi_{f,e}$  is defined in Eq.~\eqref{eq:face_edges} and the incident matrix $D_{v,e'}$
is defined in Eq.~\eqref{eq:D_matrix}. In this section, we prove that
this condition is indeed valid for the $M$ and $K$ matrices constructed in Sec.~\ref{sec:M-K}. 

To verify Eq.~\eqref{eq:gauge_invariance}, we need to prove that the relation holds for any $e$ and $v$.
Here, we classify all possible situations into three cases:
\begin{enumerate}
\item $e$ and $v$ don't belong to the same face.
\item $e$ and $v$ belong to a same face but $v$ is not an end of $e$
\item $v$ is an end of $e$
\end{enumerate}
Here we verify Eq.~\eqref{eq:gauge_invariance} for each of these
three cases.

\subsubsection{Case I}
\label{sec:case1}

The first case, where $e$ and $v$ don't belong to the same face, represents the situation
where $e$ and $v$ are separated far away from each other. It is easy to verify that in this case 
both sides of Eq.~\eqref{eq:gauge_invariance} vanish.

For the l.h.s., $M_{v,f}\ne 0$ requires $v$ being a vertex of the face $f$ and $\xi_{f,e}\ne 0$
requires $e$ being an edge of $f$. For Case I, these two conditions cannot be satisfied simultaneously,
and thus $M_{v,f}\xi_{f,e}=0$.

For the r.h.s.,  $K_{e,e'}\ne 0$ implies that $e$ and $e'$ are edges of the same face,
which will be called the face $f$. 
If $D_{v,e'}\ne 0$,  $v$ must be one end of $e'$, which means that $v$ is a vertex of $f$. As a result, 
to get a nonzero $K_{e,e'}D_{v,e'}$, both $e$ and $v$ must both belong to the same face $f$. 
This is in contradiction with the assumption of Case I, and thus we must have $K_{e,e'} D_{e',v}=0$.

Because both sides of the equation are zero, then Eq.\eqref{eq:gauge_invariance} holds for Case I.

\begin{figure}[t]
        \subfigure[~Case II]{\includegraphics[width=.45\linewidth]{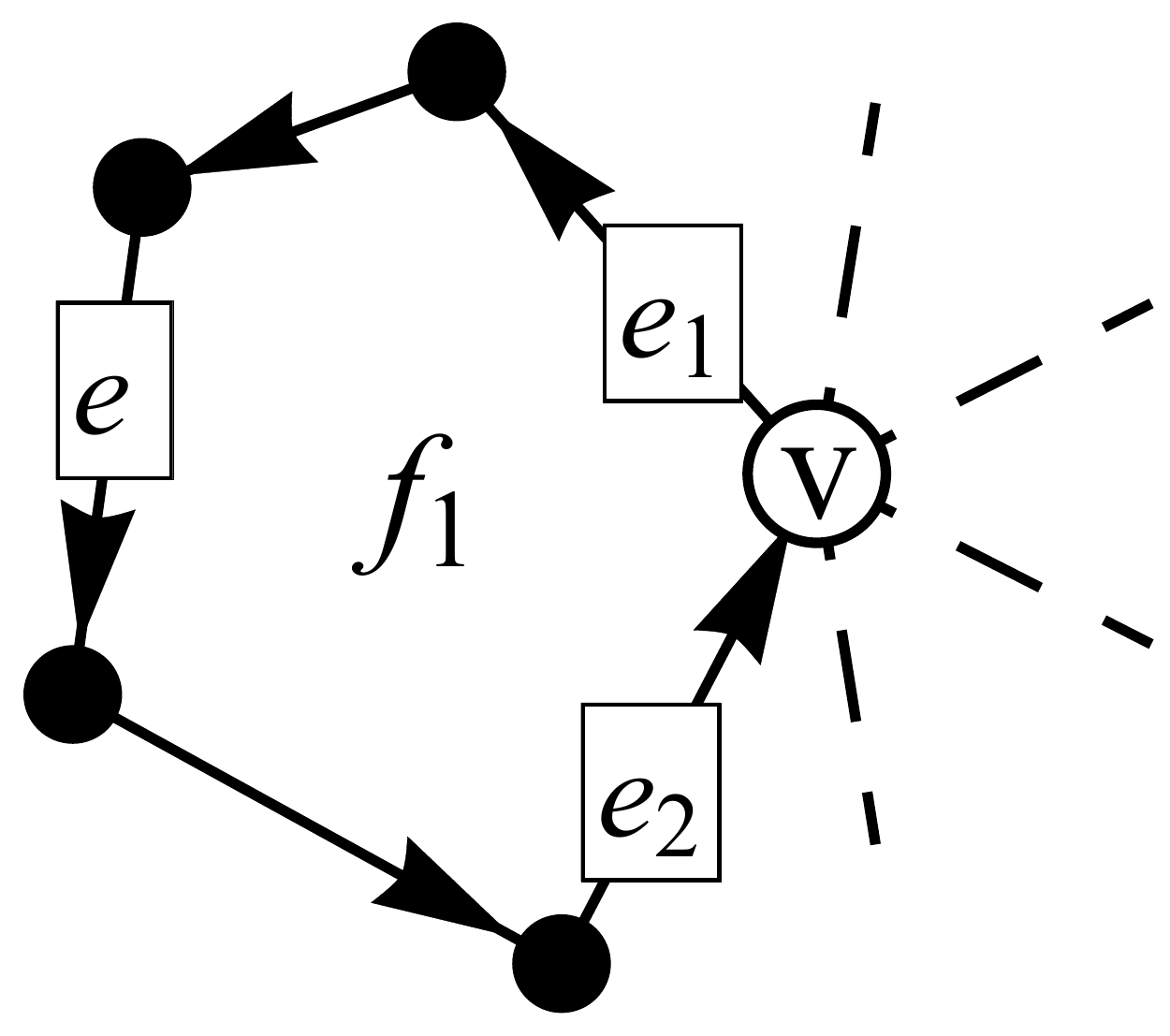}
                \label{fig:case2}}
        \subfigure[~Case III]{\includegraphics[width=.45\linewidth]{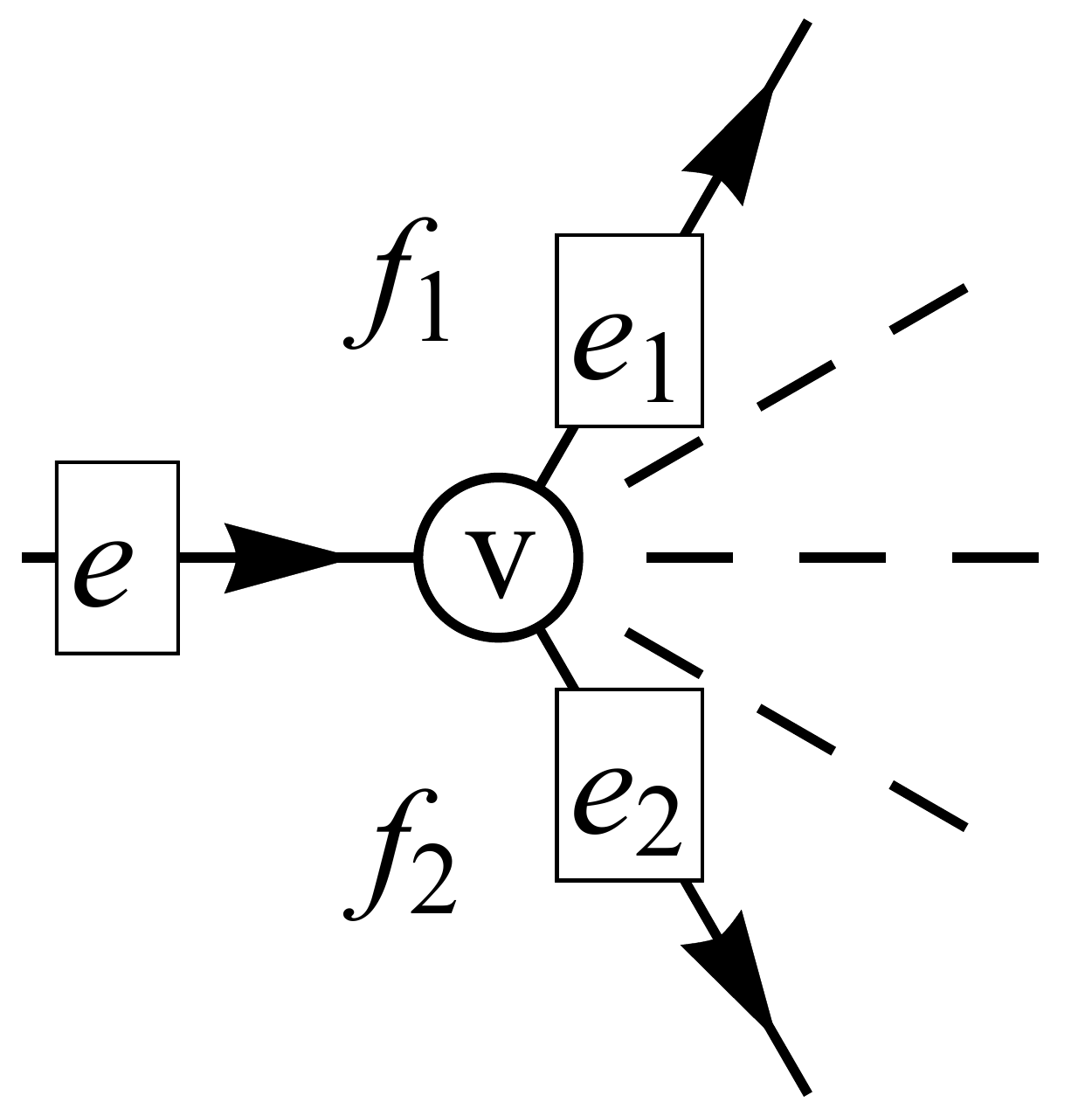}
                \label{fig:case3}}
\caption{Gauge invariance of our theory for (a) Case II and (b) Case III. In Fig.~(a), we marked two additional edges of 
$v$, $e_1$ and $e_2$, which are edges of $f_1$. In Fig.~(b), we labeled two faces $f_1$ and $f_2$ and two additional 
edges $e_1$ and $e_2$, such that $e$ is the common edge shared by $f_1$ and $f_2$, while $e_1$ and $e_2$ are two edges of $v$, 
which are adjacent to $f_1$ and $f_2$ respectively. 
Dashed lines represent (possible) additional edges of $v$, which are irrelevant for our proof and thus are not labeled. 
Although we assume a specific set of orientations for edges in these two figures, none of our final conclusions relies on the choice
of orientations for each edge, as   proven in Appendix \ref{app:gauge_symmetry}.}
\label{fig:gauge_symmetry}
\end{figure}

\subsubsection{Case II}
\label{sec:case2}

The second case, where $e$ and $v$ belong to one same face but $v$ is not an end of $e$, is shown in Fig.~\ref{fig:case2}.
In this figure, without loss of generality, we choose a specific direction for each edge.
As proved in Appendix \ref{app:edge_direction}, Eq.~\eqref{eq:gauge_invariance} is independent of 
the choice of the edge directions. Therefore, although 
we only consider one specific direction arrangement here, the conclusion is generic.

In Fig.~\ref{fig:case2}, both $v$ and $e$ belong to the same face, $f_1$. 
Because $v$ is a vertex of the face $f_1$, two of the edges of the face $f_1$ must have $v$ as their end. 
These two edges are labeled as $e_1$ and $e_2$ in Fig.~\ref{fig:case2}.

Using the edge directions shown in Fig.~\ref{fig:case2}, it is easy to verify that
\begin{align}
K_{e,e_1}=&-\frac{\eta_{1;e,e_1}\eta_{2;e,e_1}}{2}=-\frac{\eta_{1;e,e_1}}{2}
\\
K_{e,e_2}=&-\frac{\eta_{1;e,e_2}\eta_{2;e,e_1}}{2}=-\frac{\eta_{1;e,e_2}}{2}
\\
D_{v,e_1}=&-1
\\
D_{v,e_2}=&+1
\end{align}
and thus
\begin{align}
K_{e,e'}D_{v,e'}=K_{e,e_1}D_{v,e_1}+K_{e,e_2}D_{v,e_2}=\frac{\eta_{1;e,e_1}-\eta_{1;e,e_2}}{2}.
\label{eq:lhs_gauge_symmetry}
\end{align}

Here, we shall distinguish two different situations: 1) $v$ is paired up with $f_1$ according to the vertex-face correspondence, 
and  2) $v$ is not paired up with $f_1$.

If $v$ is paried up with $f_1$, 
$M_{v,f}$ vanishes for all $f$, except for $f=f_1$, and therefore,
\begin{align}
M_{v,f}\xi_{f,e}=M_{v,f_1}\xi_{f_1,e}=+1.
\end{align}
Here, we don't sum over the repeated index $f_1$ and we used the fact that $M_{v,f_1}=1$. 
For the orientation shown in Fig.~\ref{fig:case2}, $\xi_{f_1,e}=+1$.
For the r.h.s. of Eq.~\eqref{eq:gauge_invariance}, we can use Eq.\eqref{eq:lhs_gauge_symmetry}.
Here, we have $\eta_{1;e,e_1}=+1$ and $\eta_{1;e,e_2}=-1$, and thus,
\begin{align}
K_{e,e'}D_{e',v}=\frac{\eta_{1;e,e_1}-\eta_{1;e,e_2}}{2}=\frac{1}{2}+\frac{1}{2}=+1.
\end{align}
By comparing the two equations above, we find that $M_{v,f}\xi_{f,e}=K_{e,e'}D_{e',v}$

If $v$ is not paired up with $f_1$, $M_{v,f_1}\xi_{f_1,e}=0$. 
For the r.h.s. of Eq.~\eqref{eq:gauge_invariance}, it is easy to verify that $\eta_{1;e,e_1}=\eta_{1;e,e_2}$,
and thus
\begin{align}
K_{e,e'}D_{e',v}=\frac{\eta_{1;e,e_1}-\eta_{1;e,e_2}}{2}=0.
\end{align}
Again, we verified Eq.~\eqref{eq:gauge_invariance}.

\subsubsection{Case III}
\label{sec:case3}

For the last case, shown in Fig.~\ref{fig:case3},  because each edge in our graph is shared by two and only two faces (as shown above in Sec.~\ref{sec:lvfc}), 
we can label the two faces of the edge $e$ as $f_1$ and $f_2$. In addition, we also label two edges of $v$, $e_1$ and $e_2$, 
where $e_1$ is an edge of $f_1$ and $e_2$ is an edge of $f_2$. Because we have assumed that the graph and the dual graph are
both {\it simple} (see Sec.~\ref{sec:lvfc}), $e_1\ne e_2$.

Same as in Case II, here too we only need to consider one specific set of directions for the edges and the conclusion will be generic.
Using the directions shown in Fig.~\ref{fig:case3}, we have
\begin{align}
K_{e,e_1}=&-\frac{\eta_{1;e,e_1}\eta_{2;e,e_1}}{2}=-\frac{\eta_{1;e,e_1}}{2}
\\
K_{e,e_2}=&-\frac{\eta_{1;e,e_2}\eta_{2;e,e_1}}{2}=-\frac{\eta_{1;e,e_1}}{2}
\\
D_{v,e_1}=&-1
\\
D_{v,e_2}=&-1
\end{align}
and therefore
\begin{align}
K_{e,e'}D_{v,e'}=&K_{e,e_1}D_{v,e_1}+K_{e,e_2}D_{v,e_2}
\nonumber
\\=&\frac{\eta_{1;e,e_1}+\eta_{1;e,e_2}}{2}.
\end{align}

Again, we distinguish two possible situations: 1) 
$v$ is paired up with $f_1$ or $f_2$ according to the vertex-face correspondence 
2) $v$ is not paired up with either $f_1$ or $f_2$.

For the first situation, without loss of generality we assume that $v$ is paired up with $f_1$. 
Using the directions shown in Fig.~\ref{fig:case3}, we find that
\begin{align}
M_{v,f}\xi_{f,e}=M_{v,f_1}\xi_{f_1,e}=+1.
\end{align}
Here, we don't sum over the repeated index $f_1$.
In addition, we also have $\eta_{1;e,e_1}=\eta_{1;e,e_2}=+1$, 
and therefore
\begin{align}
K_{e,e'}D_{e',v}=\frac{\eta_{1;e,e_1}+\eta_{1;e,e_2}}{2}=+1.
\end{align}
So, we find $M_{v,f}\xi_{f,e}=K_{e,e'}D_{e',v}$.

If $v$ is not paired up with either $f_1$ or $f_2$, $M_{v,f}\xi_{f,e}=0$, because  it is impossible 
to make both $M_{v,f}$ and $\xi_{f,e}$ nonzero. It is also easy to verify that here 
$\eta_{1;e,e_1}=-\eta_{1;e,e_1}$ and thus
\begin{align}
K_{e,e'}D_{e',v}=\frac{\eta_{1;e,e_1}+\eta_{1;e,e_2}}{2}=0.
\end{align}
Once again, we get $M_{v,f}\xi_{f,e}=K_{e,e'}D_{e',v}$.

By summarizing all possible situations discussed above, we have verified Eq.~\eqref{eq:gauge_invariance}. 
Therefore, we conclude that our theory is invariant under local gauge transformations.

\section{Flux attachment}
\label{sec:flux-attachment}

A key property of the Chern-Simons gauge theory is the constraint of flux attachment, which binds a magnetic flux with 
each charged particle. For a point charge $q$ at location ${\bm r}_0$, the corresponding magnetic field is
\begin{align}
B({\bm r})=\frac{2\pi}{k}q\;  \delta^2({\bm r}-{\bm r}_0),
\label{eq:flux_attachment_c2}
\end{align}
In the continuum classical theory, the flux and the charge are located at the same position, as indicated by the $\delta$-function in 
Eq.~\eqref{eq:flux_attachment_c2}. In a continuum quantum gauge theory this condition is a constraint on the physical Hilbert space, and is the requirement that the quantum states be invariant under local time-independent gauge transformations,\cite{Dirac-1966} as we discussed in the Introduction, c.f. Eq.\eqref{eq:GaussCS}. This condition requires regularization (in the form of splitting the position of the charge and the flux) which leads to a  proper framing of the knots represented by Wilson loops.\cite{Witten1989,Tze1988,Polyakov-1988}  For a discrete system, however,  because electric charges live on vertices, 
while magnetic fluxes are defined on faces (which takes care of the regularization), it is necessary to specify one additional rule to dictate the location of the magnetic 
flux for charged particles at each site. This is achieved by the local vertex-face correspondence introduced in Sec.~\ref{sec:introduction}. Here too, this constraint amounts to the conditions that the states of the gauge theory be invariant under time-independent gauge transformations.\cite{Kogut-1975}

Because our action, Eq.~\eqref{eq:action}, does not contain any dynamics for the time component of the gauge field $A_v$ 
(just as in any gauge theory), $A_v$ is not a dynamical field but its role is to enforce a constraint.\cite{Dirac-1966}
By taking a variational derivative of $A_v$, we get the charge at the vertex $v$, 
\begin{align}
q_v=\frac{\delta S}{\delta A_v}=\frac{k}{2\pi} M_{v,f} \Phi_{f},
\label{eq:flux_attachment}
\end{align}
which is proportional to the magnetic flux in the face $f$. Because $M$ is an orthogonal matrix, this equation implies that
\begin{align}
\Phi_{f}=\frac{2\pi}{k} q_v M_{v,f}.
\label{eq:flux_attachment_2}
\end{align}
This equation is the discrete version of the flux attachment, analogous to Eq.~\eqref{eq:flux_attachment_c2}.

Here, we find that for a charge at a vertex $v$, a magnetic flux is bound to it and the flux is located at the face $f$, which
is the partner of $v$ according to the vertex-face correspondence. This is the physical content of the
vertex-face correspondence.

We conclude this section by emphasizing that the flux attachment rule here is {\it local}, because we have required
the vertex-face correspondence to be {\it local}, i.e., the magnetic flux attached to a charge is located on a neighboring
face. For a discrete system, this setup offers the closest analogy to the delta function 
in Eq.~\eqref{eq:flux_attachment_c2}.
 
\section{Dual graph, dual theory and the invertibility of the $K$ matrix}.
\label{sec:dual}

In this section, we verify two key (and essential) properties of the discretized Chern-Simons gauge theory:
\begin{enumerate}
\item 
The $K$-matrix is invertible
\item
 For any discretized Chern-Simons gauge theory constructed above, one can construct another discretized Chern-Simons gauge theory on the dual graph.
\end{enumerate}
Later, we will prove in Appendix~\ref{app:sec:dual} that the theory defined on the dual graph is in fact the dual theory of the 
original discretized Chern-Simons gauge theory.

As has been addressed in literature, the $K$ matrix must be nonsingular (invertible) in order to ensure the correct dynamics for 
a discretized Chern-Simons gauge theory.\cite{Eliezer1992a,Eliezer1992b}
One way to realize this is by noticing that the inverse of the $K$ matrix offers the commutator of the vector potential $A_{e}$
(see Sec.~\ref{sec:commutator} for more details),
and therefore, to avoid singularities in the commutator, the $K$ matrix must be invertible.

Here, we will first verify the second property listed above by directly constructing a Chern-Simons gauge theory on the dual graph in Secs.~\ref{sub:sec:dual_graph} and~\ref{sub:sec:K_star}.
Then, in Sec.~\ref{sub:sec:K_inverse}, we prove that $K$ is invertible by finding directly the inverse matrix of $K$, 
which is in fact the $K^*$ matrix defined on the dual graph with a minus sign.
Finally, in Sec.~\ref{sub:sec:gauge_invariance_dual}, as a consistency check, we prove that the gauge invariance condition for 
the original graph and that of the dual graph are actually equivalent to each other.

\subsection{Duality transformation}
\label{sub:sec:dual_graph}

For a planar graph $G$, one can construct the dual graph $G^*$ by putting a vertex $v^*$ in each face of $G$ 
and then connecting two vertices in $G^*$ if their corresponding faces in $G$ share a common edge.
It is easy to check that the dual of a dual graph is the original graph $(G^*)^*=G$.  
For the lattices shown in Fig.~\ref{fig:lattices_CS}, the square lattice is self-dual, while the Kagome lattice and the 
dice lattice are dual to each other.

For simplicity, we will use the same integer to label $f$ and $v^*$, if $f$ is mapped to $v^*$ under the duality transformation.
Similarly, we use the same integer to label $e$ and $e^*$ ($v$ and $f^*$ ), if they are dual to each other. 
In addition, we choose the direction for each edge in the dual graph such that ${\bm n}_e \times {\bm n}_e^*>0$,
where ${\bm n}_e$ and ${\bm n}_e^*$ are unit vectors along the direction of the edge $e$ and its dual edge $e^*$.
In other words, we rotate the edge $e$ counter-clockwise until it aligns with $e^*$, and then the direction of the rotated edge 
$e$ determines the direction of $e^*$.

With this convention, the incident matrix of the dual graph $D^*_{v^*,e^*}$ coincides with the
$\xi_{f,e}$ matrix of the original graph, Eq.~\eqref{eq:face_edges},
\begin{align}
D^*_{v^*,e^*}=\xi_{f,e}.
\label{eq:dual_of_D}
\end{align}
Similarly, the $\xi^*_{f^*,e^*}$ matrix for the dual graph is in fact the incident matrix of the original graph $D$, up to an over all minus sign
\begin{align}
\xi^*_{f^*,e^*}=-D_{v,e}.
\label{eq:dual_of_xi}
\end{align}
Here, we require $v^*=f$ and $e^*=e$ as shown in the previous graph.
The physics meaning of these two relations is that if the duality transformation maps a face $f$ of a graph $G$
into the vertex $v^*$ in the dual graph $G^*$, then a loop around the face $f$ is mapped to all the edges connected to 
vertex $v^*$, and vice versa.

It is easy to realize that under a duality transformation, the local vertex-face correspondence in the original graph is transformed
into a local vertex-face correspondence in the dual graph. As a result, we can use exactly the same construction to obtain a discretized
Chern-Simons gauge theory on the dual graph
\begin{align}
S=\frac{k^*}{2\pi} \int dt 
\Big[ A^*_{v^*} M^*_{v^*,f^*} \Phi^*_{f^*}-\frac{1}{2} A^*_{e^*} K^*_{e^*,e'^*} \dot{A}^*_{e'^*}\Big]
\label{eq:dual_action}
\end{align}
Here, $A^*_{v^*}$ and $A^*_{e^*}$ are gauge fields defined on the dual graph with $\Phi^*_f= \xi^*_{f^*,e^*} A_e^*$ 
being the magnetic flux of this gauge
field on face $f^*$. The $M^*$ and $K^*$ matrices are constructed using the same rules discussed above in Sec.~\ref{sec:M-K}.
In Appendix.~\ref{app:sec:dual}, we  show that if $k^*=-1/k$, this action is the dual theory of the original discretized Chern-Simons gauge theory,
Eq.~\eqref{eq:action}.

It is straightforward to verify that the $M^*$ matrix is the transpose of the $M$ matrix. Because $M$ is an orthogonal matrix,
it implies that $M^*$ is the inverse of $M$
\begin{align}
M^*=M^{T}=M^{-1}.
\end{align}
Below, we will study the $K^*$ matrix and prove that it is the inverse of the $K$ matrix up to an overall minus sign.

\subsection{the $K^*$ matrix}
\label{sub:sec:K_star}
In this section, we show that the $K^*$ matrix can be constructed directly in the original graph $G$, without going to the dual graph $G^*$.
This construction is fully equivalent to the dual-graph construction used in the previous section. However, as will be shown 
in the next section, by constructing the $K^*$ matrix in the original graph, it is more convenient to study
the relation between the $K$ matrix and the $K^*$ matrix.

As mentioned above, we label each edge in the dual graph using the same index as the corresponding edge in the original graph,
(i.e. $e^*=e$). Therefore, we can rewrite the $K^*$ matrix using the edge indices of the original graph ($e$ and $e'$),
\begin{align}
K^*_{e^*,e'^*}=K^*_{e,e'}
\end{align}
where $e$ and $e'$ are edges of original lattice and they are dual to $e^*$ and $e'^*$ respectively.
The matrix $K^*_{e,e'}$ is now defined on the original graph, and thus we can translate the definition of the $K^*$
matrix to the original graph. Using the original graph, it is straightforward to verify that
\begin{align}
K^*_{e,e'}=\left\{
\begin{array}{cl}
	\pm\frac{1}{2} & \textrm{if $e$ and $e'$ share a vertex} \\
	0 &  \textrm{otherwise}
\end{array}\right.
\label{eq:K_matrix_dual}
\end{align}
If $e$ and $e'$ do not share a common endpoint, $K_{e,e'}=0$. Otherwise, 
\begin{align}
K^*_{e,e'}=-\frac{\eta^*_1\times\eta^*_2}{2}=\pm \frac{1}{2}.
\end{align}
where $\eta^*_1=\pm 1$ and $\eta^*_2=\pm 1$ are two $Z_2$ integers. 
To determine the sign of $\eta^*_1$, we first label the common end of $e$ and $e'$ as 
$v$. Under the vertex-face correspondence (of the original graph),  $v$ is paired up with
a neighboring face $f$. Now, we go from the edge $e$ to the edge $e'$ by moving around 
$v$ in the counter-clockwise direction. If the path goes through the face $f$, $\eta^*_1=+1$, 
and otherwise $\eta^*_1=-1$. The sign of $\eta_2^*$ is determined by the directions of edges $e$ and $e'$. 
If both of them point toward (or away from) $v$, $\eta_2^*=+1$, and otherwise $\eta_2^*=-1$.
 
\subsection{$K^*=-K^{-1}$}
\label{sub:sec:K_inverse}
In this section, we prove that
\begin{align}
K^*=-K^{-1}
\label{eq:K_star}
\end{align}
and thus $K$ is invertible.
To prove Eq.~\eqref{eq:K_star}, we shall verify the following relations
\begin{align}
K_{e,e''}K^*_{e'',e'}=K^*_{e,e''}K_{e'',e'}=-\delta_{e,e'}.
\end{align}
where $\delta_{e,e'}$ is the Kronecker delta.
In this section, we will only demonstrate $K_{e,e''}K^*_{e'',e'}=-\delta_{e,e'}$, while one can use
the same method to prove $K^*_{e,e''}K_{e'',e'}=-\delta_{e,e'}$.

\begin{figure}[hbt]
        \subfigure[Case 1]{\includegraphics[width=.45\linewidth]{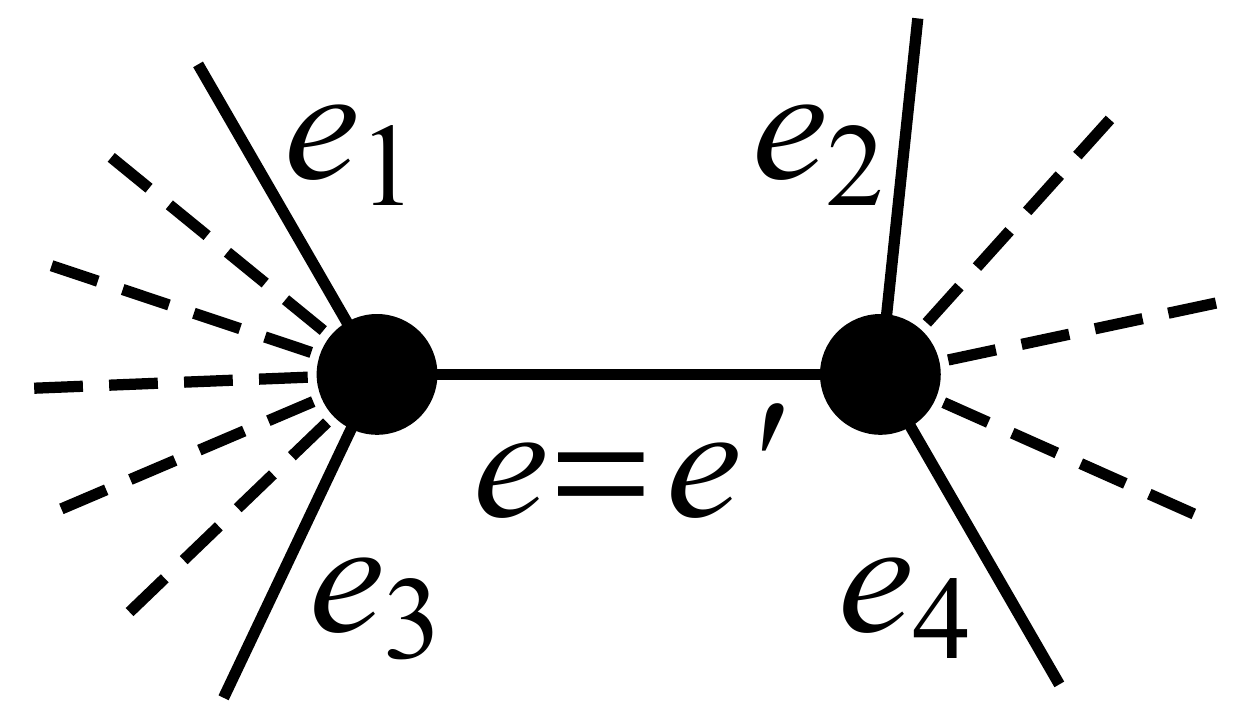}
                \label{fig:inverse_case1}}
        \subfigure[Case 2]{\includegraphics[width=.45\linewidth]{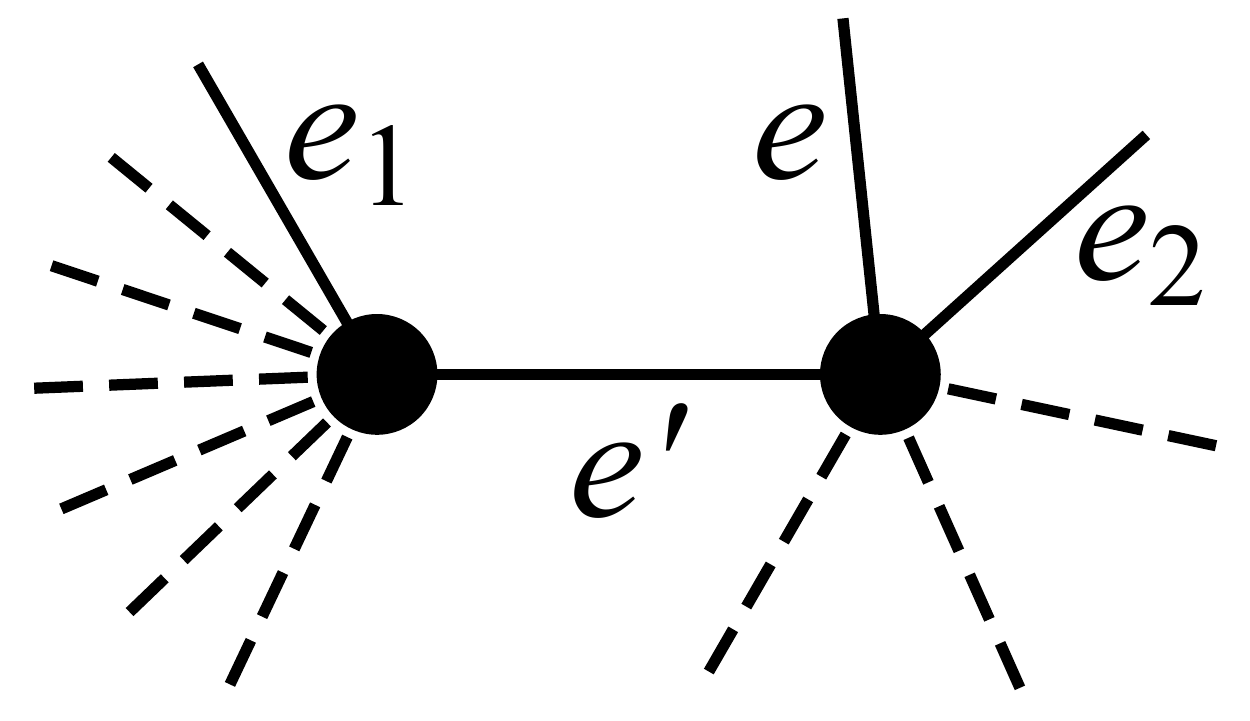}
                \label{fig:inverse_case2}}
        \subfigure[Case 3]{\includegraphics[width=.45\linewidth]{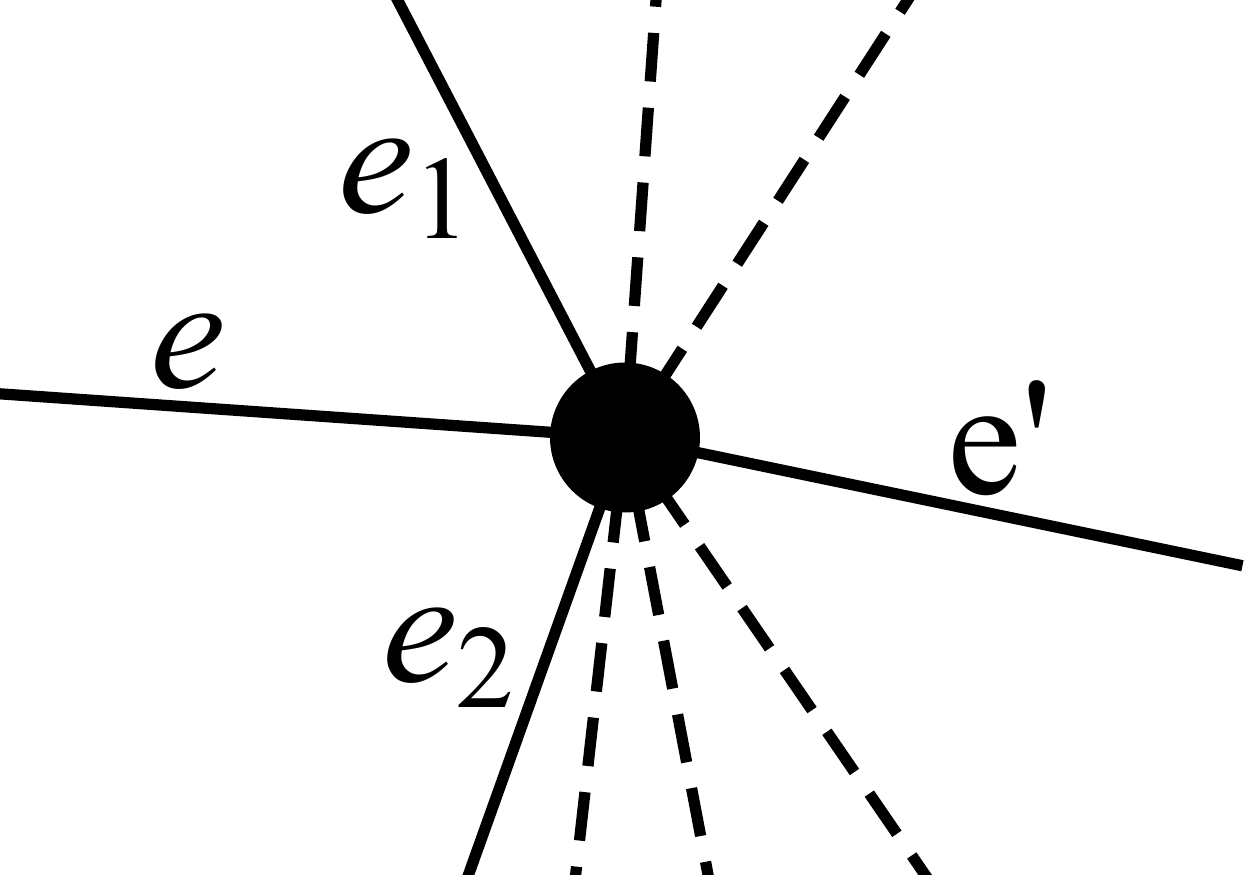}
                \label{fig:inverse_case3}}
        \subfigure[Case 4]{\includegraphics[width=.45\linewidth]{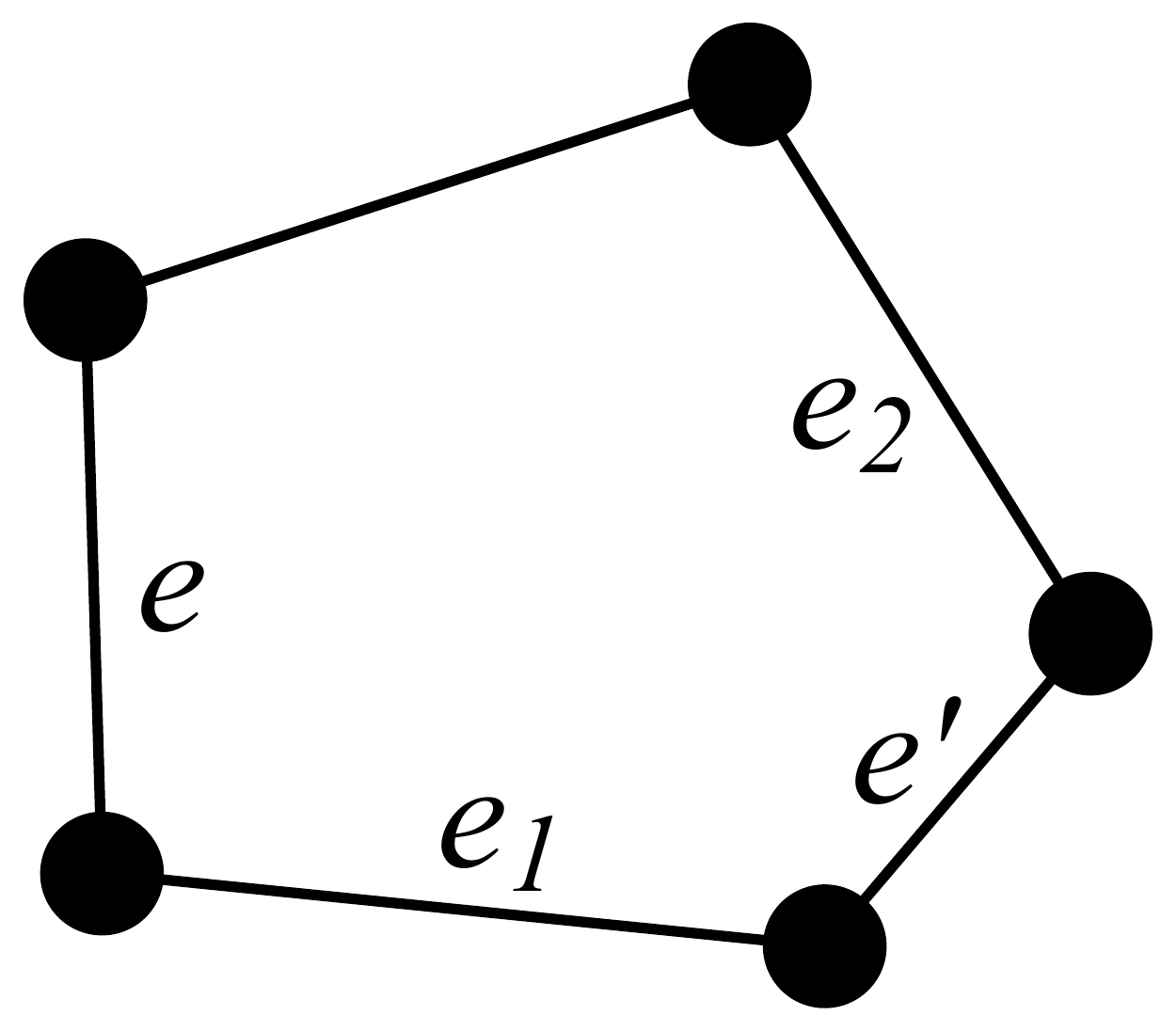}
                \label{fig:inverse_case4}}
        \subfigure[Case 5]{\includegraphics[width=.45\linewidth]{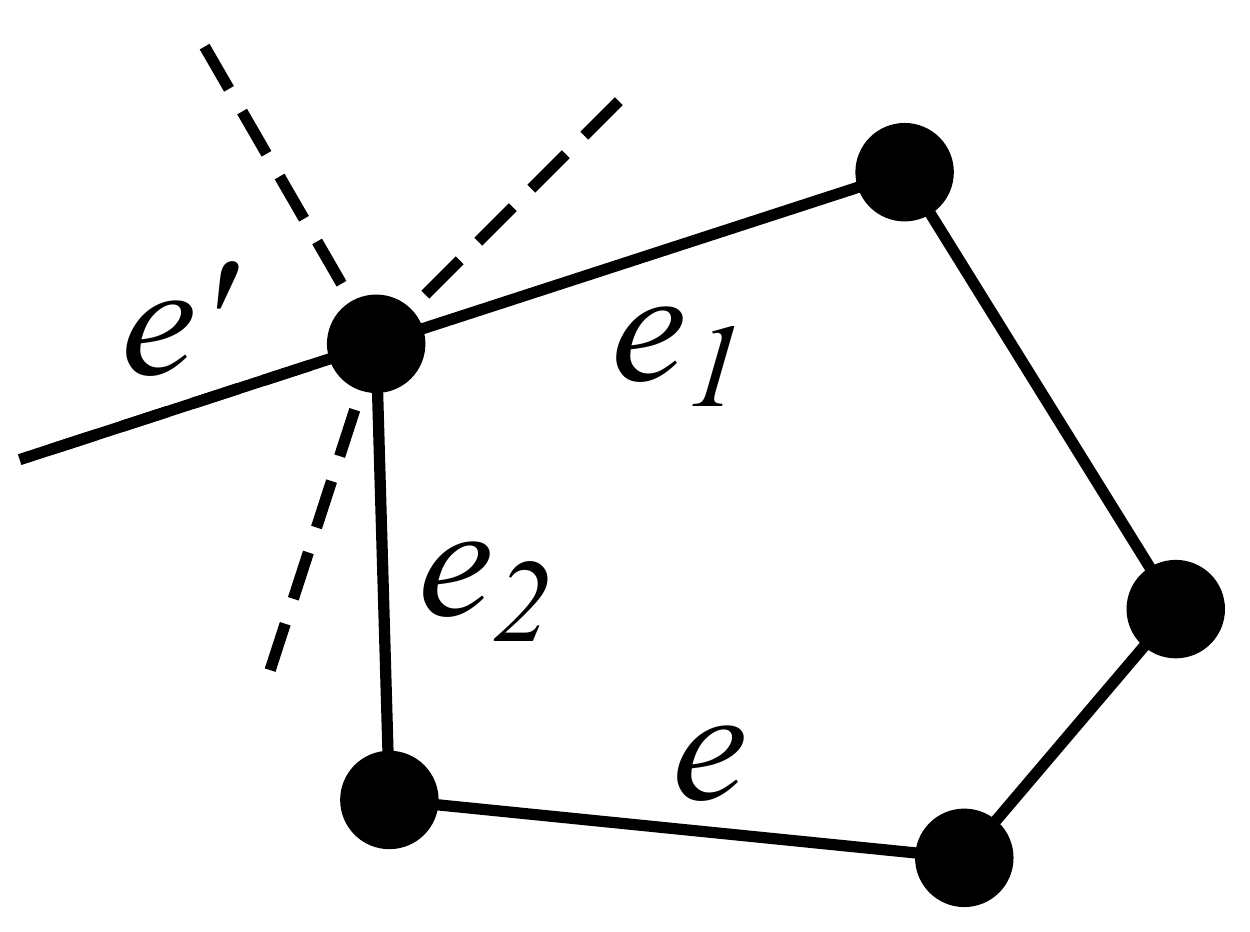}
                \label{fig:inverse_case5}}
\caption{Possible cases for computing $K_{e,e''} K^*_{e'',e'}$. Solid lines
marks the edges $e$, $e'$ and all edges that will contribute to $K_{e,e''} K^*_{e'',e'}$. 
Dashed lines are (possible) additional edges, which don't contribute to $K_{e,e''} K^*_{e'',e'}$.
}
\label{fig:inverse}
\end{figure}

Here, we need to consider six different cases,
\begin{enumerate}
\item $e=e'$.
\item $e\ne e'$, and $e$ and $e'$ share an endpoint, and $e$ and $e'$ are edges of the same face.
\item $e\ne e'$, and $e$ and $e'$ share an endpoint, but $e$ and $e'$ are not edges of the same face.
\item $e$ and $e'$ do not share any endpoint, but the belongs to the same face.
\item $e$ and $e'$ do not belong to the same face, but there is a face $f$, where $e$ is an edge of $f$
and one of the endpoints of $e'$ is a vertex of $f$.
\item otherwise 
\end{enumerate}
Among all the six cases, $\delta_{e,e'}=1$ for the first one, and $\delta_{e,e'}=0$ for all others.
In Fig.~\ref{fig:inverse}, we show the first five cases. Here, we mark $e$, $e'$ and all other edges that contribute
to $K_{e,e''} K^*_{e'',e'}$ using solid lines. Other (possible) edges, which do not contribute to $K_{e,e''} K^*_{e'',e'}$,
are labeled as dashed lines. Using Fig.~\ref{fig:inverse}, it is easy to notice that for all the first five situations
\begin{align}
K_{e,e''}K^*_{e'',e'}=&\sum_{i} K_{e,e_i}K^*_{e_i,e'}\nonumber
\\=&\sum_{i} \eta_{1;e,e_i}\eta_{2;e,e_i}
\eta^*_{1;e_i,e'}\eta^*_{2;e_i,e'}
\label{eq:K_Kinverse}
\end{align}
Here, for each $e_i$, $\eta_1$, $\eta_2$, $\eta_1^*$ and $\eta_2^*$ are obtained using the rules defined above.
Below, we compute $K_{e,e''}K^*_{e'',e'}$ for each situation using this formula.

\subsubsection{Case 1}

For the first situation [$e=e'$ as shown in Fig.~\ref{fig:inverse_case1}], we have
\begin{align}
K_{e,e''}K^*_{e'',e'=e}=\sum_{i=1}^4 K_{e,e_i}K^*_{e_i,e}. 
\end{align}
Here we don't sum over repeated indices on the r.h.s. of the equation, and the four edges $e_1$, $e_2$, $e_3$ and $e_4$ are marked in Fig.~\ref{fig:inverse_case1}.
For any $e_i$, we can verify that $\eta_1=\eta_1^*$ and $\eta_2=-\eta_2^*$.
Therefore, $K_{e,e_i}=-K^*_{e_i,e}$. As a result,
\begin{align}
K_{e,e''}K^*_{e'',e'=e}=-\sum_{i=1}^4 \left(K_{e,e_i}\right)^2
=-\sum_{i=1}^4 \left(\pm\frac{1}{2}\right)^2=-1
\end{align}
and thus, we find that $K_{e,e''}K^*_{e'',e'}=-\delta_{e,e'}$ for $e=e'$.

\subsubsection{Case 2}

For the second case, shown in Fig.~\ref{fig:inverse_case2}, it is straightforward to verify that
if $\eta_{2;e,e_1}$ and $\eta_{2;e,e_2}$ have the same sign, then $e_1$ and $e_2$
must both point towards (or away from) the two vertices shown in Fig.~\ref{fig:inverse_case2}.
As a result, $\eta^*_{2;e_1,e'}$ and $\eta^*_{2;e_2,e'}$ must have opposite sign, and thus
\begin{align}
\eta_{2;e,e_1}\eta^*_{2;e_1,e'}=-\eta_{2;e,e_2}\eta^*_{2;e_2,e'}.
\label{eq:eta2_case2}
\end{align}
Similarly, we can show that if $\eta_{2;e,e_1}=-\eta_{2;e,e_2}$, we must have
$\eta^*_{2;e_1,e'}=\eta^*_{2;e_2,e'}$. And therefore,
Eq.~\eqref{eq:eta2_case2} is aways valid for Case 2.

For $\eta_1$s, we need to examine three different cases. Here, we consider
the face $f$ formed by $e$, $e'$, $e_1$ and (possibly) other edges, and ask
whether the vertex-face correspondence pairs up
$f$ with one of these two vertices.
In general, there are three possibilities
\begin{enumerate}
\item $f$ is paired up with the vertex on the left;
\item $f$ is paired up with the vertex on the right;
\item $f$ is not paired up with either of them.
\end{enumerate}
For the first sitaution, we have $\eta_{1;e,e_1}=-1$, $\eta^*_{1;e_1,e'}=-1$ and 
$\eta_{1;e,e_2}=\eta^*_{1;e_2,e'}$.
Therefore, we find
\begin{align}
\eta_{1;e,e_1}\eta^*_{1;e_1,e'}=\eta_{1;e,e_2}\eta^*_{1;e_2,e'}.
\label{eq:eta1_case2}
\end{align}
For the second situation, we have
$\eta_{1;e,e_1}=-1$, $\eta^*_{1;e_1,e'}=+1$ and 
$\eta_{1;e,e_2}=-\eta^*_{1;e_2,e'}$. 
Therefore, Eq.~\eqref{eq:eta1_case2} is still valid.
For the third situation,  it can be shown that 
$\eta_{1;e,e_1}=+1$, $\eta^*_{1;e_1,e'}=+1$ and 
$\eta_{1;e,e_2}=\eta^*_{1;e_2,e'}$. Thus, Eq.~\eqref{eq:eta1_case2} is still valid.

In summary, we find that Eqs.~\eqref{eq:eta2_case2} and~\eqref{eq:eta1_case2} always hold 
for this case, Fig.~\ref{fig:inverse_case2}. By multiplying these
two equations together, we get 
\begin{align}
K_{e,e_1}K^*_{e_1,e'}=-K_{e,e_2}K^*_{e_2,e'}
\end{align}
Utilizing Eq.~\eqref{eq:K_Kinverse}, this relation implies that
$K_{e,e''}K^*_{e'',e'}=0$, in agreement with 
the relation $K_{e,e''}K^*_{e'',e'}=-\delta_{e,e'}$.

\subsubsection{Cases 3, 4 and 5}

Using the same approach, we can show that for the third
and the fourth cases, shown in Fig.~\ref{fig:inverse_case3} and~\ref{fig:inverse_case4},
\begin{align}
\eta_{2;e,e_1}\eta^*_{2;e_1,e'}=&\eta_{2;e,e_2}\eta^*_{2;e_2,e'},
\\
\eta_{1;e,e_1}\eta^*_{1;e_1,e'}=&-\eta_{1;e,e_2}\eta^*_{1;e_2,e'}.
\end{align}
Once again, we get $K_{e,e''}K^*_{e'',e'}=0=-\delta_{e,e'}$.

For the fifth case, shown in Fig.~\ref{fig:inverse_case5}, we have
\begin{align}
\eta_{2;e,e_1}\eta^*_{2;e_1,e'}=&-\eta_{2;e,e_2}\eta^*_{2;e_2,e'},
\\
\eta_{1;e,e_1}\eta^*_{1;e_1,e'}=&\eta_{1;e,e_2}\eta^*_{1;e_2,e'}.
\end{align}
Thus, $K_{e,e''}K^*_{e'',e'}=0=-\delta_{e,e'}$.

\subsubsection{Case 6}

The last case, 6,  is easy to verify, because here $e$ and $e'$ are
far away from each other, so that for any $e''$, either $K_{e,e''}$ or $K^*_{e'',e'}$  is zero.
Therefore, $K_{e,e''}K^*_{e'',e'}=0=-\delta_{e,e'}$.

By summarizing all the possible cases, we conclude that $K K^*=-I$.
We can use the same method to prove that $K^* K=-I$ and thus $K^*=-K^{-1}$.
This result also proves that the $K$  and $K^*$ matrices that we constructed above 
are invertible.

\subsection{Gauge invariance in the dual graph}
\label{sub:sec:gauge_invariance_dual}
As shown above in Sec.~\ref{sec:gauge_invariance}, in the original graph, our action of Eq.~\eqref{eq:action}
is  gauge invariant if and only if  
\begin{align}
	M_{v,f}  \xi_{f,e}=K_{e,e'} D_{v,e'}
\label{eq:gauge_invariance_again}
\end{align}
For the dual graph, there is a similar condition for the gauge invariance.
\begin{align}
	M^*_{v^*,f^*}  \xi^*_{f^*,e^*}=K^*_{e^*,e'^*} D^*_{v^*,e'^*}
\label{eq:gauge_invariance_dual}
\end{align}
In this section, we prove that these two conditions are in fact equivalent
as long as $M^*=M^{-1}$ and $K^*=-K^{-1}$

We start from Eq.~\eqref{eq:gauge_invariance_dual} and change the dual graph (face, edge or vertex) labels into the
corresponding labels of the original graph
\begin{align}
	M^*_{f,v}  D_{v,e}=-K^*_{e,e'}\xi_{f,e'},
\end{align}
and here we also use the relations $D^*_{v^*,e^*}=\xi_{f,e}$ and 
$\xi^*_{f^*,e^*}=-D_{v,e}$, Eqs.~\eqref{eq:dual_of_D} and~\eqref{eq:dual_of_xi}.

If $M^*=M^{-1}$ and $K^*=-K^{-1}$, the formula above implies that
\begin{align}
	M^{-1}_{f,v}  D_{v,e}=K^{-1}_{e,e'} \xi_{f,e'}
\label{eq:gauge_invariance_inverse}
\end{align}
By multiplying the matrices $M$ and $K$ on both sides, we recover the condition of gauge invariance in the original graph, Eq.~\eqref{eq:gauge_invariance}. Therefore, we find that the two gauge invariance conditions, Eq.~\eqref{eq:gauge_invariance_again} and~\eqref{eq:gauge_invariance_dual}, are equivalent.

\section{Commutation relations and the~$K^{-1}$ matrix}
\label{sec:commutator}

The Chern-Simons theory in the continuum has a very special commutation relations. In particular, the commutator
between the loop integrals of the vector potential is a topologically invariant. We will show in this 
section that our discretized theory has the same property.

\subsection{Commutators for the continuum case}

For the Chern-Simons gauge theory in the continuum, for two arbitrary curves $C$ and $C'$, we have the following 
commutation relation
\begin{align}
\left[\int_C A, \int_{C'} A\right]=\frac{2\pi i}{k}\nu\left[C,C'\right]
\label{eq:curve_commutator}
\end{align}
where $\nu[C,C']$ is the number of (oriented) intersections between the two curves, i.e. the number of right-handed interactions of $C$ and $C'$
minus the number of left-handed ones.~\cite{Eliezer1992a}

If $C$ and $C'$ are closed loops, $\nu[C,C']$ is a topologically invariant, and it is easy to verify that its value cannot change 
under any adiabatic procedures. In addition, if either $C$ or $C'$ can be contracted into a point (i.e. contractible), $\nu[C,C']=0$. 

\subsection{Canonical quantization}

Using  canonical quantization, it is straightforward to show that the conjugate field of the vector potential
field $A_{e_i}$ is 
\begin{align}
\frac{\delta S}{\delta \dot{A}_{e}}=
\frac{k}{2\pi}K_{e,e'}A_{e'} 
\end{align}
This formula implies that for our discretized Chern-Simons theory, the
vector potential $A_e$ (and linear superpositions of $A_e$'s) play both the role of the canonical coordinates and 
that of the canonical momenta. 
Because canonical coordinates and canonical momenta arise in pairs, this result requires that we must have even 
number of linear independent $A_e$s, i.e. the number of edges must be even. 
This is indeed true for any graphs considered here. Utilizing the Euler characteristic, we know that the numbers of vertices, edges and
faces must satisfy the following relation
\begin{align}
N_v-N_e+N_f=2-2g,
\end{align}
where $g$ is the genus of the underlying manifold. Because the vertex-face correspondence requires $N_v=N_f$, the number of edge is
\begin{align}
N_e=2 N_f-2+2g.
\end{align}
which is an even number.

In canonical quantization, the commutator between a canonical coordinate and the corresponding canonical momentum is $i \hbar$. 
Therefore, for our theory, we have
\begin{align}
 \left[A_e, \frac{k}{2\pi}K_{e',e''}A_{e''}\right]= i\delta_{e,e'}
\end{align}
where $\delta_{e,e'}$ is the Kronecker delta and we set $\hbar$ to unity.
Multiplying both sides by the inverse matrix of $K$, we obtain the commutation relation for the vector potential
\begin{align}
 [A_e, A_{e'}]= i \frac{2\pi}{k} K^{-1}_{e',e}=-\frac{2\pi i}{k} K^{-1}_{e,e'}
\label{eq:commutator}
\end{align}
Here, we used the fact that $K^{-1}$ is an antisymmetric matrix.

In order to ensure that the commutator $[A_e, A_{e'}]$ is nonsingular, we must require the $K$ matrix being invertible.

\subsection{Paths, contractible and noncontractible cycles}

In this section, we will introduce two concepts from the graph theory: {\it paths and cycles}, which are discrete versions of 
curves and loops, respectively.~\cite{Wilson1998}

A {\it path} is a sequence of vertices $v_0\to v_1 \to v_2 \to \ldots \to v_m$, in which any two consecutive vertices
are connected by an edge. In the literature of graph theory it is often also assumed that a path never go through 
the same vertex twice. The length of a path is the total number of edges contained in the path. 

If $v_0\ne v_m$, the path is called {\it open}. For $v_0=v_m$, the path
is {\it closed}. A closed path (with nonzero length) is also called a {\it cycle}. 
In comparison with the continuum,  it is easy to realize that open paths
are discretized open curves, while cycles (i.e., closed paths) are discretized loops (i.e., closed curves).
More precisely, a path (cycle) corresponds to a directed curve (loop), because a path (cycle)
has a natural direction built in according to its definition, i.e. $v_0\to \ldots \to v_m$.

In the continuum, loops on a 2D manifold can be classified into two categories: contractible or noncontractible,
depending on whether or not the closed curve can be adiabatically contracted to a point. For a graph,
there is a similar classification for cycles (closed paths) using a different but equivalent definition. 
We call a closed path (i.e. a cycle) contractible, if it is the boundary of some 2D area formed by a set of faces. 
Otherwise, it is noncontractible. For 2D closed and orientable surfaces in the continuum, noncontractible loops only exist for surfaces 
with nonzero genus (torus, double torus, etc.), while all loops on a genus zero surface (e.g. a sphere) are contractible.
In graph theory, the same is true for cycles. For planar graphs defined on 2D closed and orientable surfaces, 
noncontractible cycle can only exist if the genus of the underlying 2D manifold is larger than zero.

For a directed graph (or lattice), each path ($P$) can be represented by a 
$N_e$-dimensional vector,  $\xi_P$, whose $e$th component is
\begin{align}
\xi_{P,e}=\left\{
\begin{array}{cl}
	+1 & \textrm{$e\in P$ and $e$ is along the direction of $P$} \\
	-1 & \textrm{$e\in P$ and $e$ is opposite to the direction of $P$}\\
	0 &  \textrm{$e\not\in P$}
\end{array}\right.
\label{eq:path}
\end{align}
As will be shown in below, this object defines a discretized line integral. In particular, if $P$ is a cycle, $\xi_{P,e}$
provides a discretized loop integral.

\subsection{Commutators and intersections}

For a path $P$ on a graph $G$, we can define the integral (circulation) of the vector potential along this path as
\begin{align}
\mathcal{W}_P=\xi_{P, e} A_{e}
\label{eq:loop_integral}
\end{align}
This object is the discretized version of a line integral $\int_C {\bm A} \cdot d{\bm x}$ along a path $C$.

Now, we consider two different paths, $P$ and $P'$, and we define two integrals 
$\mathcal{W}_P$ and $\mathcal{W}_{P'}$ for $P$ and $P'$, respectively, using the definition of Eq.~\eqref{eq:loop_integral}. In this section, we prove that the commutator between $\mathcal{W}_P$ 
and $\mathcal{W}_{P'}$ is determined by the number of oriented intersections between the two paths $\nu[P, P']$, 
\begin{align}
[\mathcal{W}_P,\mathcal{W}_{P'}]=\frac{2\pi i}{k} \nu\left[P, {P'}\right],
\label{eq:path_commutator}
\end{align}
which is  the direct analog of the corresponding commutator of the Chern-Simons theory 
in the continuum, Eq.~\eqref{eq:curve_commutator}.

Utilizing the commutator of Eq.~\eqref{eq:commutator}, we find 
\begin{align}
[\mathcal{W}_P,\mathcal{W}_{P'}]=-\frac{2\pi i}{k} \xi_{P,e}\xi_{P',e'}  K^{-1}_{e,e'}
\label{eq:path_commutator1}
\end{align}
If the two paths $P$ and $P'$ share no common vertex, the intersection 
number is obviously zero $\nu[P,P']=0$. In the same time, $[\mathcal{W}_P,\mathcal{W}_{P'}]$ also vanishes,
because every term on the r.h.s. of Eq.~\eqref{eq:path_commutator1} is zero.

\begin{figure}[t]
        \subfigure[A right-handed intersection ($\nu=+1$)]{\includegraphics[width=.45\linewidth]{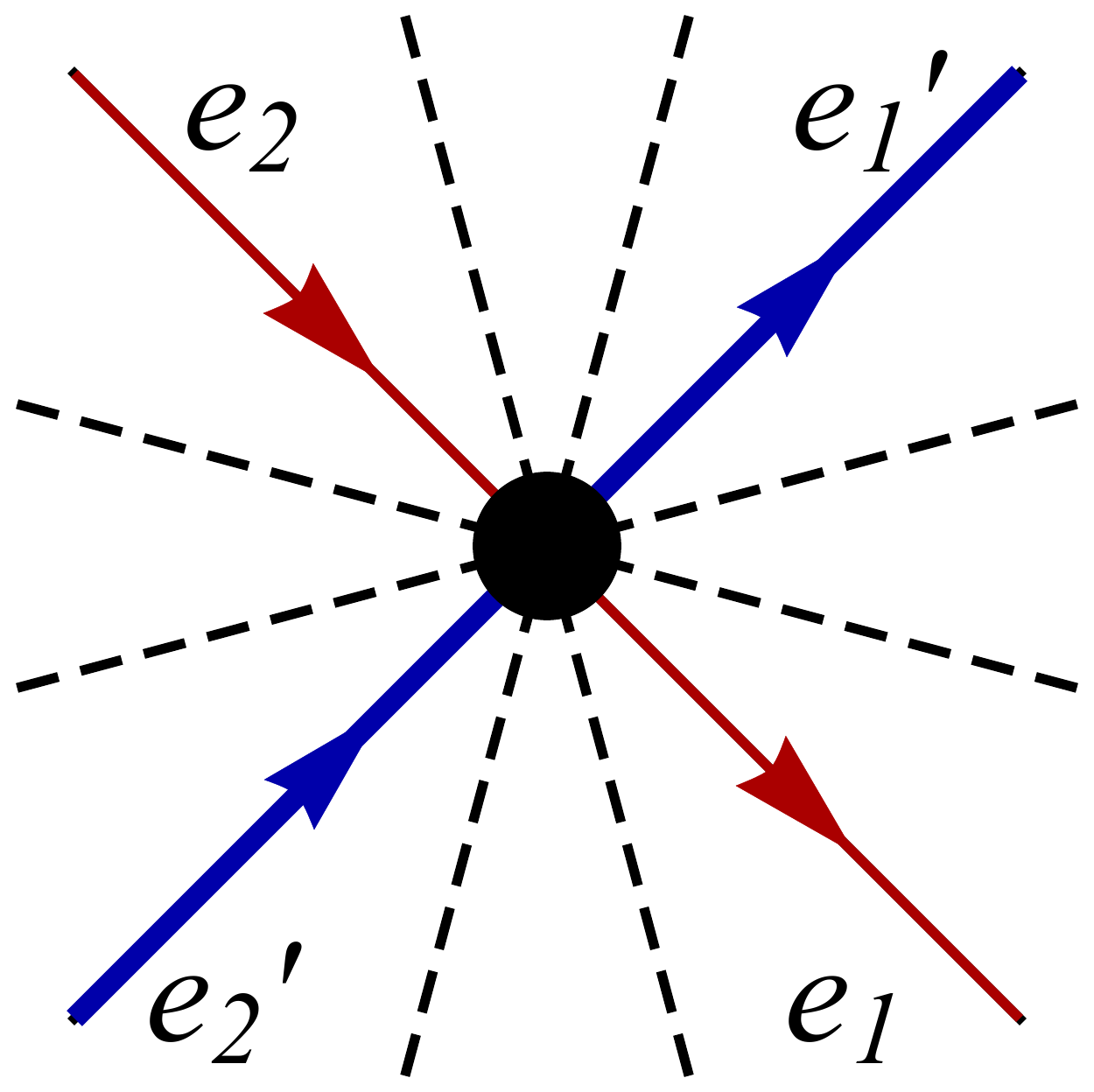}
                \label{fig:common_vertex1}}
        \subfigure[A left-handed intersection ($\nu=-1$)]{\includegraphics[width=.45\linewidth]{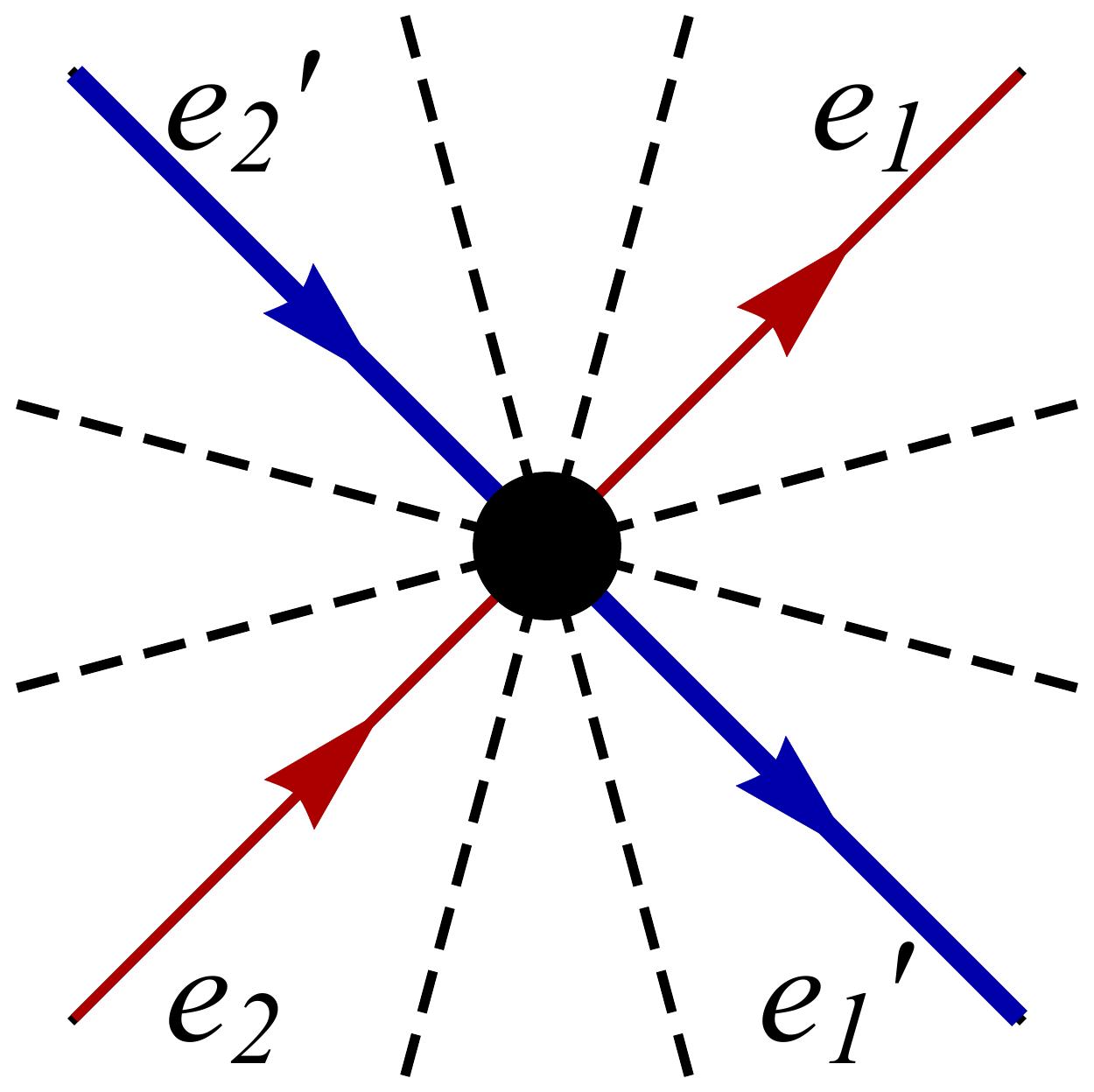}
                \label{fig:common_vertex2}}
 \subfigure[No intersection ($\nu=0$)]{\includegraphics[width=.45\linewidth]{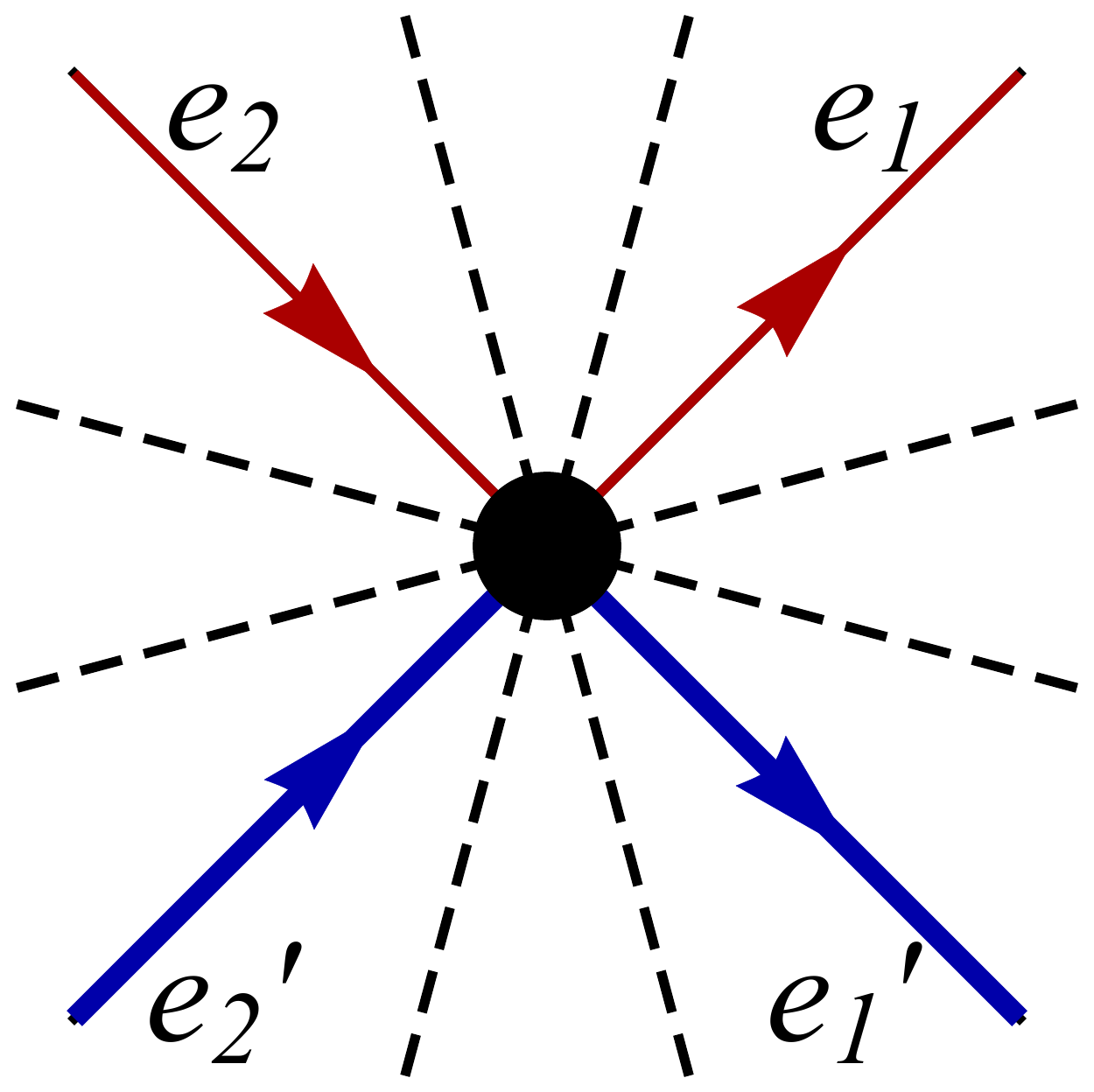}
                \label{fig:common_vertex3}}
 \subfigure[A special case ($\nu=?$)]{\includegraphics[width=.45\linewidth]{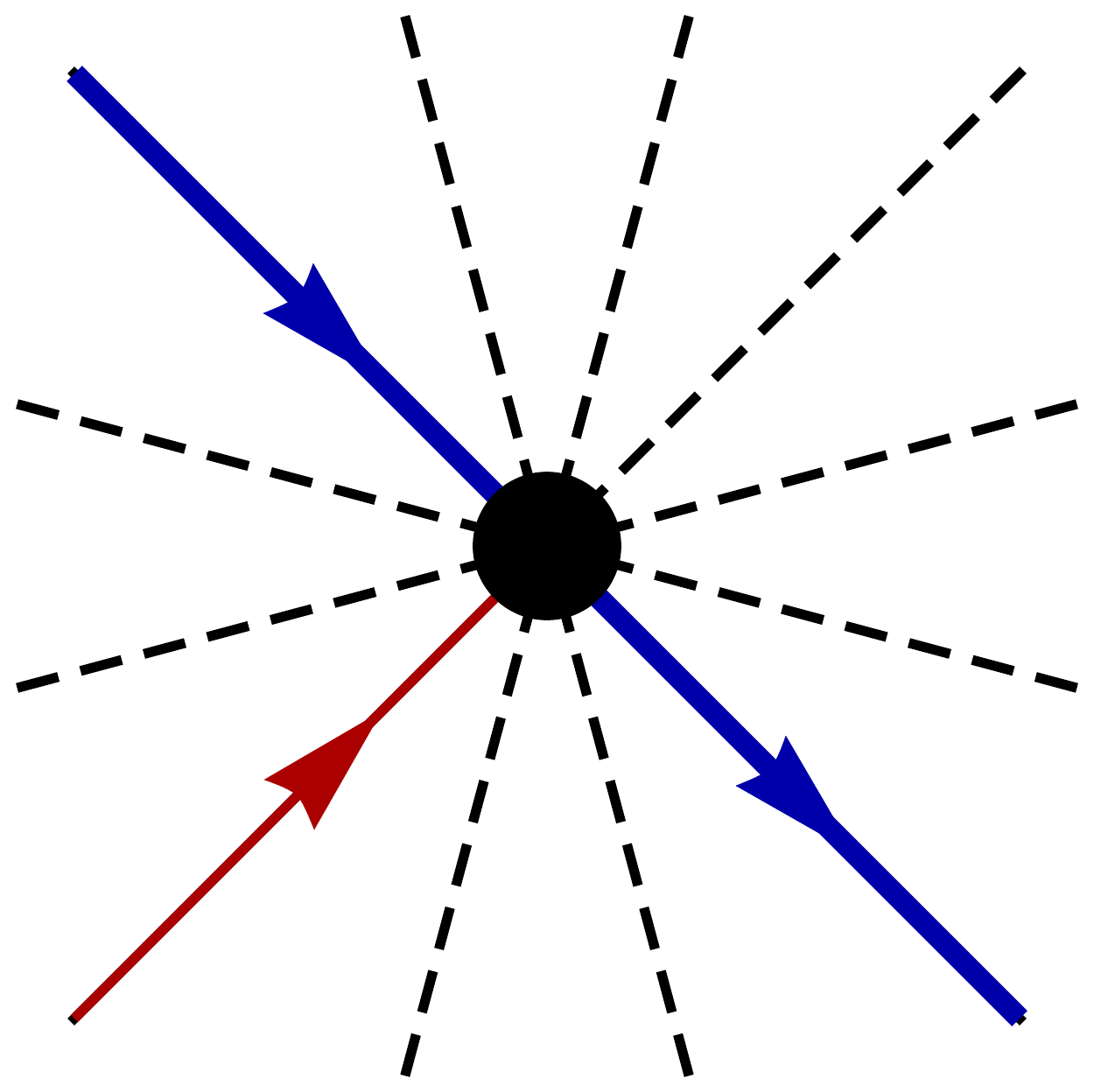}
                \label{fig:common_vertex4}}
\caption{(Color online) One common vertex shared by two paths. Here, we consider two paths $P$ (thin red solid lines) and $P'$ (thick
blue solid lines). The arrows indicate the direction of each path. The disk in the middle is one common vertex shared 
by the two paths. Dashed lines represent other (possible) edges that are connected to the vertex, and they 
don't contribte to the commutator that we want to compute. 
Figure (a) shows a right-handed intersection between $P$ and $P'$ and Fig.~(b) is a left-handed one. 
In Fig.~(c), the two paths don't intersect. Figure~(d) shows a special case, where one path terminates
at this vertex. Here, the number of intersection can be $\pm1$ or $0$ depends on microscopic details.
In Sec.~\ref{sub:sec:gauge_symmetry_commutator}, a method will be introduced 
to obtain the value of $\nu$ for Fig.~(d) by defining a dual path in the dual graph.
}
\label{fig:common_vertex}
\end{figure}

If the two paths share some common vertices, only edges connected to these common vertices contribute to 
the commutator of Eq.~\eqref{eq:path_commutator1}, because $K^{-1}_{e,e'}=0$ for all other edges. 
Therefore, we only need to consider edges adjacent to each common vertex. As shown in 
Fig.~\ref{fig:common_vertex}, we shall distinguish three different situations, shown in  
Figs.~\ref{fig:common_vertex1} to~\ref{fig:common_vertex3} respectively, depending on 
whether the common vertex is a right-handed intersection, a left-handed intersection, or not an intersection.
In Fig. \ref{fig:common_vertex} we label the edges of $P$ as $e_1$ and $e_2$, while the edges of $P'$ are called $e'_1$ and $e'_2$. 
Using Eq.~\eqref{eq:path_commutator1}, the commutator is given by
\begin{align}
[\mathcal{W}_P,\mathcal{W}_{P'}]
=&\frac{2\pi i}{k}\sum_{i=1}^2 \sum_{j=1}^2\xi_{P,e_i}\xi_{P',e'_i}  K^{*}_{e_i,e'_i}\nonumber\\
=&-\frac{2\pi i}{k}\frac{1}{2}\sum_{i=1}^2 \sum_{j=1}^2\xi_{P,e_i}\xi_{P',e'_j}  \eta^{*}_{1;e_i,e'_j}\eta^{*}_{2;e_i,e'_j}
\label{eq:path_commutator2}
\end{align}
Here, we used the fact that $K^{-1}=-K^*$ and each element of $K^*$ can be written as $-\eta^*_1 \eta^*_2/2$ as defined
in Sec.~\ref{sec:dual}.
For the first three figures in Fig.~\ref{fig:common_vertex}, it is easy to verify that
\begin{align}
\xi_{P,e_i}\xi_{P',e'_j}\eta^{*}_{2;e_i,e'_j}=
\left\{
\begin{array}{cl}
	+1 & \textrm{if $i=j$} \\
	-1 & \textrm{if $i\ne j$}\\
\end{array}\right.
\label{eq:path_commutator_eta2}
\end{align}

If the common vertex is a right-handed intersection of $P$ and $P'$, Fig.~\ref{fig:common_vertex1},  four possible cases need to be considered
depending on the location of the face that paired up with the common vertex,
i.e. 
(1) between $e_1'$ and $e_2$, (2) between $e_2$ and $e_2'$, (3) between $e_2'$ and $e_1$ and (4) between $e_1$ and $e_1'$.
For case (1), we have $\eta^*_{1;e_1,e'_1}=-1$, $\eta^*_{1;e_1,e'_2}=+1$, $\eta^*_{1;e_2,e'_1}=-1$ and $\eta^*_{1;e_2,e'_2}=-1$.
For case (2), we have $\eta^*_{1;e_1,e'_1}=-1$, $\eta^*_{1;e_1,e'_2}=+1$, $\eta^*_{1;e_2,e'_1}=+1$ and $\eta^*_{1;e_2,e'_2}=+1$.
For case (3), $\eta^*_{1;e_1,e'_1}=-1$, $\eta^*_{1;e_1,e'_2}=-1$, $\eta^*_{1;e_2,e'_1}=+1$ and $\eta^*_{1;e_2,e'_2}=-1$.
For case (4), $\eta^*_{1;e_1,e'_1}=+1$, $\eta^*_{1;e_1,e'_2}=+1$, $\eta^*_{1;e_2,e'_1}=+1$ and $\eta^*_{1;e_2,e'_2}=-1$.
Using Eqs.~\eqref{eq:path_commutator2} and~\eqref{eq:path_commutator_eta2}, we find that  for all these four cases, 
the commutator $[\mathcal{W}_P,\mathcal{W}_{P'}]=2\pi i/k$. Therefore, we find that each
right-handed intersection contribute $2\pi i/k$ to the commutator.

Using the same technique, we can prove $[\mathcal{W}_P,\mathcal{W}_{P'}]=-2\pi i/k$ for Fig.~\ref{fig:common_vertex2}, 
and $[\mathcal{W}_P,\mathcal{W}_{P'}]=0$ for Fig.~\ref{fig:common_vertex3}. In summary, we find that each right-handed 
(left-handed) intersection contribute $+2\pi i/k$ ($-2\pi i/k$) to the commutator $[\mathcal{W}_P,\mathcal{W}_{P'}]$, and thus 
we proved Eq.~\eqref{eq:path_commutator}.

\subsection{Gauge invariance and the commutation relations}
\label{sub:sec:gauge_symmetry_commutator}

In this section, we prove that the commutation relations of Eq.~\eqref{eq:path_commutator} arise naturally,  
if we require the action to be gauge invariant, Eq.~\eqref{eq:gauge_invariance}.
In addition, a by-product of this proof offers a more rigorous definition for the number of oriented intersections, which eliminates 
the ambiguity demonstrated in Fig.~\ref{fig:common_vertex4}. There, the two path $P$ and $P'$ barely touch each other. 
Shall this counts as an intersection? This question will be answered in this section.

\begin{figure}[hbt]
	\includegraphics[width=.64\linewidth]{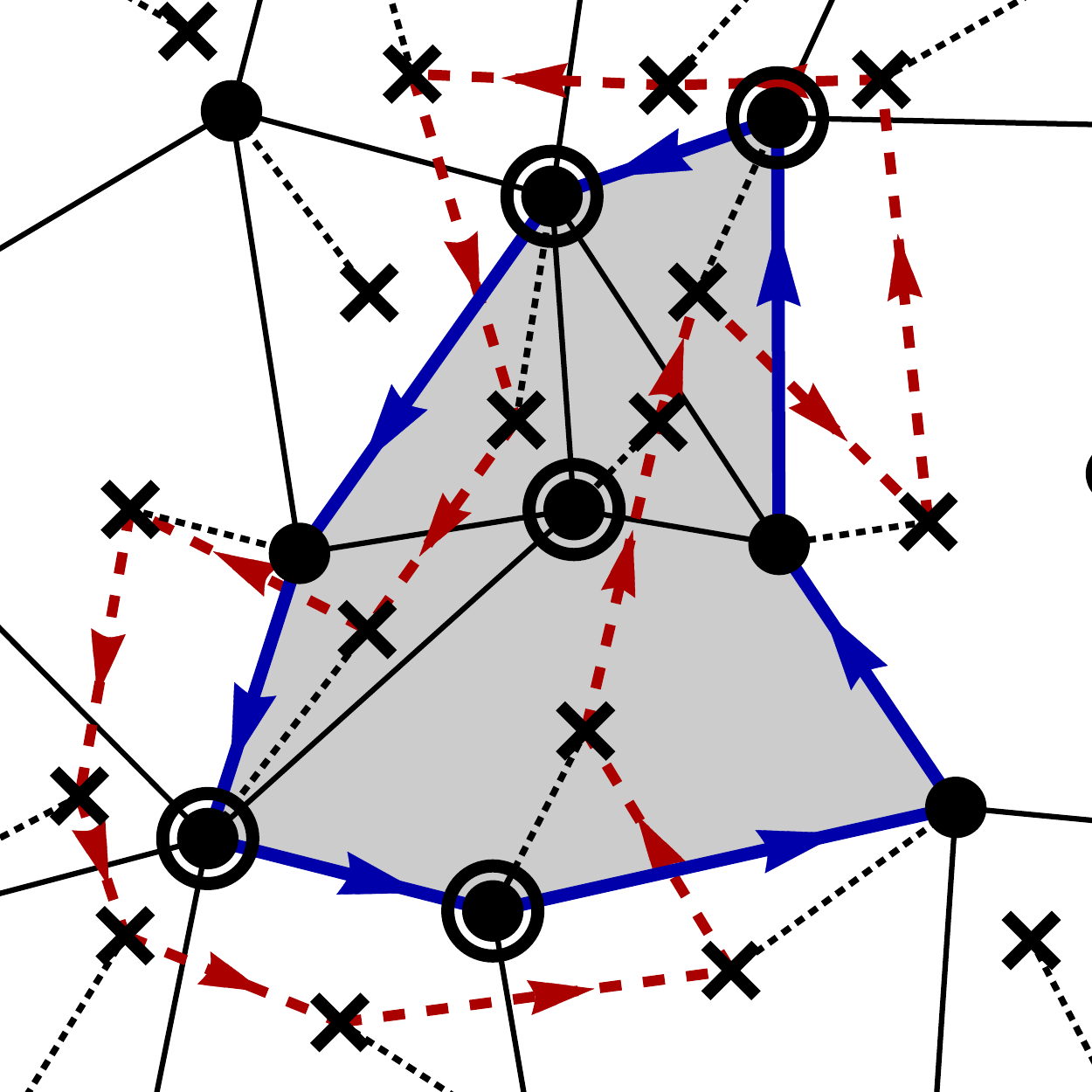}
\caption{(Color online) A cycle in a graph and the dual cycle in the dual graph. Here, we consider a planar graph with a local
face-vertex correspondence. The vertices in the original graph is marked by disks, while the crosses label the faces, i.e.
vertices of the dual graph. The local face-vertex correspondence is marked using the dotted lines, which pair up each face
with one of its neighboring vertex. The thick solid (blue) lines marks a contractible cycle (a closed path) and the orientation 
of the cycle is marked by the arrows. For a contractible cycle, its interior is formed by a set of faces (dark region). For each face
inside the dark region, we find the corresponding vertex using the local face-vertex correspondence. These vertices are marked by
circles. Then, we draw a loop in the dual graph, which encloses these vertices 
(red dashed lines connecting neighboring crosses). This loop in the dual graph
is the dual of the original loop in the original graph. And we require the two loops to have the same orientation.}
\label{fig:dual_loop}
\end{figure}

Consider two paths $P$ and $P'$. Here we assume that one of the paths is a contractible cycle ($P'$), 
while the other is an open path with two open ends ($P$). As an example, a contractible cycle $P'$ is plotted in Fig.~\ref{fig:dual_loop}. 
Because $P'$ is contractible, it is the edge of an area formed by a set of faces (the dark region in Fig.~\ref{fig:dual_loop}).
Utilizing the vertex-face correspondence, this set of faces are mapped to a set of vertices, which are marked
by circles in Fig.~\ref{fig:dual_loop}. Now we can define a cycle in the dual lattice such that the cycle encloses (and only encloses) these 
vertices (the dashed lines in Fig.~\ref{fig:dual_loop}). This new cycle will be called the dual of $P'$ and will be labeled as $P'^*$. 
Here, we choose the direction of $P'^*$ such that its orientation is the same as that of $P'$. 
Below, we will prove that the gauge invariance immediately implies the commutator
\begin{align}
[\mathcal{W}_P,\mathcal{W}_{P'}]=\frac{2\pi i}{k} \nu\left[P,P'^*\right],
\label{eq:path_commutator_dual}
\end{align}
Here, instead of the number of intersections between $P$ and $P'$, we shall count the number of intersections
for $P$ and $P'^*$. Because the cycle $P'^*$ is defined in the dual graph, the number of intersections is always 
well-defined and this eliminates the ambiguity shown in Fig.~\ref{fig:common_vertex4}.

Before proving Eq.~\eqref{eq:path_commutator_dual}, we would like to highlight that although the dual cycle $P'^*$ and 
the original cycle $P$ are not identical, the difference between them is local and microscopic. 
This comes from the fact that our vertex-face correspondence is {\it local}, where a face is paired with one of its neighboring vertex. 
If we take the continuous limit and ignore differences at the microscopic level, the differences between $P'^*$ and $P'$ vanishes,
and therefore, we recover Eq.~\eqref{eq:curve_commutator}.

Now we prove Eq.~\eqref{eq:path_commutator_dual}. First, we define a $N_f$-dimensional vector $Q_{P'}$ for the contractible cycle $P'$, 
whose the $f$th component $Q_{P',f}$ is 
\begin{align}
Q_{P',f}=\left\{
\begin{array}{cl}
	1 & \textrm{if the face $f$ is enclosed by $P'$} \\
	0 &  \textrm{if the face $f$ is outside of $P'$}
\end{array}\right.
\end{align}
With this matrix $Q_{P',f}$, the contractible cycle $P'$ can be written as
\begin{align}
\xi_{P',e}=Q_{P',f}\xi_{f,e}
\label{eq:path_face}
\end{align}
where $\xi_{f,e}$ is defined in Eq.~\eqref{eq:face_edges} and $\xi_{P,e}$ is define in Eq.~\eqref{eq:path}.
The proof for Eq~\eqref{eq:path_face} is straightforward. For the r.h.s., it is easy to notice that for any $e$ outside the
region enclosed by the cycle $P'$, 
$Q_{P',f}\xi_{f,e}=0$. For an edge inside the region enclosed by the cycle $P'$, 
it will induce two terms for the r.h.s., because each edge is shared by two faces. 
These two terms have opposite signs and thus cancel out, and thus $Q_{P',f}\xi_{f,e}=0$. The only way to have a nonzero
$Q_{P',f}\xi_{f,e}$ is to require that $e$ is an edge of the cycle $P'$. And it can be verified that the value and the sign 
of $Q_{P',f}\xi_{f,e}$ match exactly $\xi_{P',e}$.

Similarly, in the dual space, we can write down the dual cycle $P'^*$ as,
\begin{align}
\xi^*_{P'^*,e^*}=Q^*_{P'^*,f^*}\xi^*_{f^*,e^*}=-Q^*_{P'^*,v}D_{v,e}
\end{align}
where $Q^*_{P'^*,f^*}=1$ for any faces (of the dual graph) inside the dual cycle $P'^*$. 
Here, we relabeled the faces in the dual graph ($f^*$) using corresponding vertices in the original graph ($v$). 
We also used the fact that $\xi^*_{f^*,e^*}=-D_{v,e}$ as shown in Eq.~\eqref{eq:dual_of_xi}.
Because the vertices (of the original graph) enclosed by $P'^*$ are partners of the faces enclosed by $P'$, we have
\begin{align}
Q^*_{P'^*,v}=Q_{P',f}M_{v,f}
\end{align}
By combining the two equations above and relabeling $e^*$ as $e$, we find that
\begin{align}
\xi^*_{P'^*,e}=-Q_{P',f}M^{-1}_{f,v}D_{v,e}.
\label{eq:dual:cycle}
\end{align}
In the last step, we used the fact that the $M$ matrix is orthogonal, $M_{v,f}=M_{f,v}^{-1}$.

By substituting Eq.~\eqref{eq:path_face} into Eq.~\eqref{eq:path_commutator1}, we find that 
\begin{align}
[\mathcal{W}_P,\mathcal{W}_P']
=&-\frac{2\pi i}{k} \xi_{P,e} K^{-1}_{e,e'}\xi_{P',e'}
\nonumber\\
=&-\frac{2\pi i}{k} \xi_{P,e}  K^{-1}_{e,e'} Q_{P',f}\xi_{f,e'}
\nonumber\\
=&-\frac{2\pi i}{k} \xi_{P,e} Q_{P',f} M^{-1}_{f,v}  D_{v,e}.
\end{align}
Here, we utilized the condition of gauge invariance $M^{-1}_{f,v}  D_{v,e}=K^{-1}_{e,e'} \xi_{f,e'}$,
Eq.~\eqref{eq:gauge_invariance_inverse}.
Using Eq.~\eqref{eq:dual:cycle}, the r.h.s. can be written as
\begin{align}
[\mathcal{W}_P,\mathcal{W}_P']
=\frac{2\pi i}{k} \xi_{P,e} \xi^*_{P'^*,e}.
\end{align}
It is easy to verify that only intersections between $P$ and $P'^*$ contribute to the r.h.s. of the equation. At a right-handed/left-handed 
intersection, $\xi_{P,e} \xi^*_{P'^*,e}=\pm 1$, and thus 
\begin{align}
\left[\mathcal{W}_P,\mathcal{W}_P'\right]=\frac{2\pi i}{k} \nu\left[P,P'^*\right].
\end{align}


\section{Wilson loops for non-contractible cycles}
We start this section by considering a planar graph embedded on a 2D torus (with genus $g=1$). For this graph, there are two independent
non-contractible cycles (i.e. discretized counterparts of the two non-contractible loops on a torus), which will be labeled as $C$ and $C'$ in this section. These two cycles intersect once with each other. Without loss of generality, we choose the oriented intersection number to be $+1$, instead of $-1$, 
i.e., $\nu[C,C']=+1$.
As we  proved above in Eq.~\eqref{eq:path_commutator}, 
the commutator $[\mathcal{W}_C,\mathcal{W}_{C'}]=2\pi i/k$.

Here we define Wilson loops for the two non-contractible cycles $C$ and $C'$ of the torus
\begin{align}
W_C=&\exp(i \mathcal{W}_C)
\\
W_{C'}=&\exp(i \mathcal{W}_C')
\end{align}
Because the commutator $[\mathcal{W}_C,\mathcal{W}_{C'}]=2\pi i/k$ is a complex number (i.e. is proportional to the identity operator), 
it commutes with both $\mathcal{W}_C$ and $\mathcal{W}_C'$. Hence, using the Baker-Hausdorff-Campbell formula it follows that
\begin{align}
e^{i \mathcal{W}_C}e^{i \mathcal{W}_C'}=
e^{i \mathcal{W}_C'}e^{i \mathcal{W}_C} e^{[i \mathcal{W}_C,i \mathcal{W}_{C'}]}
\end{align}
and thus
\begin{align}
W_{C} W_{C'}=W_{C'} W_{C} e^{-2\pi i/k}
\label{eq:wilson_loop_commutation}
\end{align}

If we consider an eigenstate of $W_{C}$ with eigenvalue $w$,
\begin{align}
W_{C}|\Psi\rangle= w |\Psi\rangle
\end{align}
where $w$ is a complex number, utilizing Eq.~\eqref{eq:wilson_loop_commutation},
it is straightforward to show that $W_{C'} |\Psi\rangle$ is also an eigenstate of $W_{C}$ and its
eigenvalue is $w  e^{-2\pi i/k}$ 
\begin{align}
W_{C} (W_{C'}  |\Psi\rangle)= w e^{-2 \pi i/k} (W_{C'} |\Psi\rangle)
\end{align}
In other words, we can consider $W_{C'}$ as a raising/lowering operator for the operator $W_{C}$, and vice versa. 
Starting from the eigenstate $|\Psi\rangle$, eigenstates of $W_{C}$ can be generated
by applying this raising/lowering operator,
\begin{align}
W_{C} (W_{C'}^n  |\Psi\rangle)= w e^{-2 n \pi i/k} (W_{C'}^n |\Psi\rangle)
\end{align}
 i.e. $W_{C'}^n |\Psi\rangle$ is an eigenstate with eigenvalue $w e^{-2n \pi i/k}$

For an integer $k$, it is easy to note that when $n=k$, the state $W_{C'}^k |\Psi\rangle$ has the same eigenvalues as $|\Psi\rangle$. 
If $W_{C'}^k |\Psi\rangle$ and $|\Psi\rangle$ are the same quantum state, $W_{P'}^n |\Psi\rangle$ generates $k$ different
eigenstates of $W_{P}$. From this results it follows the well known result  that a Chern-Simons gauge theory has a $k$-fold topological degeneracy on
a torus. This conclusion is well known in the continuum. Our discussion above shows that the same is true in our
discretized theory.

It is straightforward to generate the discussion above to other 2D manifolds with different genus. For a planar graph defined on a 2D surface with 
genus $g$, there are $2 g$ independent non-contractible cycles. As will be discussed in Sec.~\ref{sec:Nv_Nf} (Fig.~\ref{fig:tri_torus}), 
we can choose $g$ of these cycles such that they don't intersect with each other, $C_1$, $C_2$, \ldots, $C_g$. The other $g$ non-contractible 
cycles will be labeled $C'_1$, $C'_2$, \ldots, $C'_g$. The absence of intersection for cycles $C_i$ ($i=1,2, \ldots, g$) implies that the Wilson loops 
defined on this cycles commute with each other, and thus we can consider common eigenstates for 
$W_{C_1}$, $W_{C_2}$, $\ldots$, $W_{C_g}$. The Wilson loops
for the other $g$ non-contractible cycles serve as raising and lowering operators. Starting from one common eigenstate of all $W_{C_i}$s, we can use $W_{C'_i}$
to generate $k^g$ eigenstates (including the original one), which reflects the topological degeneracy of the Chern-Simons gauge theory on a surface
with genus $g$.


\section{Locality}
\label{sec:locality}

In this section, we verify that our theory is {\it local}. More precisely, we prove that
(1) our action is local, i.e. the action does not  have any coupling between fields that are not around the same face,
(2) the flux attachment is local, i.e. for a charge at the vertex $v$, its magnetic flux must be located on a neighboring face,
and (3) the commutator between vector potentials is {\it local}, i.e. for any two edges that do share a common vertex, 
the vector fields defined on them commute with each other.

\subsection{The action}

In the discretized action of Eq.~\eqref{eq:action} there are no long range couplings beyond edges and vertices of the same face.
In the first term in Eq.~\eqref{eq:action}, because $M_{f,v}$ vanishes unless $f$ and $v$ are adjacent to each other, 
the action only contains couplings between nearby $A_v$ and $A_e$ (i.e. $e$ and $v$ must belong to the same face).
For the second term, we know that $K_{e,e'}=0$, if $e$ and $e'$ do not belong to the same face, and therefore, 
only short-range coupling (for edges of the same face) is included in this term.

\subsection{Flux attachment}

For the Chern-Simons gauge theory in the continuum, the flux attachment is {\it local}, i.e. for a point charge at $r_0$,
the magnetic field is a delta function $B \propto \delta(\vec{r}-\vec{r}_0)$, and the $B$ field vanishes when we
more away from the point charge.

In our discretized theory, this condition of locality is preserved to the maximum extent. As shown in Eq.~\eqref{eq:flux_attachment_2},
for a charge at the site $v$, the corresponding magnetic field only is present inside {\it a single face}, which is the closest analog of 
a delta function in a discrete setup. As for the relative locations of the charge and its flux, because these two objects  on different parts
of the graph (charges on vertices and fluxes on faces), it is impossible to require their location to coincide. Instead,
we require the charge and the flux to be adjacent to each other.

We emphasize that  this locality condition plays a very important role, if we use the Chern-Simons gauge theory as a statistical field 
to change the statistics for matter fields coupled to it. To ensure that all particles have the correct statistics, when
we move a particle $A$ around another particle $B$, $A$ must feel all the statistical field of $B$.
In other words, no matter which path we choose, as long as $A$ moves around $B$, the magnetic flux attached to $B$ must be 
enclosed by the path of $A$. For our theory (and for the continuous Chern-Simons gauge theory), 
this is always true. However, if one were to violate the locality condition by putting the magnetic flux in a face not adjacent to the charge, 
it would be possible to move $A$ around $B$ without enclosing the flux inside the path. 
As a result, the statistics of the matter field would become ill defined.

\subsection{The commutation relations}

As shown in Eq.~\eqref{eq:commutator}, the commutator for the vector potential is determined by the inverse of the $K$ matrix 
(or say the dual matrix $K^*$). As proved in Sec.~\ref{sec:dual}, 
the $K^{-1}$ matrix is also local, where the matrix element $K^{-1}_{e,e'}=0$, if $e$ and $e'$ does not  share a common vertex.

This results implies that a nonzero commutator can only arise for two neighboring edges, while for two edges separated away from each other 
(i.e. not sharing a common vertex), the vector potential always commutes with each other.

\section{Why $N_v=N_f$?}
\label{sec:Nv_Nf}

Above, we have shown that the existence of a local vertex-face correspondence is sufficient for the discretization of
the Chern-Simons gauge theory. In this section, we prove that this condition is also necessary, 
if we want to preserve  key properties of the Chern-Simons theory.

Let us consider a generic discretized action of gauge fields $A_v$ and $A_e$. Just as in the Chern-Simons gauge theory in the continuum,
we assume that the action does not  contain time derivatives of the time component of the gauge field $A_v$, and that  $A_v$ plays the role of a Lagrange multiplier field that enforces a constraint on the local flux. For the coupling among the components of the gauge fields $A_e$ on different edges, the action only contains product between $A_e$ and $\partial_t A_{e'}$.
We ignore possible terms with higher orders in time derivatives, which are less relevant in the sense of the renormalization group.
In addition, we will only keep terms to the leading order in our action.

With these assumptions, the most generic action that one can write down is
\begin{align}
S=\frac{k}{2\pi} \int dt \left[A_{v} M_{v,f} \xi_{f,e} A_e-\frac{1}{2} A_{e} K_{e,e'} \dot{A}_{e'}\right]
\label{eq:action_generic}
\end{align}
This action is very similar to the action we constructed above in Eq.~\eqref{eq:action}. However, we must emphasize that
here $M$ and $K$ are generic matrices, and that so far we are not putting any constraints on them.
Most importantly, now we don't require the graph to support a local vertex-face correspondence. 
Instead, we will consider generic situation and show that if we want the action to take this form, then the local vertex-face correspondence
will arise naturally.

In Sec.~\ref{sub:sec:edge_space}, we first introduce some mathematical tools from algebraic graph theory. Then,
in Sec.~\ref{sub:sec:N_v>=N_f}, we will prove that the number of faces 
cannot exceed the number of vertices ($N_v\ge N_f$), and otherwise the theory will be singular.
Then, in Sec.~\ref{sub:sec:N_v<=N_f}, we show that
the flux attachment requires the number of vertices not to exceed the number of faces ($N_v\le N_f$). Combining these
two conclusions together, we find that the graph must have the same number of vertices and faces ($N_v=N_f$).
Finally, in Sec.~\ref{sub:sec:local_correspondence}, we prove that a local vertex-flux correspondence is necessary, if
 we further require the flux attachment to be local.

\subsection{Edge-space, Cut-space and Loop-space}
\label{sub:sec:edge_space}

Here we introduce some concepts from the algebra graph theory,~\cite{Biggs1974} that will be used later.
In algebraic graph theory, an $N_e$-dimensional vector  represents each edge $e$ of a graph $G$, 
\begin{align}
\epsilon_e=(0,0,\ldots,1,\ldots, 0),
\label{eq:basis}
\end{align}
where the $e$th component of the vector is $1$ and all other components are $0$.
These vectors form the basis of a $N_e$-dimensional linear space, which is called the {\it edge space} of the graph $G$.
It is easy to realize that the $K$-matrix defined above is a rank-$2$ tensor in this linear space.

For a directed graph (i.e., a digraph), each (contractible or non-contractible) cycle $C$ can be represented as an 
$N_e$-dimensional vector,  $\xi_C$, whose $e$th component is
\begin{align}
\xi_{C,e}=\left\{
\begin{array}{cl}
	+1 & \textrm{$e\in C$ and $e$ is along the direction of $C$} \\
	-1 & \textrm{$e\in C$ and $e$ is opposite to the direction of $C$}\\
	0 &  \textrm{$e\not\in C$}
\end{array}\right.
\end{align}
These vectors  span a linear space, which is a subspace of the edge space. In algebraic graph theory, this subspace is known as 
the {\it circuit-subspace}.

A {\it cutset} is a set of edges, where if we cut all the edges in a cutset, the graph is cut into two disconnected pieces.
A more rigorous definition of a cutset relies on a partition of vertices. If $V$ is the set of all vertices of a graph $G$, 
we can separate these vertices into two subset $V_1$ and  $V_2$, such that $V_1\cup V_2=V$
and $V_1 \cap V_2=0$. This is called a partition of the set $V$. 
For each partition of $V$, we can define a cutset by collecting all edges of $G$ that have one end in $V_1$ and the other in $V_2$.
For a digraph, one can choose one of the two possible orientations for a cutset by specifying the vertices in $V_1$ (or $V_2$)
to be the positive ends, while the other to be negative. If an edge in the cutset points to the positive end of the cutset,
it is a positive edge in this cutset. Otherwise, it is a negative edge.

Similar to cycles discussed above, each cutset can also be represented by 
an $N_e$-dimensional vector $\xi_H$, whose $e$th component is
\begin{align}
\xi_{H,e}=\left\{
\begin{array}{cl}
	+1 & \textrm{$e\in H$ and $e$ is a positive edge} \\
	-1  & \textrm{$e\in H$ and $e$ is a negative edge} \\
	0   & \textrm{$e\not\in H$}
\end{array}\right.
\end{align}
The linear space spanned by these vectors is known as the {\it cut-subspace}, which is also a subspace of 
the edge space. For a planar graph, each cutset corresponds to a contractible cycle in the dual graph.

In algebraic graph theory, it is shown that the edge space is the direct sum of the circuit-subspace and the cut-subspace.
In Appendix \ref{app:edge_space} we provide a proof for the planar graphs considered here.
This result implies that for the edge space, instead of using the basis shown above in Eq.~\eqref{eq:basis},
we can choose a new basis for the edge space by selecting a complete basis of the circuit-subspace and 
a complete basis of the cut-subspace.

For planar graphs, we can use all independent (contractible or noncontractible) cycles to form a basis for the circuit-subspace.
For the cut-subspace, all independent contractible cycles in the dual graph forms a complete basis. 
Therefore, we can span the edge space using these loops.
Using this new basis, we can rewrite all tensors defined on the edge space, including the $K^{-1}$ matrix, which will be done
in the next section.

\subsection{$N_v \ge N_f$}
\label{sub:sec:N_v>=N_f}

We will now prove that for  the $K$ matrix to be nonsingular and the discretized theory to preserve
the correct commutation relation of Eq~\eqref{eq:curve_commutator}, the number of faces can never exceed
the number of vertices. 
Using the generic action shown in Eq.~\eqref{eq:action_generic} (remember that $K$ and $M$ are now two arbitrary matrices),
we find that for the generic setup, we shall still expect the commutation relation
\begin{align}
 [A_e, A_{e'}]=-\frac{2\pi i}{k} K^{-1}_{e,e'}
\label{eq:commutator_generic}
\end{align}
Because singularities in the commutation relations must be avoided, the $K$ matrix must be invertible.
In addition, if we consider two cycles (loops) $C$ and $C'$, we shall expect the commutation relation
\begin{align}
[\mathcal{W}_C,\mathcal{W}_{C'}]=\frac{2\pi i}{k} \nu\left[C, C'\right],
\label{eq:cycle_commutator_generic}
\end{align}
As shown above, this commutator is a topological invariant and it is one of the key feature of the
Chern-Simons gauge theory. Thus, we will require Eq.~\eqref{eq:cycle_commutator_generic} for our discretized theory.

Below, we prove that if we assume the topologically invariant commutation relation, Eq.~\eqref{eq:cycle_commutator_generic}, 
then the $K$ matrix must be singular if $N_v<N_f$. Therefore, we must have $N_v\ge N_f$.
We will start from a genus zero surface and then expand the conclusion to other surfaces with nonzero genus.

\subsubsection{Graphs on a genus zero surface}

Here, we consider graphs defined on a genus zero surface (a sphere). 
Instead of directly showing that the $K$ matrix is singular for $N_v<N_f$, here we take a different but equivalent approach.
We will start by assuming the $K$ matrix is invertible and work with the $K^{-1}$ matrix. Then, using the commutation
relation, we will show that the determinant of $K^{-1}$ matrix is zero for $N_v<N_f$, and thus the $K$ matrix is singular.

Using Eqs.~\eqref{eq:commutator_generic} and~\eqref{eq:cycle_commutator_generic},
we know that
\begin{align}
K^{-1}_{e,e'} \xi_{C,e} \xi_{C',e'}=-\nu\left[C, C'\right]
\label{eq:crossing_generic}
\end{align}

Here, we choose a new  basis set for the edge space. Instead of using the vectors shown in Eq.~\eqref{eq:basis},
we use a set of vectors $\xi_i$ with $i=1,2,\ldots,N_e$. 
For $i=1,2,\ldots N_f-1$, $\xi_i$  are independent cycles, i.e. they form a complete basis of the circuit-subspace.
For $i=N_f, N_f+1, \ldots, N_e$, the corresponding $\xi_i$ are independent cutsets, i.e. they are a complete basis of 
the cut-subspace.  Using this new basis, we can define a $\tilde{K}^{-1}$ matrix as
\begin{align}
\tilde{K}_{i,j}^{-1}= K^{-1}_{e,e'} \xi_{i,e} \xi_{j,e'},
\end{align}
For $i$ and $j$ smaller than $N_f$, $\xi_{i}$ and $\xi_{j}$ are contractible cycles of the graph (for a planar defined on 
a closed orientable 2D surface with genus zero, all cycles are contractible). Using Eq.~\eqref{eq:crossing_generic},
it is easy to realize that $\tilde{K}^{-1}_{i,j}=0$ for $i$ and $j$ smaller than $N_f$. (As shown above, the number of 
oriented intersection for 
contractible loops is always zero). Therefore, we can write the $\tilde{K}^{-1}$ matrix in a block form\begin{align}
\tilde{K}^{-1}=
\left(\begin{matrix}
0 & A \\
-A^T & B
\end{matrix}
\right)
\label{eq:block_g=0}
\end{align}
Here the first block $0$ is a $(N_f-1)\times (N_f-1)$ zero matrix and $B$ is a $(N_e-N_f+1)\times (N_e-N_f+1)$
matrix. Using the Euler characteristic $N_v-N_e+N_f=2-2g$, we can rewrite the dimensions of $B$
as $(N_v-1)\times (N_v-1)$, since we have assumed the genus being zero, $g=0$. 
The block $A$ has dimension$(N_f-1)\times(N_v-1)$ and $A^T$ is the transpose of $A$. 

For a matrix with a block of zeros as shown in Eq.~\eqref{eq:block_g=0}, the determinant of the matrix must be zero, 
if the zero block is larger than the $B$ block (see Appendix Sec.~\ref{app:zero_det} for a proof). 
Therefore, if $N_v < N_f$, $\det \tilde{K}^{-1}=0$. Because $\xi_{i}$
is a complete basis for the edge space, this implies that $\det K^{-1}=0$ and thus $K$ is a singular matrix.

\subsubsection{Surfaces with nonzero genus}
\begin{figure}[t]
\includegraphics[width=.7\linewidth]{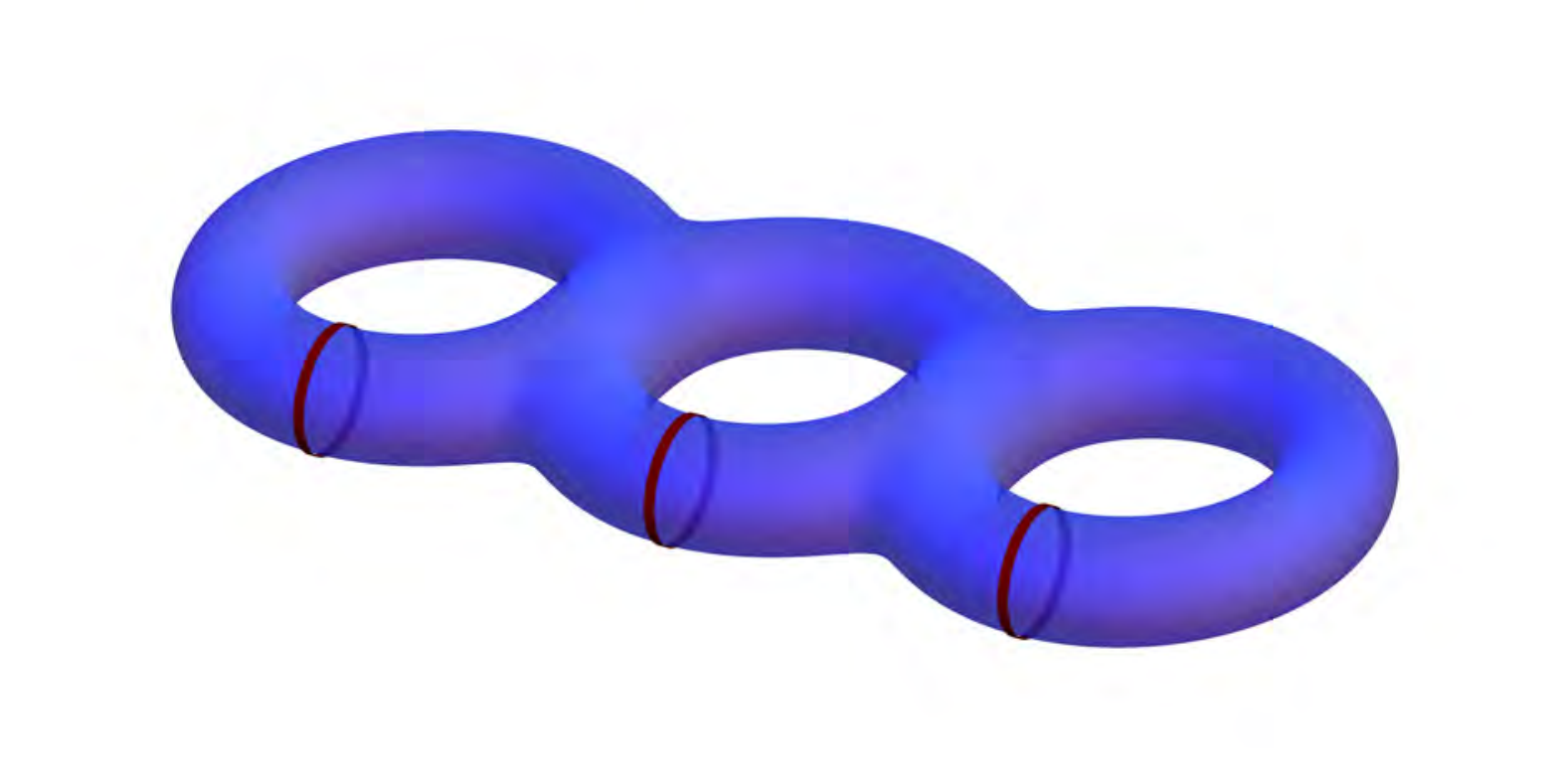}
\caption{(Color online) Non-contractible cycles on a surface with nonzero genus. For a genus $g$ surface, we can 
choose $g$ independent non-contractible cycles, which do not intersect with one another. These $g$ cycles will
be used as vector $\xi_i$ with $i=N_f, N_f+1, \ldots, N_f+g-1$, in our complete basis for the edge space. Here, we show 
an example with $g=3$. The three red loops marks three independent noncontractible cycles without any intersections.}
\label{fig:tri_torus}
\end{figure}

For a surface with nonzero genus, the same conclusion can be proved. Here, 
we choose the following basis of the edge space ${\xi_1,\xi_2,\ldots \xi_{N_e}}$. For  $i=1,2,\ldots N_f-1$, $\xi_i$ are $N_f-1$
independent contractible cycles. Then, for $i=N_f, N_f+1, \ldots, N_f+2 g-1$, $\xi_i$ are independent noncontractible cycles.
For these noncontractible cycles, we choose to have the first $g$ noncontractible cycles ($N_f \le i \le N_f+g-1$) not to cross 
with one another as shown in Fig.~\ref{fig:tri_torus}.
It is easy to realize that the first $N_f+2g-1$ vectors here form a basis of the circuit subspace, 
while the rest are chosen to be a complete basis of the cut-subspace.

Using this new basis, we can define a $\tilde{K}^{-1}$ matrix as
\begin{align}
\tilde{K}_{i,j}^{-1}=K^{-1}_{e,e'} \xi_{i,e} \xi_{j,e'},
\end{align}
For $i < N_f$ and $j< N_f+g$ (or $i< N_f+g$ and $i < N_f$), $\tilde{K}_{i,j}^{-1}=0$. This is because here 
$\xi_i$ and $\xi_j$ are to cycles of the graph $G$ and at least one of them is contractible. 
According to Eq.~\eqref{eq:crossing_generic},  $\tilde{K}_{i,j}^{-1}=0$, because the number of oriented
intersections vanish when one of the cycle is contractible.
For $N_f \le i \le N_f+g-1$ and $N_f \le j \le N_f+g-1$, $\xi_i$ and $\xi_j$ are two noncontractible cycles, 
but we have required that these cycles do not cross each other, i.e., $\nu[\xi_i,\xi_j]=0$, and thus, $\tilde{K}_{i,j}^{-1}=0$.

With this knowledge, we can write the matrix $\tilde{K}_{i,j}$ in this block form
\begin{align}
\tilde{K}_{i,j}=
\left(\begin{matrix}
0 & A \\
-A^T & B
\end{matrix}
\right)
\end{align}
Here, the upper-left conner to be a zero matrix with dimension $(N_f+g-1)\times (N_f+g-1)$. The dimension of the $B$ matrix is
$(N_e-N_f-g+1)\times (N_e-N_f-g+1)$. Utilizing the Euler characteristic $N_v-N_e+N_f=2-2g$,
we find that the dimensions of $B$ is in fact $(N_v+g-1)\times (N_v+g-1)$.

If $N_f>N_v$, again, we find that the $0$ block is larger than the block of $B$, and therefore, $\det \tilde K^{1}=0$
(see Appendix \ref{app:zero_det} for a proof).
Because $\{\xi_i\}$ is a compete basis for the edge space, this implies that $\det K^{-1}=0$, and thus $K$ is a singular matrix.

In summary, we proved that in order to preserve the commutation relations, Eq.~\eqref{eq:cycle_commutator_generic},
we must have $N_v \ge N_f$. Otherwise the $K$ matrix would be singular.

\subsection{Flux attachment and $N_v \le N_f $}
\label{sub:sec:N_v<=N_f}

Let us now prove that the flux attachment also requires $N_v\le N_f$.
Flux attachment implies that for each charge distribution, there is a corresponding unique distribution for magnetic fluxes. 
Because charge can be distributed on $N_v$ sites, to ensure that there is a corresponding flux distribution for every charge configuration,
we must have equal number or more faces to put the fluxes.

A more rigorous proof can be formulated by taking a functional derivative to the generic action Eq.~\eqref{eq:action_generic}, $\delta S/\delta A_v$,
which result in the flux attachment condition
\begin{align}
q_v=\frac{k}{2\pi} M_{v,f} \Phi_{f},
\label{eq:flux_attachment_generic}
\end{align}
If we want the flux attached to a charge to be local (i.e. the flux for a point charge occupies only a single face), 
for each vertex $v$, the $M_{v,f}$ is nonzero only for one value of $f$. As a result, the $M$ matrix defines a mapping from $v$ to $f$.

This mapping must be injective. Namely, for two different vertices, their corresponding faces must be different. This is so because if two
different vertices $v$ and $v'$ are mapped to the same face $f$, then Eq.~\eqref{eq:flux_attachment_generic} 
will require that $q_v=q_v'$, i.e. two different vertices always have the same charge, which is obviously not a physically necessary constraint.
Thus, for a injective mapping from vertices to faces, we must have $N_v\le N_f$.

\subsection{local vertex-surface correspondence}
\label{sub:sec:local_correspondence}

In the previous two subsections we proved that $N_v\le N_f$ and $N_v \ge N_f$ must hold simultaneously. Therefore, the graph must have
the same numbers of vertices and faces $N_v=N_f$.
With $N_v=N_f$, the mapping from vertices to faces discussed above become a one-to-one correspondence between
vertices and faces. As addressed in Sec.~\ref{sec:locality},
it is important to ensure that this correspondence is local. As a result, the local vertex-face correspondence arises
naturally, when we try to ensure the theory being nonsingular and the key properties of the Chern-Simons gauge theory is preserved.


\section{The Chern-Simons gauge theory on a tetrahedron}
\label{sec:tetra}

\begin{figure}[hbt]
\includegraphics[width=.68\linewidth]{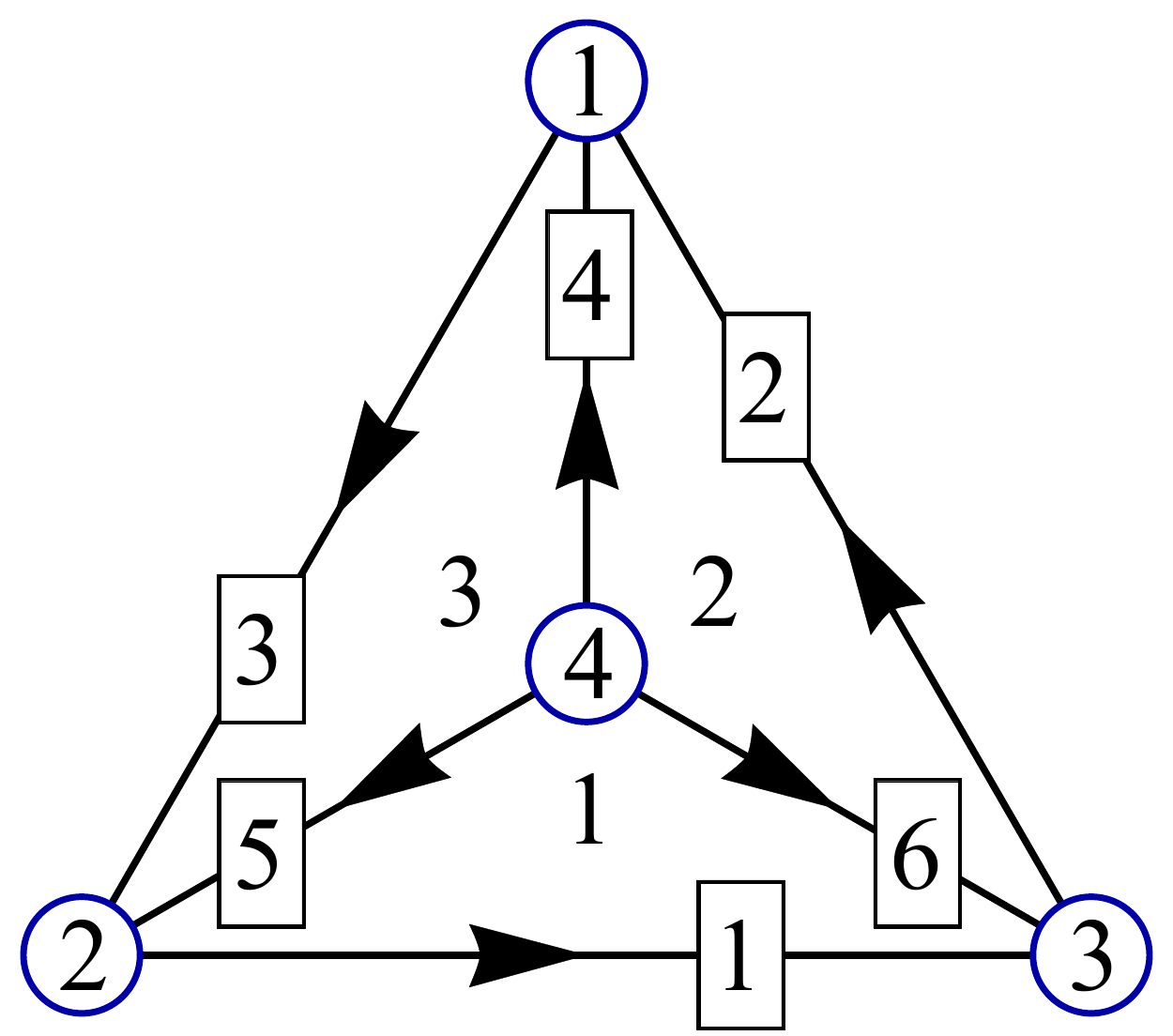}
\caption{(Color online) A tetrahedron viewing from the top. The circles represent the vertices, and the arrows are the edges (direction assigned). 
Here, we label each vertices, edges and faces using integers. The last face (i.e. the face $4$) is on the other side of the 
tetrahedron invisible from the current view point.}
\label{fig:tetrahedron}
\end{figure}

In this section, we demonstrate our generic theory by presenting a specific example, i.e. by discretizing the Chern-Simons gauge theory
on a tetrahedron. A tetrahedron is a planar graph defined on a manifold with $g=0$ (a sphere). In addition, it is easy to verify that 
a tetrahedron satisfies the cretieron presented in Sec.~\ref{sec:lvfc}, and thus a discretized Chern-Simons gauge theory can be 
constructed. It is also worthwhile to emphasize that a tetrahedron is self dual (i.e. the dual graph is also a tetrahedron). This is also the simplest setup
for discretizing the Chern-Simons gauge theory.

\subsection{The action}
We label the vertices, edges and faces of a tetrahedron as shown in Fig.~\ref{fig:tetrahedron}. In this convention, the
incident matrix $D_{v,e}$ [Eq.~\eqref{eq:D_matrix}] is
\begin{align}
D=
\left(\begin{matrix}
0 & +1 & -1 & +1 & 0 & 0  \\
-1 & 0 & +1 & 0 & +1 &0 \\
+1 & -1 & 0 & 0 & 0 & +1 \\
0 & 0 & 0 & -1 & -1 & -1 \\
\end{matrix}
\right)
\end{align}
and the $\xi_{f,e}$ matrix defined in Eq.~\eqref{eq:face_edges} is
\begin{align}
\xi=
\left(\begin{matrix}
+1 & 0 & 0 & 0 & +1 & -1  \\
0 & +1 & 0 & -1 & 0 & +1 \\
0 & 0 & +1 & +1 & -1 & 0 \\
-1 & -1 & -1 & 0 & 0 & 0 \\
\end{matrix}
\right)
\end{align}

We choose the local vertex-face correspondence such that vertices $1$, $2$, $3$ and $4$ pairs up with faces $2$, $3$, $4$ and $1$
respectively. Therefore,  the $M_{v,f}$ matrix is
\begin{align}
M=
\left(\begin{matrix}
0 & 1 &0 & 0  \\
0 & 0 &1 & 0 \\
0 & 0&0 & 1 \\
1 & 0 &0 & 0
\end{matrix}
\right)
\end{align}
Using this vertex-face correspondence, we can get the $K_{e,e'}$ matrix following the procedure described in Sec.~\ref{sec:action}, which is
\begin{align}
K=
\frac{1}{2}\left(\begin{matrix}
0 & +1 & +1 & 0 & -1 & -1  \\
-1 & 0 & -1 & +1 & 0 & -1  \\
-1 & +1 & 0 & -1 & +1 & 0  \\
0 & -1 & +1 & 0 & +1 & -1  \\
+1 & 0 & -1 & -1 & 0 & -1  \\
+1 & +1 & 0 & +1 & +1 & 0
\end{matrix}
\right)
\end{align}
Using these two matrices, the action can be written down as shown in Eq.~\eqref{eq:action}.

Here, we can verify easily that the $K$ matrix is invertible. In addition, it is straightforward to show that 
$M^{T}=M^{-1}$ and $K^{T}=-K$ (i.e. $M$ is an orthogonal matrix and $K$ is antisymmetric), in agreement
with the generic result proved above.

In addition, it is also straightforward to verify that the matrices satisfy the gauge invariance condition,
Eq.~\eqref{eq:gauge_invariance}, because $M \xi=D K^{T}$.

\subsection{Dual graph}

In the dual graph, it is straightforward to get the dual of the incident matrix and that of the $\xi$ matrix.
\begin{align}
D^*=
\left(\begin{matrix}
+1 & 0 & 0 & 0 & +1 & -1  \\
0 & +1 & 0 & -1 & 0 & +1 \\
0 & 0 & +1 & +1 & -1 & 0 \\
-1 & -1 & -1 & 0 & 0 & 0 \\
\end{matrix}
\right)
\end{align}
and
\begin{align}
\xi^*=
\left(\begin{matrix}
0 & -1 & +1 & -1 & 0 & 0  \\
+1 & 0 & -1 & 0 & -1 &0 \\
-1 & +1 & 0 & 0 & 0 & -1 \\
0 & 0 & 0 & +1 & +1 & +1 \\
\end{matrix}
\right)
\end{align}
By comparing the $D$ and $\xi^*$ ($\xi$ and $D^*$) matrices, we find that 
$D^*=\xi$ and $\xi^*=-D$, which verify Eqs.~\eqref{eq:dual_of_D} and~\eqref{eq:dual_of_xi}.

In the dual graph, if we use the same vertex-face correspondence, we get the $M^*$ matrix
\begin{align}
M^*=
\left(\begin{matrix}
0 & 0 & 0 & 1  \\
1 & 0 & 0 & 0 \\
0 & 1 & 0 & 0 \\
0 & 0 & 1 & 0
\end{matrix}
\right)
\end{align}
and the $K^*$ matrix is
\begin{align}
K^*=
\frac{1}{2}\left(\begin{matrix}
0 & +1 & +1 & 0 & -1 & -1  \\
-1 & 0 & -1 & +1 & 0 & -1  \\
-1 & +1 & 0 & -1 & +1 & 0  \\
0 & -1 & +1 & 0 & +1 & -1  \\
+1 & 0 & -1 & -1 & 0 & -1  \\
+1 & +1 & 0 & +1 & +1 & 0
\end{matrix}
\right)
\end{align}
Using the matrices $M^*$, $K^*$ and $\xi^*$, we can write down the action for the discretized Chern-Simons
gauge theory in the dual graph using Eq.~\eqref{eq:dual_action}.

Here, we can verify that $K^*=-K^{-1}$ and $M^*=M^{-1}$, as well as the gauge invariance condition
$M^* \xi^*= D^* (K^*)^{T}$, Eq.~\eqref{eq:gauge_invariance_dual}.

\section{Conclusions and Discussion}
\label{sec:discussion}

In this paper, we proved that the Chern-Simons gauge theory can be discretized for generic planar graphs 
on arbitrary 2D  closed orientable manifold as long as a local vertex-face correspondence can be defined on the graph. 
This condition is also necessary, if we want the theory to be nonsingular and to preserve some key properties of the 
Chern-Simons gauge theory.  In particular, we showed that the gauge invariance of the discretized theory requires that the vertex-face correspondence to be strictly enforced.

We also find a necessary and sufficient condition, which an be used to determine whether such a correspondence
can be defined on a particle graph or not, based on  the number of faces and vertices in this graph and its subgraphs.

The generalized discretized Chern-Simons gauge theory that we presented here has a number of interesting applications. One direction of further research is to consider the fractional quantum Hall effect on lattices, a problem that has not attracted much attention so far.\cite{Kol1993,Moller2009} A more general theory of the fractional quantum Hall effect on lattices is of interest in the context of fractionalized time-reversal breaking topological insulators so far as adiabatic continuity holds.\cite{Wu2012,Wu2015} These methods are also relevant to frustrated quantum antiferromagnets as we showed recently.\cite{Kumar-2014}

There are several open as yet unsolved issues. One is to relax somewhat the vertex-face correspondence. Since, as we showed, this  is required by gauge invariance, any violation of this correspondence is equivalent  to either the insertion of background static charges or background static fluxes. This viewpoint may offer a way to generalize this construction to other lattices (e.g.  triangular and  honeycomb) as well as to investigate the role of lattice topological defects such as dislocations and disclinations of time-reversal breaking fluids, including quantum Hall fluids, where the role of geometry has been focus of recent interest.\cite{Read2009,Hughes2011,Bradlyn2012,Hughes2013,Cho2014,You2014,Fremling2014,Bradlyn2015,Gromov2015,Geracie2015}

As a side comment, it is also worthwhile to note that two of the graphs shown in Fig.~\ref{fig:lattices_CS}
(the Kagome lattice and the dice lattice) belong to the family of {\it isostatic lattices}. The terminology of isostatic lattices is developed in
the study of mechanical stability transition~\cite{Maxwell1865} and recently, topologically nontrivial elastic modes are observed in
some of these isostatic systems, including protected zero-energy edges states, 
nontrivial topological indices and topological zero-energy solitons~\cite{Sun2012,Kane2014,Chen2014}.
Although the topological nature of those isostatic elastic systems are very 
different from a Chern-Simons gauge theory,
it is not an accident that same lattices arises in these two seemingly unrelated areas. 
As shown in Appendix (Sec.~\ref{app:isostatic}), 
the isostatic condition is closely related with (and slightly stronger than) the criterion for the existence of local
vertex-face correspondence, which is the fundamental reason why some lattices can be used for both studies.

\begin{acknowledgments}
The work was supported in part by the National Science Foundation, under grants PHY-1402971 at the University of Michigan (KS) and DMR-1408713 at the University of Illinois (EF), and and by the U.S. Department of Energy, Division of Materials Sciences under Awards No. 
DE-FG02-07ER46453 and  DE-SC0012368 through the Frederick
Seitz Materials Research Laboratory of the University of Illinois at Urbana-Champaign.
\end{acknowledgments}

\appendix

\section{Simple graph}
\label{app:simple}

\begin{figure}[hbt]
        \subfigure[]{\includegraphics[width=.48\linewidth]{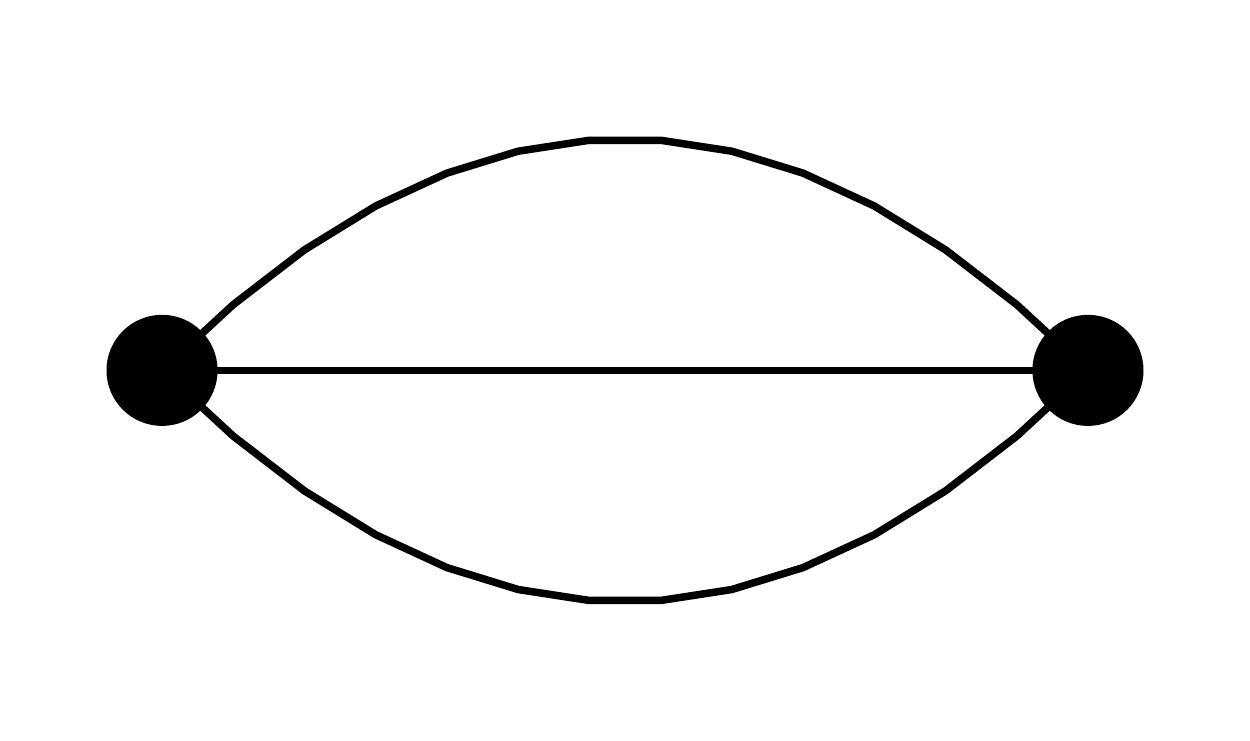}}
        \subfigure[]{\includegraphics[width=.48\linewidth]{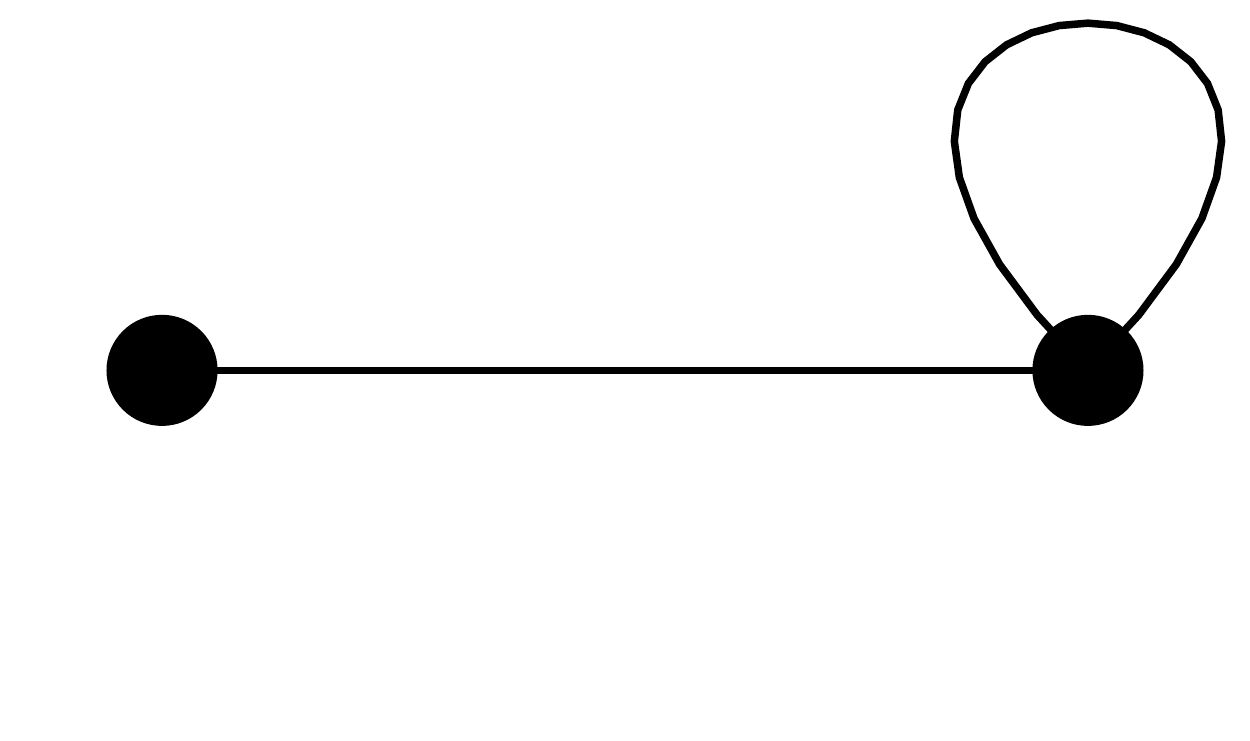}}
\caption{Examples of graphs that are not simple. The figures demonstrate situations that are not allowed
for a simple graph. Figure~(a), shows a pair of sites connected by three different edges. In Fig.~(b), one of the edges connects 
a site with itself (i.e. the two ends of a edge coincide).}
\label{app:fig:not_simple_graphs}
\end{figure}

Here, we demonstrate the definition of a simple graph by presenting situations that are not allowed in a simple graph,
as shown in Fig.~\ref{app:fig:not_simple_graphs}.

\section{Local vertex-face correspondence}
\label{app:criterion:lvfc}

In this section,we prove that criterion presented in Sec.~\ref{sec:lvfc} is a sufficient and necessary  condition for a graph to have
a local vertex-face correspondence by mapping this problem to {\it Hall's marriage problem}~\cite{Wilson1998}.

The {\it marriage problem} considers a finite set of girls, each of whom knows several boys, 
and the task is to find the sufficient and necessary condition, under which all the girls can marry the boys in such a way that each girl marries 
a boy she knows (marriage is assumed to be one-to-one here).
The solution to the {\it marriage problem} lies in {\it Hall's marriage theorem}, which states that a necessary and sufficient condition 
for the existence of such a matching is that each set of $k$ girls collectively knows at least $k$ boys.

Our goal here is to identify the sufficient and necessary for a graph satisfies the local-flux-attachment condition. 
And this problem can be mapped to the marriage problem by mapping faces into girls and vertices as boys. 
If the vertex $v$ is the vertex of the face $f$,
the corresponding boy and girl know each other. 
Under this mapping, the local-flux-attachment condition is exactly the marriage problem.

Here, we first prove that the criterion presented in Sec.~\ref{sec:lvfc} is a necessary condition. 
We consider a subgraph of $G$. By mapping to the marriage problem, all the faces in the subgraph forms a subset of girls, and the boys that they know
are included by the set of vertices of the subgraph, i.e. if the subgraph contains $N_f'$ faces (girls) and $N_v'$ vertices (boys), the
number of boys that these girls know is equal to or smaller than $N_v'$. Therefore, based on {\it Hall's marriage theorem}, 
we must have $N_v'\ge N_f'$ in every subgraph, if every face can marry a vertex that is adjacent to it. In other words,
this criterion is necessary for the existence of a local vertex-face correspondence.

We can also prove that the criterion is sufficient by considering subgraphs that satisfying the following condition: every vertex in the subgraph
is adjacent to at least one face of the subgraph. (This condition does not  hold for all subgraphs. For example, if a subgraph contains dangling bonds, 
the vertex located at the free end of a dangling bond is not adjacent to any faces in the subgraph). For these subgraphs (in which all vertices are 
adjacent to at least one face of the subgraph), the number of vertices ($N_v'$) equals to the number of boys that are known by the girls (faces) in the subgraph. And therefore, if the criterion in Sec.~\ref{sec:lvfc} is satisfied, the marriage theorem ensures immediately the existence
of (at least) one local vertex-face correspondence.

\section{Gauge symmetry}
\label{app:gauge_symmetry}

Here, we prove that Eq.~\eqref{eq:gauge_invariance} is the sufficient and necessary condition to maintain the gauge symmetry
in our theory [Eq.~\eqref{eq:action}]

First, we substitute the magnetic flux in Eq.~\eqref{eq:action} by Eq.~\eqref{eq:flux},
\begin{align}
S=\frac{k}{2\pi} \int dt (A_{v} M_{v,f}  \xi_{f,e} A_e-\frac{1}{2} A_{e_i} K_{e,e'} \dot{A}_{e'})
\label{app:eq:action_1}
\end{align}
Under the gauge transformation
\begin{align}
A_{v} \to &  A_{v} - \partial_0 \phi_{v} 
\label{app:eq:gauge_transformation_A_v}\\
A_{e} \to & A_{e} - D_{v,e} \phi_{v}
\label{app:eq:gauge_transformation_A_e}
\end{align}
the action Eq~\eqref{app:eq:action_1} is transfered to
\begin{widetext}
\begin{align}
S\to S+\frac{k}{2\pi} \int dt \left(-\dot \phi_{v} M_{v,f}  \xi_{f,e} A_e
+\frac{1}{2} D_{v,e} \phi_{v} K_{e,e} \dot{A}_{e'}+\frac{1}{2} A_{e} K_{e,e'} D_{v,e'}\dot{\phi}_{v}\right)
-\frac{k}{4\pi} \int dt \left(D_{v,e} \phi_{v} K_{e,e'} D_{v',e'}\dot{\phi}_{v'}\right)
\end{align}
\end{widetext}
Here, the second term on the r.h.s. is linear in $\phi$, while the last term is $O(\phi^2)$. In order
to preserve the gauge symmetry, we need both these two terms to vanish, i.e.
\begin{align}
\int dt \left(\dot \phi_{v} M_{v,f}  \xi_{f,e} A_e
-A_{e} K_{e,e'} D_{v,e'}\dot{\phi}_{v}\right)=0
\label{app:eq:gauge_symmery_1}
\\
\int dt \left(D_{v,e} \phi_{v} K_{e,e'} D_{v',e'}\dot{\phi}_{v'} \right)=0
\label{app:eq:gauge_symmery_2}
\end{align}
In Eq.~\eqref{app:eq:gauge_symmery_1}, we used the fact that
\begin{align}
\int dt  D_{v,e} \phi_{v} K_{e,e'} \dot{A}_{e'}=\int dt A_{e} K_{e,e'} D_{v,e'}\dot{\phi}_{v}
\end{align}
which can be proved via integrating by part and realizing that the $K$ matrix is antisymmetric.

Equations~\eqref{app:eq:gauge_symmery_1} and~\eqref{app:eq:gauge_symmery_2} imply that 
\begin{align}
M_{v,f}  \xi_{f,e}=K_{e,e'} D_{v,e'}
\label{app:eq:gauge_symmery_3}
\\
D_{v,e} K_{e,e'} D_{v',e'}=0
\label{app:eq:gauge_symmery_4}
\end{align}
These two conditions are not independent to each other. In fact, Eq.~\eqref{app:eq:gauge_symmery_3}
automatically implies Eq.~\eqref{app:eq:gauge_symmery_4}.
This can be realized by noticing that according to Eq.~\eqref{app:eq:gauge_symmery_3}, we have
\begin{align}
D_{v,e} K_{e,e'} D_{v',e'}=D_{v,e}M_{v',f}  \xi_{f,e}.
\end{align}
The r.h.s. of this equation is zero because $D_{v,e} \xi_{f,e}=0$, and thus Eq.~\eqref{app:eq:gauge_symmery_4}
arises automatically.

Here, we explain why $D_{v,e} \xi_{f,e}=0$. For any fixed $f$, $\xi_{f,e}$ represent an loop in the graph.
If $v$ is not a vertex on this loop, $D_{v,e} \xi_{f,e}$ must vanish, because $D_{v,e}=0$. 
If the loop paths through $v$, there must be two edges along these loop that are connected to $v$, which we will
call $e_1$ and $e_2$. It is easy to realize that according to the definition of $\xi$ and $D$,
$D_{v,e_1} \xi_{f,e_1}=-D_{v,e_2} \xi_{f,e_2}$ (here, we don't sum over repeated indices $e_1$ and $e_2$). 
Therefore, the contributions to $D_{v,e} \xi_{f,e}$ cancels out, i.e.  $D_{v,e} \xi_{f,e}=0$.
This relation can also be written in a matrix form and the same is true for the dual graph
\begin{align}
D \xi^T=D^*(\xi^*)^T=0
\label{app:eq:D_xi_matrix}
\end{align}
which will be used below in Sec.~\ref{app:sec:dual}. Here $\xi^T$ represents the transpose matrix of $\xi$.

\section{The directions of edges}
\label{app:edge_direction}
In this section, we prove that the condition of gauge invariance [Eq.~\eqref{eq:gauge_invariance}]
is independent of the choice on the edge directions.

As shown in the main text, we assign a direction for each edge in order to define the vector potential on a graph.
These directions can be assigned in arbitrary ways and the choice of directions will not have any impact for 
any physics properties. For the condition of gauge invariance [Eq.~\eqref{eq:gauge_invariance}], 
this statement is also true.

To prove this statement, we flip the direction of an arbitrary edge $e_0$ and consider two different situations $e_0=e$
and $e_0 \ne e$.

If $e_0=e$, as we flip the direction assigned to the edge $e_0$, the l.h.s. of Eq.~\eqref{eq:gauge_invariance} changes
sign, because $M_{v,f}\to M_{v,f}$ and $\xi_{f,e_0} \to -\xi_{f,e_0}$. The r.h.s. of the equation also flips sign, since
$K_{e_0,e'} \to -K_{e_0,e'}$ and $D_{v,e'}\to D_{v,e'}$. Because both sides of the equation flips sign when we flip the direction
of $e_0$, the equation remains invariant and thus is independent of the choice of the direction of $e_0$

If $e\ne e_0$, the l.h.s. of Eq.~\eqref{eq:gauge_invariance} remains invariant, because 
neither $M_{v,f}$ nor $\xi_{f,e}$ relies on the direction of $e_0$. For the r.h.s., because both 
$K_{e,e_0}$ and $D_{e_0,v}$ flip signs as we flip the direction of $e_0$, their product remains the same.
As a result, the equation is again independent of the direction of $e_0$. 

This conclusion implies that in order to prove Eq.~\eqref{eq:face_edges}, it is sufficient to verify 
the formula for just one specific choice of edge directions.


\section{Lattice duality}
\label{app:sec:dual}
In this section, we prove that the discretized Chern-Simons theory on the original lattice and the dual lattice
are dual to each other by coupling the Chern-Simons gauge theory with gauge fields on the dual lattice.
\begin{align}
S=S_{\textrm{CS}}+S_{\textrm{coupling}}
\end{align}
Here, the first term is our discrete Chern-Simons gauge theory
\begin{align}
S_{\textrm{CS}}=&\frac{k}{2\pi} \int dt \left(A_{v} M_{v,f} \Phi_{f}-\frac{1}{2} A_{e} K_{e,e'} \dot{A}_{e'}\right)
\end{align}
and the second term couples the Chern-Simons field $A$ with gauge fields on the dual graph $a^*$
\begin{align}
S_{\textrm{coupling}}=\int \frac{dt}{2\pi} (&\xi^*_{f^*,e^*}a^*_{e^*} A_v\delta_{f^*,v}+D^*_{v^*,e^*}a^*_{v^*}A_e \delta_{e,e^*}
\nonumber\\&-\partial_0 a^*_{e^*} A_e \delta_{e,e^*})
\label{eq:gauge_coupling}
\end{align}
As will be shown in  Sec.~\ref{app:sub:sec:gauge_dual_lattice}, this coupling is gauge invariant.

Below, in Sec.~\ref{app:sub:sec:gauge_dual_lattice}, we first prove that same as in the continuum, the dual gauge field $a^*$ 
can be used to describe the charge and current on the original lattice and we will also show that $S_{\textrm{coupling}}$ 
is gauge invariant.  Then, in Sec.~\ref{app:sub:sec:dual}, we show that by integrating out the $A$ field, 
the dual theory is obtained, which matches exactly the discrete Chern-Simons field on the dual lattice (but with a different coupling constant $k^*=-1/k$).
Because our action describes a quadratic theory, this calculation is exact.

\subsection{Gauge field on the dual lattice}
\label{app:sub:sec:gauge_dual_lattice}
Same as in the continuum, we can consider the dual gauge field $a^*$ (defined on the dual lattice) as a description for the charge and current 
on the original lattice. Here, the charge that resides at each vertex is called $\rho_v$ and the current on each edge is labeled as $j_e$.
\begin{align}
\rho_v=&\frac{1}{2\pi}\xi^*_{f^*,e^*}a^*_{e^*}
\label{eq:charge}
\\
j_e=&\frac{1}{2\pi} \left(D^*_{v^*,e^*}a^*_{v^*}-\partial_0 a^*_{e^*}\right)
\label{eq:current}
\end{align}
In Eq.~\eqref{eq:charge}, we choose $f^*=v$, and for Eq.~\eqref{eq:current}, $e=e^*$.
It is easy to verify that the charge and current are gauge invariant and satisfy the charge conservation law (i.e. the continuity equation)
\begin{align}
\partial_t \rho_{v}-D_{v,e} j_e=0
\end{align}
where $\partial_t$ is the time derivative and $D_{v,e}$ (the incident matrix) plays the role of a discretized divergence (multiplied by -1).

To prove the gauge invariance, we perform the gauge transformation for $a^*$
\begin{align}
a^*_{v^*} \to & a^*_{v^*}-\partial_0 \phi^*_{v^*}
\label{eq:gauge_transfer_a_star_v}
\\
a^*_{e^*} \to & a^*_{e^*}-D^*_{v^*,e^*} \phi^*_{v^*}
\label{eq:gauge_transfer_a_star_e}
\end{align}
Under this transformation, $\rho_v$ and $j_e$ are invariant
\begin{align}
\rho_v \to& \rho_v - \frac{1}{2\pi}\xi^*_{f^*,e^*} D^*_{v^*,e^*} \phi^*_{v^*}=\rho_v
\label{eq:charge_gauge}
\\
j_e \to & j_e-\frac{1}{2\pi}\left(D^*_{v^*,e^*} \partial_0 \phi^*_{v^*}-\partial_0 D^*_{v^*,e^*} \phi^*_{v^*}\right)=j_e
\label{eq:current_gauge}
\end{align}
For Eq.~\eqref{eq:charge_gauge}, 
we used the fact  $\xi^*_{f^*,e^*} D^*_{v^*,e^*}=0$. This relation is proved in Eq.~\eqref{app:eq:D_xi_matrix} and it is a discretized version 
of the formula $\nabla\times\nabla=0$. 
In Eq.~\eqref{eq:current_gauge}, $D^*_{v^*,e^*} \partial_0 \phi^*_{v^*}$ and $-\partial_0 D^*_{v^*,e^*} \phi^*_{v^*}$
cancels out because the incident matrix  $D^*_{v^*,e^*}$ is time-independent and thus commute with the time derivative $\partial_0$.
 
The continuity equation can be proved using the following two equations 
\begin{align}
\partial_0 \rho_v&= \frac{1}{2\pi} \xi^*_{v,e^*} \partial_0 a^*_{e^*}
\\
-D_{v,e} j_e&= \frac{1}{2\pi}\xi^*_{f^*,e^*}(D^*_{v^*,e^*} a^*_{v^*}-\partial_0 a^*_{e^*})
=-\frac{1}{2\pi}\xi^*_{f^*,e^*}\partial_0 a^*_{e^*}
\end{align} 
In the second equation here, we used the fact that $D_{v,e}=-\xi^*_{f^*,e^*}$ [Eq.~\eqref{eq:dual_of_xi}] and $\xi^*_{f^*,e^*}D^*_{v^*,e^*}=0$ [Eq.~\eqref{app:eq:D_xi_matrix}].
By adding the two questions together, the continuity equation is obtained. 

Using Eqs.~\eqref{eq:charge} and~\eqref{eq:current}, the coupling between the $A$ and $a^*$ fields [i.e. Eq.~\eqref{eq:gauge_coupling}] 
can be rewritten as
\begin{align}
S_{\textrm{coupling}}=\int dt \left(\rho_v A_v+j_e A_e\right).
\label{app:eq:coupling:density_charge}
\end{align}
Because both $\rho_v$ and $j_e$ are gauge invariant, the coupling term must also be gauge invariant.

The coupling $S_{\textrm{coupling}}$ is also invariant under gauge transformation
\begin{align}
A_{v} \to &  A_{v} - \partial_0 \phi_{v} 
\\
A_{e} \to & A_{e} - D_{v,e} \phi_{v}
\end{align}
Using Eq.~\eqref{app:eq:coupling:density_charge}, we find that the gauge transformation turns $S_{\textrm{coupling}}$ into
\begin{align}
S'_{\textrm{coupling}}&=S_{\textrm{coupling}}-\int dt \left(\rho_v  \partial_0 \phi_{v} +j_e  D_{v,e} \phi_{v}\right)
\nonumber\\
&=S_{\textrm{coupling}}+\int dt \left( \partial_0\rho_v-j_e  D_{v,e}\right) \phi_{v}
\end{align}
After a integration by part (for $t$), the last term in this formula vanishes
due to the continuity equation, and thus $S_{\textrm{coupling}}$ is gauge invariant.

\subsection{Duality transformation}
\label{app:sub:sec:dual}

In the path integral approach, a gauge fixing term $S_{\textrm{gauge fixing}}$ needs to be introduced
\begin{align}
S=S_{\textrm{CS}}+S_{\textrm{coupling}}+S_{\textrm{gauge fixing}}
\label{eq:cs_with_gauge_fixing}
\end{align}
Without loss of generality, here we choose
\begin{align}
S_{\textrm{gauge fixing}}=\frac{\alpha}{2} \int \frac{dt}{2\pi} \left(\frac{d A_{v}}{dt}\frac{d A_{v}}{dt}\right)
\end{align}
In the frequency space, the action of Eq.~\eqref{eq:cs_with_gauge_fixing} takes the following form
\begin{widetext}
\begin{align}
S=&S_{CS}+S_{\textrm{coupling}}+S_{\textrm{gauge fixing}}\\
=&\frac{k}{2}\sum_{\omega}
\left(\begin{matrix}
\mathbf{A_{v}}(\omega) & \mathbf{A_{e}}(\omega)
\end{matrix}
\right)
\left(\begin{matrix}
\alpha \omega^2 /k &  M \xi \\
(M \xi)^T & -i \omega K
\end{matrix}
\right)
\left(\begin{matrix}
\mathbf{A_{v}}(-\omega)\\
\mathbf{A_{e}}(-\omega)
\end{matrix}
\right)
+\sum_{\omega} 
\left(\begin{matrix}
\mathbf{a^*_{v^*}}(\omega) & \mathbf{a^*_{e^*}}(\omega)
\end{matrix}
\right)
\left(\begin{matrix}
0 & D^* \\
(\xi^*)^T & i \omega
\end{matrix}
\right)
\left(\begin{matrix}
\mathbf{A_{v}}(-\omega) \\ 
\mathbf{A_{e}}(-\omega)
\end{matrix}
\right)
\label{eq:action_matrix}
\end{align}
Here, we write the Lagrangian as block matrices. 
Bold letters in this equation are vectors. For example,
$\mathbf{a^*_{e^*}}$ represents a $N_{e^*}$-dimensional vector, whose components are 
$a^*_{e^*}$ on each edge. The same is true for $\mathbf{a^*_{v^*}}$, $\mathbf{A_{e}}$, or $\mathbf{A_{v}}$.
The first matrix in the equation above contains $S_{\textrm{CS}}$ and $S_{\textrm{gauge fixing}}$, while the second matrix
is for $S_{\textrm{coupling}}$.

By integrating out the $A$ field, we obtain a dual gauge theory for $a^*$ on the dual graph. For a
quadratic theory as shown above, this can be done exactly.
\begin{align}
S=-\frac{1}{2k}\sum_{\omega} 
\left(\begin{matrix}
\mathbf{a^*_{v^*}}(\omega) & \mathbf{a^*_{e^*}}(\omega)
\end{matrix}
\right)
\left(\begin{matrix}
0 & D^* \\
(\xi^*)^T & i \omega
\end{matrix}
\right)
\left(\begin{matrix}
\alpha \omega^2 /k &  M \xi \\
(M \xi)^T & -i \omega K
\end{matrix}
\right)^{-1}
\left(\begin{matrix}
0 & \xi^* \\
(D^*)^T & -i \omega
\end{matrix}
\right)
\left(\begin{matrix}
\mathbf{a^*_{v^*}}(-\omega) \\
\mathbf{a^*_{e^*}}(-\omega)
\end{matrix}
\right)
\end{align}
The inverse matrix in the equation above can be computed using the technique of {\it blockwise inversion} 
\begin{align}
\left(\begin{matrix}
A & B \\
C & D
\end{matrix}
\right)^{-1}
=\left(\begin{matrix}
(A-B D^{-1} C)^{-1}& -(A-B D^{-1} C)^{-1} B D^{-1} \\
-D^{-1} C (A-B D^{-1} C)^{-1} & D^{-1}+ D^{-1} C (A-B D^{-1} C)^{-1}B D^{-1}
\end{matrix}
\right)
\end{align}
where $A$, $B$, $C$ and $D$ are matrix sub-blocks. For our matrix inverse, the block $A$ is
an identity matrix (multiply by a real number $\alpha \omega^2/k$), and we have
$B=C^T=M \xi$ and the block $D$ is $-i \omega K$. Using the commutation relation Eq~\eqref{eq:path_commutator1}
and the fact that this commutator is zero for two contractible loops, it is easy to show that 
\begin{align}
\xi (K)^{-1} (\xi)^{T}=0
\label{app:eq:xi_K_inverse_xi}
\end{align} 
i.e., $B D^{-1} C=0$. Therefore, we find that 
\begin{align}
\left(\begin{matrix}
\alpha \omega^2 /k &  M \xi \\
(M \xi)^T & -i \omega K
\end{matrix}
\right)^{-1}
=\left(\begin{matrix}
\frac{k}{\alpha \omega^2} & -\frac{i k}{\alpha \omega^3} M \xi  K^{-1} \\
-\frac{i k}{\alpha \omega^3} K^{-1} \xi^T M^T   & -\frac{1}{i \omega}K^{-1}
- \frac{k}{\alpha \omega^4} K^{-1} 
\xi^T M^T  M \xi K^{-1} 
\end{matrix}
\right)
\end{align}
Now, we will use Eq.~\eqref{eq:gauge_invariance_again}, which tells that
\begin{align}
M \xi K^{-1}=-D=\xi^*\\
K^{-1} \xi^T M^T  =D^T=-(\xi^*)^T
\end{align}
Here, we also used the fact that $K$ is an anti-symmetric matrix and Eq.~\eqref{eq:dual_of_xi} ($D=-\xi^*$).
Using these two relations, we find that
\begin{align}
\left(\begin{matrix}
\alpha \omega^2 /k &  M \xi \\
(M \xi)^T & -i \omega K
\end{matrix}
\right)^{-1}
=\left(\begin{matrix}
\frac{k}{\alpha \omega^2} & -\frac{i k}{\alpha \omega^3} \xi^*\\
\frac{i k}{\alpha \omega^3} (\xi^*)^T   & -\frac{1}{i \omega}K^{-1}
+\frac{k}{\alpha \omega^4}  (\xi^*)^T \xi^*
\end{matrix}
\right)
\end{align}
And therefore,
\begin{align}
S=&-\frac{1}{2 k}\sum_{\omega} 
\left(\begin{matrix}
\mathbf{a^*_{v^*}}(\omega) & \mathbf{a^*_{e^*}}(\omega)
\end{matrix}
\right)
\left(\begin{matrix}
0 & D^* \\
(\xi^*)^T & i \omega
\end{matrix}
\right)
\left(\begin{matrix}
\frac{k}{\alpha \omega^2} & -\frac{i k}{\alpha \omega^3} \xi^*\\
\frac{i k}{\alpha \omega^3} (\xi^*)^T   & -\frac{1}{i \omega}K^{-1}
+\frac{k}{\alpha \omega^4}  (\xi^*)^T \xi^*
\end{matrix}
\right)
\left(\begin{matrix}
0 & \xi^* \\
(D^*)^T & -i \omega
\end{matrix}
\right)
\left(\begin{matrix}
\mathbf{a^*_{v^*}}(-\omega) \\
\mathbf{a^*_{e^*}}(-\omega)
\end{matrix}\right)
\nonumber\\
=&-\frac{1}{2 k}\sum_{\omega} 
\left(\begin{matrix}
\mathbf{a^*_{v^*}}(\omega) & \mathbf{a^*_{e^*}}(\omega)
\end{matrix}
\right)
\left(\begin{matrix}
0  &  -\frac{1}{i \omega}D^* K^{-1}\\
0 &  -K^{-1}
\end{matrix}
\right)
\left(\begin{matrix}
0 & \xi^* \\
(D^*)^T & -i \omega
\end{matrix}
\right)
\left(\begin{matrix}
\mathbf{a^*_{v^*}}(-\omega) \\
\mathbf{a^*_{e^*}}(-\omega)
\end{matrix}\right)
\nonumber\\
=&-\frac{1}{2 k}\sum_{\omega} 
\left(\begin{matrix}
\mathbf{a^*_{v^*}}(\omega) & \mathbf{a^*_{e^*}}(\omega)
\end{matrix}
\right)
\left(\begin{matrix}
0 &  D^* K^{-1}\\
-K^{-1} (D^*)^T  & i \omega K^{-1}
\end{matrix}
\right)
\left(\begin{matrix}
\mathbf{a^*_{v^*}}(-\omega) \\
\mathbf{a^*_{e^*}}(-\omega)
\end{matrix}\right)
\end{align}
Here, we used the fact that $D^*(\xi^*)^T=0$ [Eq.~\eqref{app:eq:D_xi_matrix}] and $D^* K^{-1}  (D^*)^T=\xi K^{-1}  \xi^T=0$ 
[Eqs.~\eqref{eq:dual_of_D} and~\eqref{app:eq:xi_K_inverse_xi}].

Because $K^{-1}=-K^*$ and $D^* K^*=-M^* \xi^*$, we get
\begin{align}
S=&-\frac{1}{2 k}\sum_{\omega} 
\left(\begin{matrix}
\mathbf{a^*_{v^*}}(\omega) & \mathbf{a^*_{e^*}}(\omega)
\end{matrix}
\right)
\left(\begin{matrix}
0 &  -D^* K^{*}\\
 -(D^*K^{*})^T  & -i \omega K^{*}
\end{matrix}
\right)
\left(\begin{matrix}
\mathbf{a^*_{v^*}}(-\omega) \\
\mathbf{a^*_{e^*}}(-\omega)
\end{matrix}
\right)
\nonumber\\
=&-\frac{1}{2 k}\sum_{\omega} 
\left(\begin{matrix}
\mathbf{a^*_{v^*}}(\omega) & \mathbf{a^*_{e^*}}(\omega)
\end{matrix}
\right)
\left(\begin{matrix}
0 &  M^* \xi^{*}\\
 (M^*\xi^{*})^T  & -i \omega K^{*}
\end{matrix}
\right)
\left(\begin{matrix}
\mathbf{a^*_{v^*}}(-\omega) \\
\mathbf{a^*_{e^*}}(-\omega)
\end{matrix}
\right)
\nonumber\\
\end{align}
\end{widetext}
By transferring from the frequency space $\omega$ back to time $t$, we find
\begin{align}
S=\frac{-1/k}{2\pi} \int dt \left(a^*_{v^*} M^*_{v^*, f^*} \Phi^*_{f^*}-\frac{1}{2} a^*_{e^*} K^*_{e^*,e'^*} \dot{a}^*_{e'^*}\right)
\end{align}
This is exactly our discrete Chern-Simons gauge theory defined on the dual graph [Eq.~\eqref{eq:dual_action}] with topological index $k^*=-1/k$.

\section{Edge space,  circuit-subspace and the cut-subspace}
\label{app:edge_space}
In this section, we prove that the edge space is the direct sum of the circuit-subspace and the cut-subspace.
Although this conclusion applies generically to planar and non-planar graphs, we will only discuss planar graphs here
for simplicity, since the manuscript only consider planar ones.

We first prove that the circuit-subspace and the cut-subspace are orthogonal to each other. This can be verified easily 
by noticing that vectors from these two spaces ($\xi_{C}$ and $\xi_{H}$) are orthogonal to each other (i.e. their dot product is zero)
\begin{align}
\xi_{C,e},\xi_{H,e}=0.
\end{align}

In addition, we can prove that the dimension of the circuit-subspace plus the dimension of the cut-subspace coincides
with the dimension of the edge space. Combined with the orthogonality proved above, this conclusion implies that 
the direct sum of the circuit-subspace and the cut-subspace is the edge space.

As mentioned in the main text, the basis of the circuit-subspace can be formed by all independent (contractible and non-contractible) cycles 
of a graph. For a planar graph with $N_f$ faces defined on a manifold with genus $g$, there are $N_f-1$ independent
contractible cycles and $2g$ independent non-contractible cycles, i.e. $N_f-1+2g$ independent loops in total. 
Therefore, the dimensionality of the of circuit-subspace is $N_f-1+2g$. 

For a planar graph $G$, cutsets corresponds to contractible cycles in the dual graph $G^*$. 
Because the dual graph has $N_v$ faces, same as the number of vertices in the original graph, 
the number of independent cutsets (i.e. the number of independent contractible cycles in the dual graph) is $N_v-1$.

If we add the dimensions of the cut subspace and the circuit subspace together, we get $N_f+N_v-2+2g$, which coincides with
the number of edges $N_e$, i.e. the dimension of the edge space. Here we utilized the fact that a closed orientable surface 
with genus $g$, the Euler characteristic is $2-2g$
\begin{align}
N_v-N_e+N_f=2-2g.
\end{align}

Because the circuit-subspace and the cut-subspace are two orthogonal subspaces of the edge space,
and the total dimensions of these two subspaces match the dimensions of the edge space,
we proved that the direct sum of these two subspace is the edge space.

\section{the Determinant of  a block matrix}
\label{app:zero_det}
Here, we consider a $(N+M)\times (N+M)$ matrix
\begin{align}
\mathcal{M}=\left(\begin{matrix}
0 & C \\
D & B
\end{matrix}
\right)
\label{app:eq:block_matrix}
\end{align}
where each letter in the matrix represents a block matrix and the $0$ ($B$) matrix has dimensions $N\times N$ ($M\times M$).
We will prove below that the determinant of this matrix is zero when the size of the $0$ block is larger than the $B$ block (i.e. $N>M$).

First, we define a set of $N+M$ vectors $e_i=(0,0,\ldots,1,\ldots,0)$, such that only the $i$th component of the vector 
$e_i$ is nonzero, while $i=1,2,\ldots, N+M$. Here, these vectors span a $N+M$-dimensional linear space. This linear space is the direct sum of two subspace $V_1\oplus V_2$, where $V_1$ is spanned by the vectors $e_i$ with $1\le i \le N$ and $V_2$ by those with $N+1\le i \le N+M$, and it is easy to check that
$V_1$ and $V_2$ are orthogonal to each other.

Because the upper-left block of the matrix only contains zeros, $e_i \mathcal{M} e_j=0$ for $i\le N$ and $j \le N$. It implies that for any vector $e_j$
with $j\le N$, the vector $\mathcal{M} e_j$ is orthogonal to any vectors in the subspace $V_1$. In other words, $\mathcal{M} e_j$ is a vector in the subspace $V_2$.
For $j=1,2,\ldots,N$,  $\mathcal{M} e_j$ generates $N$ vectors in the space of $V_2$. If $N>M$, some of these $N$ vectors must be linearly dependent,
because the space $V_2$ can only have $M$ independent vectors.
Therefore, we can construct (at least) one zero vector using these $N$ vectors
$\mathcal{M} e_j$:
\begin{align}
\sum_{j=1}^N a_j \mathcal{M} e_j=0
\end{align}
where $a^j$ are $N$ numbers.
This implies immediately that the $\mathcal{M}$ matrix has (at least) one null vector $v=\sum_{j=1}^N a_j e_j$
\begin{align}
\mathcal{M} v=\sum_{j=1}^N  \mathcal{M} a_j e_j=0
\end{align}
Having a zero eigenvalue implies that the determinant of $\mathcal{M}$ must be zero.


\section{isostatic condition, elasticity and local vertex-face correspondence}
\label{app:isostatic}
In this section, we reveal the connection between the isostatic condition and the criterion for the existence of a
 local vertex-face correspondence.

The idea of isostaticity plays an important role in the study of mechanics stability. It comes from the counting argument developed by Maxwell~\cite{Maxwell1865}. 
If we construct an elastic system by connecting beads with rigid rods (and allow the rods to rotate freely around each joint), the rigidity
of the system can be determined by comparing the total number of constrains and the total number of degrees of freedom.
In 2D, the total number of degrees here is 2 times the number of beads, because each bead has two degrees of freedom in 2D, 
while the number of contains is the number of rods, since each rod enforces one constrain by fixing the distance between two beads.
If we consider such a system as a graph (i.e. beads as vertices and rods as edges), the number of degrees of freedom is $2 N_v$, while the
number of constrain is $N_e$. The isostatic condition requires these two numbers to coincide. If
all the constrains are independent (i.e. no redundancy), this condition represents the verge of a mechanical stability (i.e. a phase transition point).
If we add/remove one edge (rod) to the system, the system becomes stable/floppy. The rigorous formula for the isostatic condition in 2D is
\begin{align}
2N_v= N_e+3
\end{align}
Here, a extra number $3$ is introduced to the r.h.s. to represent the trivial global degrees of freedom
(two translations and one rotation), which will alway arise. Using the Euler characteristic ($N_v-N_e+N_f=2-2g$), 
we can rewrite the isostatic condition as
\begin{align}
N_v=N_f+1+2g,
\end{align}
where $g$ is the genus of the underlying manifold.
In the thermal dynamic limit ($N_v\to \infty$ and $N_f\to \infty$), we can ignore the finite part $1+2g$ 
and therefore, the condition coincides with the our requirement of $N_v=N_f$.

In addition, for an isostatic system, to ensure that there is no redundant constrains, one shall require that
for any subsystem (subgraph), the total number of degrees of freedom always exceeds (or equal to) the number of constrains (plus three)
\begin{align}
2N'_v\ge N'_e+3
\end{align}
where $N'_v$ and $N'_e$ are number of vertices and edges in a subgraph, while $3$ on the r.h.s. comes from global
translations and rotations.
If a subgraph has the topology of a disk (i.e. the Euler characteristic  is $N_v-N_e+N_f=1$),
we can rewrite the condition as
\begin{align}
N'_v\ge N'_f+2
\end{align}
which is very similar to but slightly stronger than our criterion of local vertex-face correspondence ($N'_v\ge N'_f$).

Because our criterion is slightly weaker, some of the lattices/graph that are not isostatic can still be used to
construct a discretized Chern-Simons gauge theory, e.g., Fig.~\ref{fig:honeycomb1}.


\begin{thebibliography}{64}
\bibitem [{not()}]{note_orientable}%
  \BibitemOpen
  \href@noop {} {}\bibinfo {note} {For 2D orientable surfaces in 3D Euclidean
  space, an orientation can be chosen using the right-hand rule to define a
  ``clockwise'' direction for loops in the surface. This orientation (i.e.
  clockwise or counter-clockwise) cannot be globally defined for non-orientable
  surfaces, e.g. a M\"{o}bius stripe. For the Chern-Simons gauge theory,
  orientation is required, and thus we only study orientable
  surfaces.}\BibitemShut {Stop}%
\end{thebibliography}
%

\end{document}